# Bayesian Optimization for Self-Driving Materials Laboratories: From Algorithms to Physics-Informed Workflows

Yuki K. Wakabayashi[*a] and Takuma Otsuka[*b]

[a]Basic Research Laboratories, NTT, Inc., 3-1 Morinosato Wakamiya, Atsugi, Kanagawa 243-0198, Japan

[b]Communication Science Laboratories, NTT, Inc., 2-4 Hikaridai, Seika, Soraku, Kyoto 619-0237, Japan

*E-mail: yuuki.wakabayashi@ntt.com, takuma.otsuka@ntt.com

**Abstract**

Self-driving laboratories (SDLs) are transforming materials research by closing the loop among synthesis, characterization, data analysis and experimental decision making. Bayesian optimization (BO) is a decision engine for these loops because it can select experiments from scarce and noisy data while balancing exploitation of promising regions with exploration of uncertain ones. Yet real materials campaigns frequently deviate from the standard black-box optimization setting. They involve failed syntheses, missing or delayed measurements, equipment drift, mixed continuous and categorical variables, hard and uncertain constraints, multiple objectives, variable cost and fidelity, the reuse of historical data from related experiments, batch and asynchronous experimentation, and prior physics knowledge. This review presents BO for materials SDLs through the lens of these practical challenges. We first summarize Gaussian-process-based BO and the formulation of materials goals as quantitative objectives. We then discuss major BO design choices in surrogate modelling and acquisition to address these challenges. Particular emphasis is placed on physics-informed Bayesian optimization (PIBO), in which domain knowledge is incorporated into the optimization loop through representations, priors, kernels, acquisition functions, and constraints. We further survey representative achievements enabled by BO and related active-learning approaches across semiconductors, catalysis, chemical reactions, batteries, alloys, other functional materials and quantum materials, highlighting how these approaches have contributed beyond parameter optimization to substantive advances in materials science, including the realization of new materials and synthesis routes, improvements in functional performance, and the generation of reusable scientific knowledge. We conclude by outlining open problems that will shape the next generation of BO-driven materials SDLs, including nonstationarity, multimodal observations, adaptive problem formulation and scientific reasoning by humans, large language models and research agents. Addressing these challenges may advance SDLs beyond efficient optimization toward interpretable and knowledge-generating experimentation.

## 1. Introduction

Many central problems in materials and nanoscale science can be viewed as optimization problems in which the relevant objectives are expensive, noisy and only partially observable. In practice, a synthesis temperature may change phase purity, defect chemistry and morphology;[1,2] a precursor or gas-flow ratio may alter stoichiometry, carrier density or reaction selectivity;[3,4] and the choice of solvent or additive, heat-treatment conditions, or other post-processing steps may improve one desirable property while degrading another.[5,6] In thin-film and inorganic materials, for example, annealing can improve crystallinity or carrier mobility while also promoting elemental volatility, interdiffusion, surface roughening or interface degradation.[7–9] Such competing effects create trade-offs among desirable properties and make optimization more complex than following a single monotonic route to improvement. In solution-processed materials, catalysts, batteries, polymers, alloys and thin films, the possible combinations of compositions, recipes and process histories rapidly become too numerous to explore exhaustively. As a result, obtaining a desired material or device within a limited number of trials and under constraints on time and resources has traditionally depended strongly on domain knowledge, physical insight, empirical experience and the intuition of individual researchers and engineers. These materials-optimization challenges create a practical need for experimental decision-making frameworks that can learn from very few experiments, handle noise, failures and constraints, incorporate prior knowledge in a controlled way, and select the next experiment within a closed-loop workflow.

Bayesian optimization (BO) addresses this need by coupling a probabilistic surrogate model with an acquisition function. Its conceptual roots lie in probabilistic approaches to noisy sequential extremum seeking,[10,11] which later developed into Bayesian global optimization and modern Gaussian-process (GP)-based optimization with criteria such as expected improvement (EI).[12–16] In this framework, the surrogate summarizes what is currently known about the objective and its uncertainty, while the acquisition function converts this information into a priority for the next experiment. For materials researchers, the key attraction of BO is not only that it is a powerful optimizer, but also that it provides a principled framework for sequential experimental design under uncertainty. This perspective becomes especially powerful when the experimental cycle is closed through automation.

Self-driving laboratories (SDLs) extend this idea by embedding sequential algorithmic decision-making into closed-loop experimental workflows, as demonstrated in closed-loop experimental platforms and workflows for autonomous carbon-nanotube synthesis,[17] spin-coated organic thin films,[18] epitaxial thin-film growth,[19,20] robotic chemistry,[21] battery protocol optimization,[22] autonomous inorganic synthesis[23] and autonomous phase mapping.[24] In highly automated implementations, a typical SDL combines an optimizer with robotic or automated synthesis or processing, automated characterization, data processing, and a database or orchestration layer. For example, the Autonomous Research System (ARES) integrated automated chemical vapor deposition (CVD) growth, in situ Raman feedback and an artificial-intelligence planner to iteratively select carbon nanotube (CNT) growth conditions, providing an early example of task-specific autonomous closed-loop materials experimentation.[17] Such implementations span diverse experimental modalities, including solution-based robotic chemistry, powder-based inorganic synthesis and thin-film processing (Fig. 1). These platforms differ substantially in their automation challenges: solution-based SDLs may require mobile robotic access to distributed instruments and autonomous liquid handling;[18,21] powder-based SDLs require automated precursor handling, furnace synthesis and X-ray diffraction (XRD) analysis;[23,25] and thin-film SDLs often require high-temperature vacuum processing and automated wafer or substrate transfer under vacuum.[26–29] Despite these hardware differences, materials SDLs share a common closed-loop structure in which experimental outcomes are converted into subsequent decisions by an algorithmic decision layer.[30–33] In such systems, BO often serves as this decision layer by using accumulated data, uncertainty estimates and constraints to propose actionable synthesis, characterization or measurement decisions. The practical success of this loop depends on how well the objective, surrogate model, acquisition function and constraints represent the realities of materials experiments. This decision-layer view of BO is summarized schematically in Fig. 2, where practical requirements in materials experiments and physics-informed knowledge shape the BO layer within a closed synthesis–characterization–optimization loop.

The standard textbook recipe of a GP with EI is therefore only a starting point.[13–15] This review begins with the basic BO loop and then discusses how it can be extended to address practical features of materials experiments, including failed or missing observations,[34] feasibility constraints,[35–38] competing objectives,[39–42] variable cost and fidelity,[22,43–46] batch and asynchronous operation,[47–51] transfer from historical data[52–55] and domain-specific physics knowledge.[56–58] We aim to provide practical guidelines for combining these components when designing BO workflows for real materials laboratories. A useful way to understand the role of BO in materials SDLs is to map common experimental difficulties to corresponding design choices in the optimization loop (Table 1). Local stagnation on rugged response surfaces motivates acquisition strategies and prior designs that maintain global exploration.[16,59] Historical data from related experiments, substrates, instruments or material families motivate warm starts, transferred priors, learned representations and multi-task surrogate models.[52–55] Competing material requirements are often better formulated as multi-objective optimization and Pareto-front discovery than as a single scalarized objective.[39–42] Variable experimental cost and fidelity motivate cost-aware and multi-fidelity BO, while parallel synthesis or characterization requires batch and asynchronous strategies.[22,43–51] Failed syntheses and missing measurements motivate failure-aware and constrained BO.[34–38] Finally, prior physics knowledge, such as thermodynamic stability, stoichiometry, lattice matching or kinetic growth windows, can be introduced through representations, prior means, kernels, acquisition functions, or constraints.[56–58]

Following this roadmap, this review is organized as follows. Section 2 introduces the basic BO loop and the translation of materials objectives into surrogate targets. Sections 3 and 4 discuss surrogate models, acquisition functions and experimental-design strategies, including multi-objective, cost-aware, multi-fidelity, batch and asynchronous BO. Section 5 focuses on practical complications in real materials experiments that strongly influence BO design, including failures, missing data, constraints, mixed-type experimental variables and transfer across related experiments. Section 6 then develops physics-informed Bayesian optimization (PIBO) as a unifying framework for incorporating domain knowledge into representations, priors, kernels, acquisition functions and constraints. Section 7 surveys representative materials-science and chemical achievements enabled by BO and related active-learning approaches. Section 8 provides practical guidance for designing, implementing, validating and reporting BO-driven SDL campaigns. Section 9 concludes with an outlook on the open challenges that will shape the next generation of BO-driven materials SDLs.

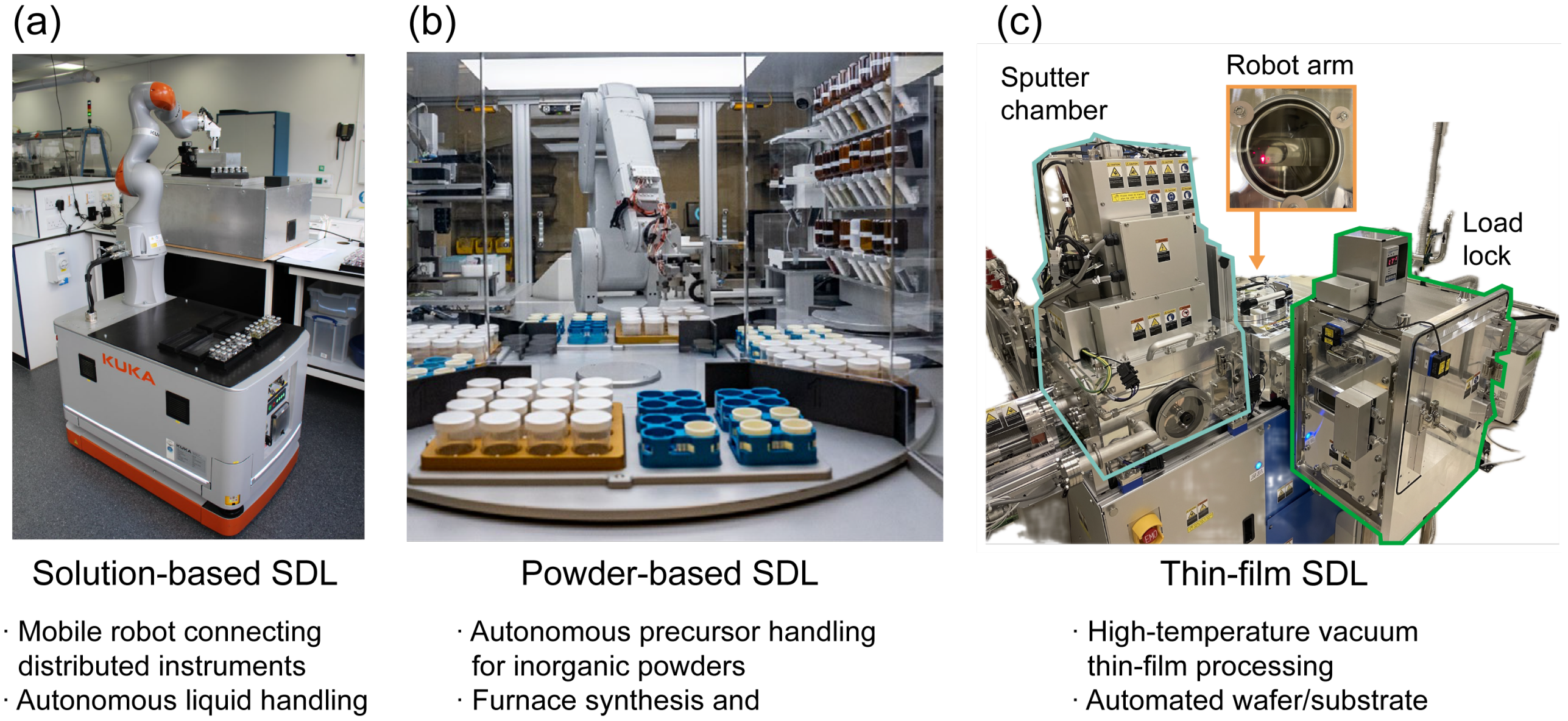

Figure 1. Representative SDL implementations across solution, powder and thin-film materials systems. These examples highlight the diversity of automation technologies required for different experimental modalities, while sharing a common closed-loop structure linking experiment execution, characterization, data analysis and algorithmic decision making. Image in panel (a) reproduced from Ref. 21 with permission from Springer Nature, copyright 2020. Images in panels (b) and (c) reproduced from Refs. 23 and 29, respectively, under a CC BY 4.0 license.

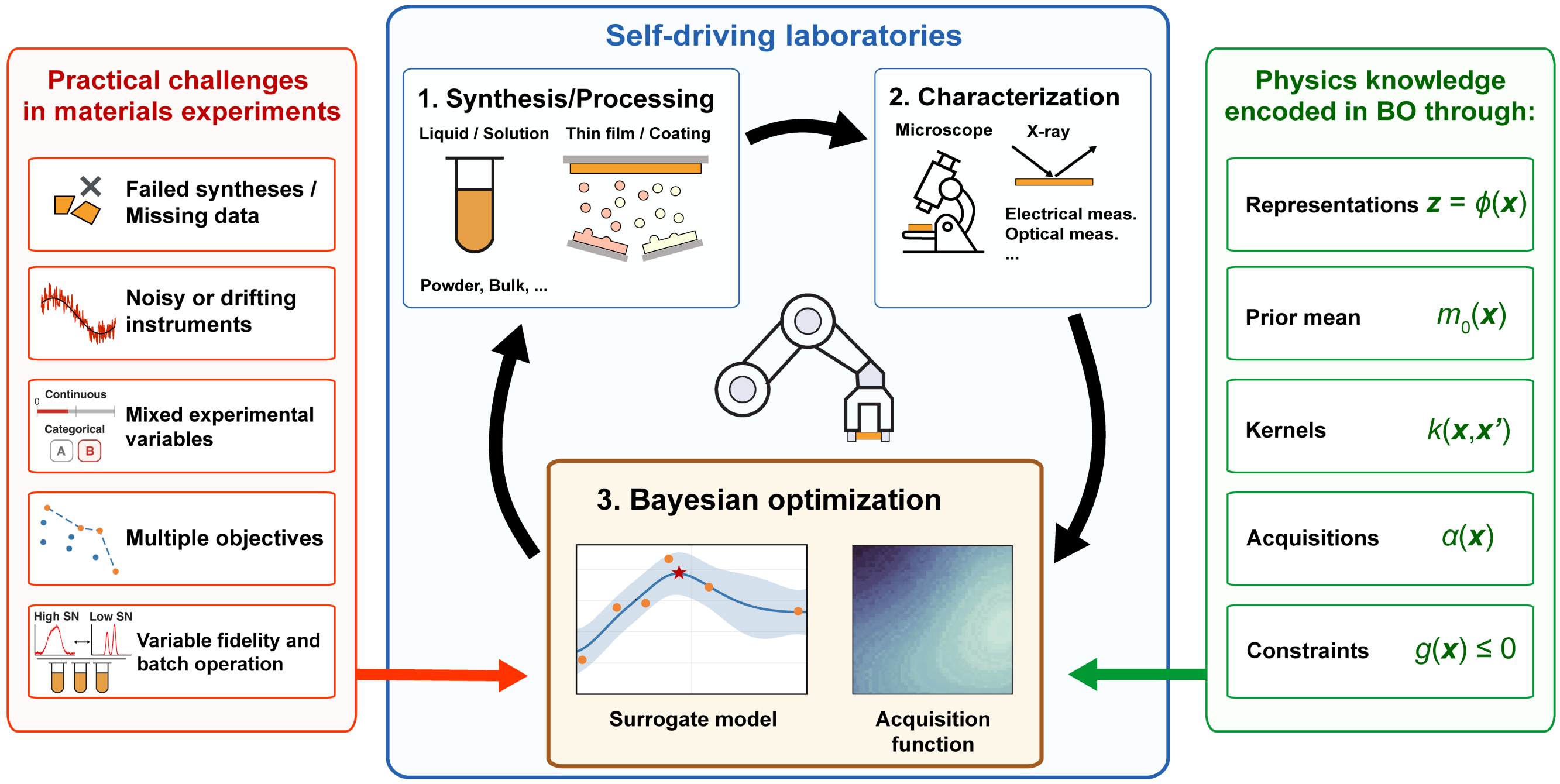

Figure 2. Bayesian optimization as a decision layer in self-driving materials laboratories. A materials SDL closes the loop among synthesis or processing, characterization and algorithmic decision making, with BO using surrogate models and acquisition functions to select subsequent experiments from accumulated observations. Practical requirements in materials experiments motivate extensions beyond standard black-box BO. PIBO incorporates domain knowledge through representations, priors, kernels, acquisitions and constraints.

| Experimental issue | Typical case | BO design choice | Main section |
|---|---|---|---|
| Local stagnation | Early exploitation around a suboptimal region of the experimental search space | Adaptive priors and exploration-weighted acquisition functions | Sections 3 and 4 |
| Multiple objectives | Several objective values are evaluated for each candidate condition | Multi-objective BO and Pareto-front acquisition functions such as expected hypervolume improvement (EHVI) | Section 4 |
| Variable cost or fidelity | Candidate conditions can be screened by cheap proxy measurements before expensive high-fidelity tests | Cost-aware and multi-fidelity BO | Section 4 |
| Parallel experiments | Multiple candidate conditions are synthesized or characterized at the same time | Batch BO, asynchronous BO and pending-point treatment | Section 4 |
| Failed or infeasible candidate conditions | A proposed condition leads to no target phase, film peeling, shorted devices, failed analysis, not-a-number (NaN) outputs or unsafe operation | Failure-aware and constrained BO | Sections 4 and 5 |
| Mixed-type experimental variables | Continuous and categorical variables in one experimental search space | Mixed-variable encodings, mixed-input kernels and tree-based surrogates | Section 5 |
| Transfer across different experiments | Historical data from related experimental search spaces are available | Warm starts, transferred priors, and multi-task GPs | Section 5 |
| Prior physics knowledge | Domain-specific physics knowledge, such as thermodynamic stability, is available for the experimental search space | Physics-informed representations, priors, kernels, acquisition weights and constraints | Section 6 |

Table 1. Materials-science challenges that motivate extensions of BO beyond the standard setting.

## 2. Fundamentals of Bayesian Optimization (BO) for Materials Research
### 2.1 Materials objectives and the BO loop

In this section, we introduce the basic concepts of BO in the context of materials research. The purpose is not to cover all variants of BO, but to explain the simplest and most widely used setting: sequential optimization of an expensive-to-evaluate black-box objective using a GP surrogate and an acquisition function.[12–15]

A materials experiment can be represented in a simplified form as follows. We choose a $D$-dimensional vector of experimental conditions $\boldsymbol{x} \in \mathbb{R}^D$, carry out the experiment, and obtain an output value $y$ through characterization or measurement. The vector $\boldsymbol{x}$ may include temperature, pressure, gas flow, elemental flux, composition, annealing time, solvent ratio, precursor concentration, substrate choice, or other controllable variables. The output $y \in \mathbb{R}$ is a numerical score that represents the quality or performance of the material. For example, $y$ may be residual resistivity ratio, carrier mobility, catalytic activity, photoluminescence (PL) intensity, phase purity, device efficiency, or the negative absolute or squared deviation from a target lattice constant, composition or band gap.

The optimization problem is to find the experimental condition $\boldsymbol{x}$ that gives the best value of $y \approx f(\boldsymbol{x})$. This can be simply written as

$$\boldsymbol{x}^* = \underset{\boldsymbol{x} \in \mathcal{X}}{\operatorname{argmax}} f(\boldsymbol{x}), \tag{2-1}$$

where $f(\boldsymbol{x})$ represents the relationship between the experimental condition $\boldsymbol{x}$ and the output $y$, and $\mathcal{X}$ is the experimentally allowed search space. In materials synthesis and processing, however, the explicit form of $f(\boldsymbol{x})$ is usually unknown. Growth, reaction and processing outcomes are determined by complicated thermodynamics, chemical reactions, and instrumental factors. Therefore, this is a black-box optimization problem: we can observe the output $y$ by performing an experiment at $\boldsymbol{x}$, but we do not know the analytical form of the function in advance.

BO combines two steps: probabilistic prediction and selection of the next experimental condition. Suppose that $M$ experiments have already been performed and the dataset is

$$\mathcal{D}_M := \{(\boldsymbol{x}_i, y_i)\}_{i=1}^{M}. \tag{2-2}$$

In real materials experiments, the observed value is not the underlying objective function itself, but the result of experimental and analysis noise added to the underlying response. We therefore write the observation at the $i$-th condition as

$$y_i = f(\boldsymbol{x}_i) + \epsilon_i, \tag{2-3}$$

where $f(\boldsymbol{x}_i)$ is the underlying material response and $\epsilon_i$ represents observational noise. In the simplest Gaussian-noise model,

$$\epsilon_i \sim \mathcal{N}(0, \sigma_{\text{noise}}^2), \tag{2-4}$$

where $\mathcal{N}(\mu, \sigma^2)$ denotes a Gaussian distribution with mean $\mu$ and variance $\sigma^2$. The parameter $\sigma_{\text{noise}}^2$ is the noise variance. This term may include measurement uncertainty, sample-to-sample variation, instrumental fluctuation and uncertainty introduced by automated data analysis. For example, the value extracted from an XRD pattern, optical spectrum or electrical measurement may vary even when the nominal synthesis condition is repeated. For this reason, a BO algorithm may sometimes propose repeating an already tested condition. Such a repeated measurement can be useful for estimating noise, checking reproducibility or confirming an apparent optimum, although previous conditions can also be explicitly excluded when the experimental goal is to explore only new recipes. Using this dataset, BO first constructs a probabilistic model that predicts the output at untested conditions. This model is called a surrogate model. The surrogate estimates not only the expected value of $y$, but also the uncertainty of the prediction. The uncertainty is essential to encourage exploration because untested regions in the search space with greater uncertainty may contain better conditions than those already examined.

Next, BO uses an acquisition function to decide which experimental condition should be tested next. The acquisition function gives a score to each candidate condition by using the predicted mean and uncertainty. A condition with a high predicted value is attractive because it may directly improve the best result. A condition with large uncertainty is also attractive because it may reveal an unexplored high-performance region. Thus, the acquisition function is designed to balance exploitation of promising regions and exploration of uncertain regions. The basic BO loop is therefore:

1. prepare initial experimental data;

2. train a surrogate model using the observed data;

3. calculate an acquisition function over the search space;

4. select the next experimental condition $\boldsymbol{x}'$ that maximizes the acquisition function;

5. perform the experiment and obtain a new value of $y'$;

6. add the new data point ($\boldsymbol{x}'$, $y'$) to the dataset and update the surrogate model.

This loop is repeated until the objective no longer improves or a predefined performance criterion is met.

### 2.2 Gaussian process (GP) surrogates and expected improvement (EI)

The most common surrogate used in BO is a GP regression model.[13,14] A GP defines a probability distribution over possible functions rather than fitting a single deterministic curve to the observed data. This probabilistic formulation naturally provides predictive uncertainty, which is essential for uncertainty-guided decision making in BO. In this review, we use the term simply "GP" or "GP surrogate" when the context is clear, to refer to the GP regression model used in BO. In the standard formulation used here, the GP models the underlying material-response function $f(\boldsymbol{x})$, while the observed values include experimental noise. Assuming Gaussian observation noise, this is accounted for by adding the noise variance to the diagonal of the kernel covariance matrix. The resulting GP posterior therefore represents our uncertainty about the underlying response function given the noisy observations.

Before observing data, the unknown function $f(\boldsymbol{x})$ is assumed to follow

$$f(\boldsymbol{x}) \sim \mathcal{GP}(m(\boldsymbol{x}), k(\boldsymbol{x}, \boldsymbol{x}')), \tag{2-5}$$

where $m(\boldsymbol{x})$ is the prior mean function and $k(\boldsymbol{x}, \boldsymbol{x}')$ is the kernel function. The prior mean gives the expected value of the function before considering the observed data. The kernel defines how strongly the function values at two experimental conditions $\boldsymbol{x}$ and $\boldsymbol{x}'$ are correlated.

Individual curves sampled from the GP prior can be viewed as possible material-response functions before any experiment is observed, as illustrated in Fig. 3a. For example, if the prior mean is set to a constant value, the prior expresses the idea that the objective is initially expected to fluctuate around that baseline. The kernel determines both the prior uncertainty and the smoothness of these possible functions. The kernel value evaluated at identical inputs, $k(\boldsymbol{x}, \boldsymbol{x})$, specifies how much function value $f(\boldsymbol{x})$ may vary at each experimental condition $\boldsymbol{x}$, whereas $k(\boldsymbol{x}, \boldsymbol{x}')$ with $\boldsymbol{x} \neq \boldsymbol{x}'$ specifies how strongly function values $f(\boldsymbol{x})$ and $f(\boldsymbol{x}')$ at different conditions are correlated. Thus, the prior distribution encodes assumptions about the scale and smoothness of the unknown response surface before data are available. In practice, the kernel encodes the relationship between the closeness in the experimental condition space and similarity of their results. Namely, nearby points in the experimental space are assumed to give similar output values. This assumption is often reasonable for continuous variables such as temperature, pressure, flux or composition. The kernel length scale determines the distance over which two points are regarded as similar. A small length scale allows the response to vary rapidly, whereas a large length scale imposes a smoother response over a wider region. The meaning of "nearby" depends on how the experimental variables are represented and scaled; therefore, input variables are often normalized before fitting a GP with a distance-based kernel.

After the observed data $\mathcal{D}_M$ are obtained, a GP gives a predictive distribution $p(f(\boldsymbol{x})|\mathcal{D}_M)$ for the underlying response at an arbitrary condition $\boldsymbol{x}$:

$$p(f(\boldsymbol{x})|\mathcal{D}_M) = \mathcal{N}(\mu(\boldsymbol{x}), \sigma^2(\boldsymbol{x})). \tag{2-6}$$

Here, $\mu(\boldsymbol{x})$ is the predicted mean and $\sigma(\boldsymbol{x})$ is the predicted standard deviation. The mean $\mu(\boldsymbol{x})$ represents the expected material response, while the standard deviation $\sigma(\boldsymbol{x})$ represents the uncertainty of the prediction. The posterior distribution of the underlying response can be interpreted as the set of response functions that remain plausible once conditioned on the observed data (Fig. 3b). Near measured conditions, these functions concentrate on the actual data and the predictive uncertainty becomes small. In unexplored regions, many different functions remain plausible and the uncertainty remains large. This interpretation is particularly useful in materials optimization because it separates what has been learned from what remains unknown.

After GP predicts $\mu(\boldsymbol{x})$ and $\sigma(\boldsymbol{x})$, BO must decide which condition to test next. A widely used acquisition function is expected improvement (EI).[15] EI evaluates how much improvement can be expected over the current best observation. For a maximization problem, let the best observed value after $M$ experiments be

$$y_{\text{best}} = \max_{i=1,\dots,M} y_i. \tag{2-7}$$

If the true output at a candidate condition $\boldsymbol{x}$ is $f(\boldsymbol{x})$, the improvement is

$$I(\boldsymbol{x}) = \max\{0, f(\boldsymbol{x}) - y_{\text{best}}\}. \tag{2-8}$$

Because $f(\boldsymbol{x})$ is a black-box function, BO calculates the expectation value of this improvement using the predictive distribution from GP:

$$\text{EI}(\boldsymbol{x}) = \mathbb{E}_{f(\boldsymbol{x})|\mathcal{D}_M}[I(\boldsymbol{x})], \tag{2-9}$$

where $\mathbb{E}_{f(\boldsymbol{x})|\mathcal{D}_M}$ denotes expectation over the posterior predictive distribution of $f(\boldsymbol{x})$ given the observed data $\mathcal{D}_M$. The next experimental condition is then selected as

$$\boldsymbol{x}_{M+1} = \underset{\boldsymbol{x}\in\mathcal{X}}{\text{argmax}}\,\text{EI}(\boldsymbol{x}). \tag{2-10}$$

EI becomes large in two cases (Fig. 3c). The first is when the predicted mean $\mu(\boldsymbol{x})$ is higher than or close to the best value observed so far. This corresponds to exploitation, because the algorithm refines a region that already appears promising. The second is when the uncertainty $\sigma(\boldsymbol{x})$ is large. Even if the predicted mean is not very high, a highly uncertain point may still produce a value better than the current best. This corresponds to exploration. In this way, EI balances exploitation and exploration and is therefore often used as the standard starting point for BO-guided materials optimization.

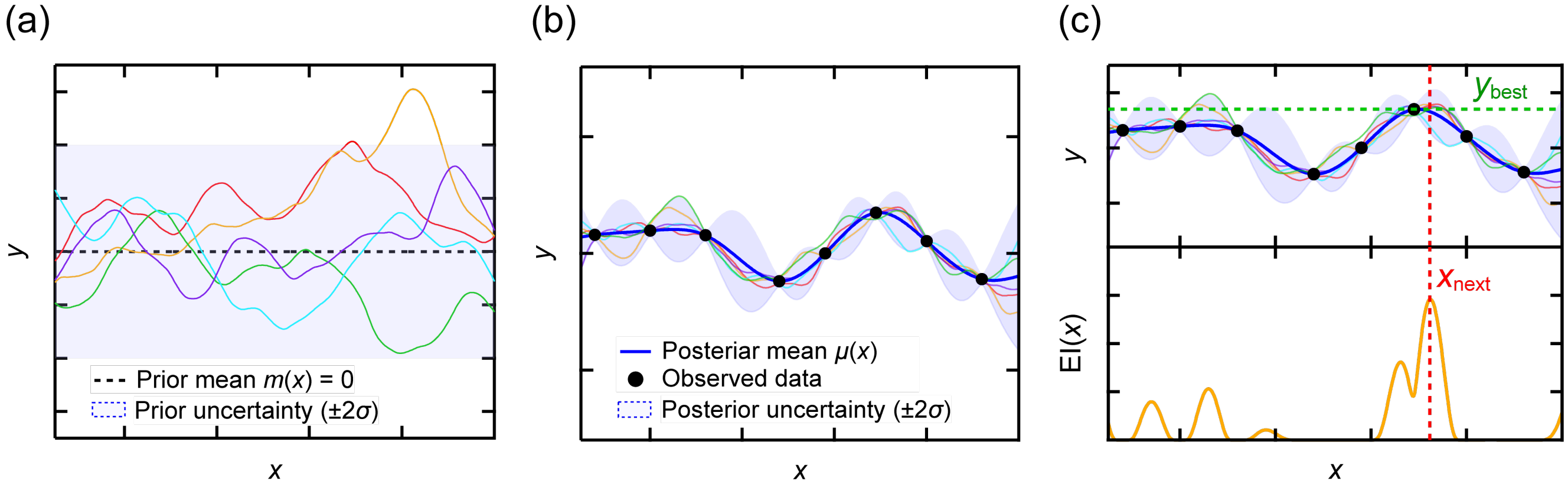

Figure 3. Schematics of GP-based BO. (a) GP prior over the unknown objective function before experimental observations, showing the prior mean, prior uncertainty, and representative sampled functions (colored curves). (b) GP posterior conditioned on observed data points (black circles), showing the posterior mean, posterior uncertainty, and representative sampled functions (colored curves). (c) EI (lower panel) calculated from the posterior distribution (upper panel).

### 2.3 Deploying BO in materials research

To apply BO to a real materials experiment, the first practical step is to define the objective value $y$. This step is sometimes more important than the choice of the BO algorithm itself. BO can only optimize the quantity that is provided to the surrogate model. In the simplest case, $y$ is a directly measured material property. For example, one may maximize conductivity in metals, carrier mobility in semiconductors, catalytic current density in electrocatalysts, PL intensity in optical materials, or power-conversion efficiency in solar cells. In such cases, the measured property can be used directly as the objective. In other cases, the goal is to approach a target value rather than simply maximize or minimize a property. For example, one may want to obtain a target lattice constant, target composition, target band gap, target emission wavelength, or target film thickness. Such problems can be converted into optimization problems by defining a deviation from the target. For example, if $q(\boldsymbol{x})$ is a measured quantity and $q_{\text{target}}$ is the desired value, one may define

$$y(\boldsymbol{x}) = -\left|q(\boldsymbol{x}) - q_{\text{target}}\right| \tag{2-11}$$

or

$$y(\boldsymbol{x}) = -\left|q(\boldsymbol{x}) - q_{\text{target}}\right|^2. \tag{2-12}$$

Then maximizing $y(\boldsymbol{x})$ is equivalent to minimizing the deviation from the target. Though this negative-deviation objective works in practice, one should note that such negative-deviation objectives are bounded above by zero, whereas a standard GP assigns nonzero probability outside this range. One way to address this mismatch is to evaluate the acquisition function using a truncated Gaussian distribution, in which predicted probability for $y > 0$ is set to 0.

In many materials workflows, the evaluation value is extracted from characterization data. X-ray diffraction patterns may be converted into peak positions, peak widths or phase purity. Optical spectra may be converted into band gaps, absorption tails or defect-related signals. Microscopy images may be converted into particle size, roughness or morphology descriptors. Electrical measurements may be converted into mobility, leakage current, resistivity or device performance. Therefore, the data-analysis procedure that converts raw measurements into $y$ is an essential part of the BO loop.

The simple GP–EI framework described here provides a useful starting point for materials optimization. As a representative example of a GP-based BO loop in materials synthesis, Fig. 4a shows BO-guided CVD growth of CNTs. Following CVD growth under BO-selected conditions, the CNT growth rate is extracted from in situ Raman spectra and added to the BO dataset. The updated dataset is then used to refine the surrogate model, and the acquisition function is evaluated to select subsequent CVD growth conditions. This closed loop searched a four-dimensional CVD parameter space, defined by ethylene partial pressure, hydrogen partial pressure, water-vapor amount and temperature, and identified high-growth-rate conditions within approximately 100 BO-guided CVD experiments (Fig. 4b).[60] More broadly, BO and uncertainty-aware active-learning and adaptive-design formulations, which combine surrogate modelling with improvement- or uncertainty-based selection criteria, have already been applied to a wide range of materials research.[61–74] Early methodological work framed BO as a general strategy for materials design,[61] and uncertainty-guided adaptive design was demonstrated for computational materials screening.[62] In computational materials discovery, BO has been used to search for low-thermal-conductivity compounds and to design nanostructures for phonon transport.[63,64] In compositional spaces, adaptive-design strategies accelerated the search for shape-memory alloys with targeted properties,[65] while BO identified unexpected compositions in mesoporous PtPdAu electrocatalyst films.[66] In powder and ceramic materials, active-learning and adaptive-design workflows closely related to BO have accelerated the discovery of $BaTiO_3$-based piezoelectrics with large electrostrain.[67] BO software libraries such as COMBO have also helped establish practical workflows for materials-science optimization.[68] For thin films, BO has been applied to epitaxial $SrRuO_3$ growth,[19,20] superconducting TiN thin films,[69] and III–V thermoelectric thin films.[70] Related active-learning and BO workflows have also been applied to magnetic-material processing, such as Nd-Fe-B magnets,[71] and benchmark studies have compared BO performance across multiple experimental materials domains, including CNT polymer blends, silver nanoparticles, lead-halide perovskites and additively manufactured polymer structures.[72] These examples show that BO and related surrogate-guided sequential optimization workflows are not tied to a particular synthesis modality or material system, but provide a general decision-making template for scarce-data materials optimization.

However, real materials experiments often include complications beyond the simple GP–EI setting. Some conditions may fail to produce the target phase, causing missing data. Some experiments may involve multiple objectives, such as activity and stability, conductivity and transparency, or performance and process temperature. Some variables may be categorical, such as solvent, substrate or precursor type. Some measurements may have different costs or fidelities. Prior physics knowledge may also be available, such as stoichiometric constraints, phase-stability windows, lattice matching or monotonic relations between process variables and composition. These complications do not necessarily invalidate the basic BO loop. Rather, they motivate various directions for framework extensions. The following sections discuss how the surrogate model, acquisition function, constraint handling and physics-informed design can be modified for realistic materials and SDL workflows.

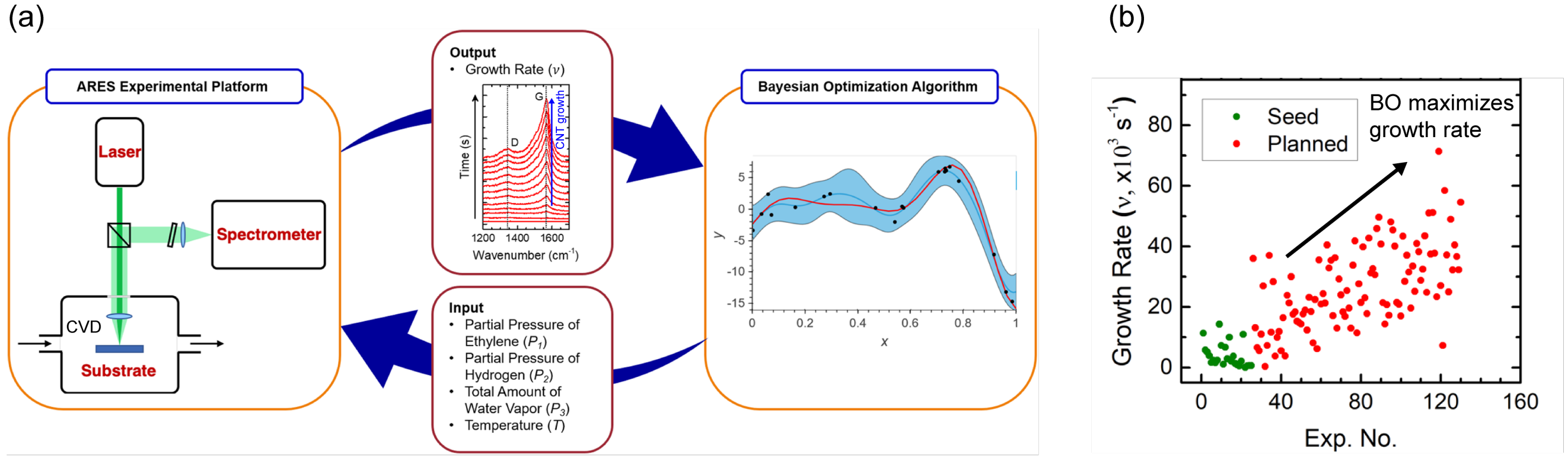

Figure 4. Representative example of a GP-based BO loop in materials synthesis: BO-guided CVD growth of CNTs. (a) Closed-loop workflow combining CVD growth, in situ Raman analysis and BO-based experimental selection. After each CVD growth experiment, CNT growth rates are extracted from Raman spectra and incorporated into the BO dataset. The updated dataset is used to refine the GP surrogate, and the acquisition function is evaluated to select subsequent CVD growth conditions. (b) BO-guided optimization result showing the evolution of CNT growth rate over sequential experiments. The study used a Matérn 5/2 GP surrogate with an upper confidence bound (UCB) acquisition function, illustrating the same surrogate–acquisition loop introduced above using EI as a simple pedagogical example. Adapted from Ref. 60 under a CC BY 4.0 license.

## 3. Surrogate Models

Section 2 introduced BO in its simplest form, where a scalar objective is modelled by a probabilistic surrogate and an acquisition function uses the surrogate prediction and uncertainty to select the next experiment. Section 3 focuses on the modelling choices underlying this surrogate. In practical materials SDLs, one must decide how to incorporate prior information, define similarity between experimental conditions, represent observational noise and predictive uncertainty, and extend the surrogate to workflows involving multiple outputs, fidelities or experimental stages. We first discuss GP surrogates, including prior mean functions, kernels and noise models, and then introduce structured GP models and alternative tree-based or neural surrogates.

### 3.1 GP as the standard surrogate

GP remains the standard starting point for BO in materials research.[13,14] It is particularly useful for materials experiments with few observations, expensive evaluations and a need for uncertainty-guided decision making.[75,76] In such low-data settings, GP provides a flexible nonparametric surrogate together with predictive uncertainty, making it a natural baseline for low-dimensional continuous process variables such as temperature, pressure, flux, composition, annealing time or gas-flow ratio. However, the performance and uncertainty calibration of a GP depend strongly on its prior mean, kernel, noise model and hyperparameter estimation. The prior mean specifies the baseline response expected before observing data, while the kernel specifies how similarity in the experimental space is translated into correlation in the response. These choices are especially important in small datasets, where hyperparameters can be poorly identified and posterior uncertainty can be over- or under-estimated. We therefore next discuss the roles of prior means and kernels in GP-based materials BO.

### 3.2 Prior mean functions

In GP-based materials BO, the prior mean defines the baseline materials response expected before observing data. This baseline is important because, in regions with little or no experimental data, the GP prediction tends to revert toward the prior mean with a typical choice of distance-based kernel. This section therefore focuses on how the choice of prior mean influences BO behavior. The most common choice is a zero or constant prior mean,

$$m(\boldsymbol{x}) = c, \tag{3-1}$$

where $c$ is a constant usually chosen after scaling the objective values. The zero-mean prior is obtained by setting $c$ = 0. This choice is attractive because it is simple, weakly informative and does not encode a priori trend across the search space.[13,14,77] It is appropriate when no reliable physical, empirical or historical baseline is available, or when the campaign is intended to let the observed data dominate the posterior. Even in this simple setting, however, the value of a constant prior mean can influence the exploration–exploitation balance for acquisition functions where predictions in unexplored regions are mostly determined by the constant. Figure 5 illustrates this effect for the same observed data but different constant prior means. Near the observations, the posterior means are strongly concentrated on the data and therefore remain similar. Away from the observations, however, the posterior means revert toward their respective prior means. As a result, a larger constant prior mean ($c = m^{+}$) can make unobserved regions appear more promising and thus promote exploration, whereas a smaller constant prior mean ($c = m^{-}$) can make those regions appear less attractive and thereby bias the search toward exploitation. This effect can be especially pronounced in the early stage of BO, when data are scarce and the acquisition function is sensitive to the predicted mean in unobserved regions.

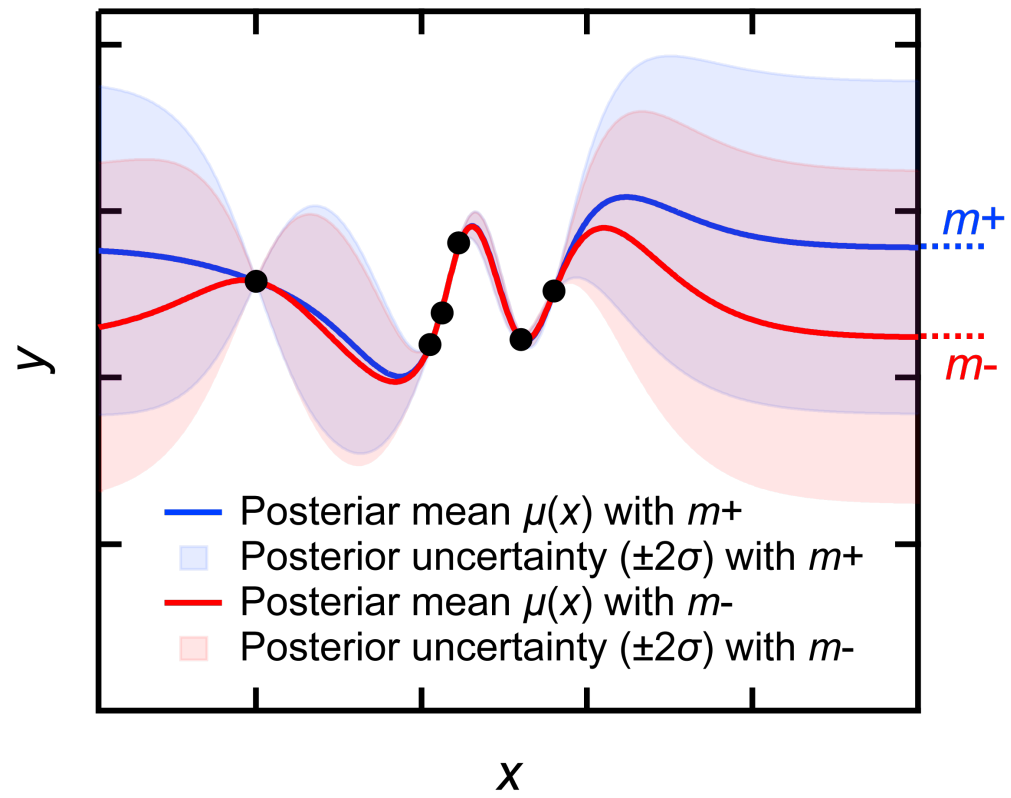

Figure 5. Schematic illustration showing that different constant prior means ($m^{+}$) and ($m^{-}$) lead to different posterior means in unobserved regions for the same observed data (black circles).

This sensitivity motivates adaptive or randomized prior-mean strategies, which modify the prior mean during the campaign to reduce dependence on the initial data and reduce the risk of becoming trapped in local optima.[59] In Ref. 59, such a strategy was implemented by resampling the prior-mean constant at each BO iteration from a uniform distribution bounded by the minimum and maximum of the observed objective values,

$$m_0^{n'} \sim \mathcal{U}\left(\min_{1 \le i < n'} y_i, \max_{1 \le i < n'} y_i\right), \tag{3-2}$$

where $m_0^{n'}$ is the constant prior mean used at the $n'$-th iteration and $y_i$ denotes the objective value observed in the $i$-th experiment. This randomized prior-mean strategy alters the baseline to which the GP posterior reverts in unexplored regions, thereby modifying the acquisition landscape and enabling the search to escape from locally optimal regions. Figures 6a and b illustrate this mechanism using a benchmark function with multiple local optima.[59] In Fig. 6a, the EI maximum (red star) remains near a suboptimal peak, reflecting exploitation around a locally promising region while the global-optimum region indicated by the dashed circle remains unexplored. In Fig. 6b, increasing the prior mean reshapes the predicted mean and EI landscape, shifting the EI maximum toward the unexplored region containing the global optimum. In the same study, this adaptive-prior-mean strategy was implemented in molecular beam epitaxy (MBE) for $SrTiO_3$ (STO) thin-film growth and enabled stoichiometric, high-quality epitaxial STO films (Fig. 6c) to be obtained after only 44 growth runs.

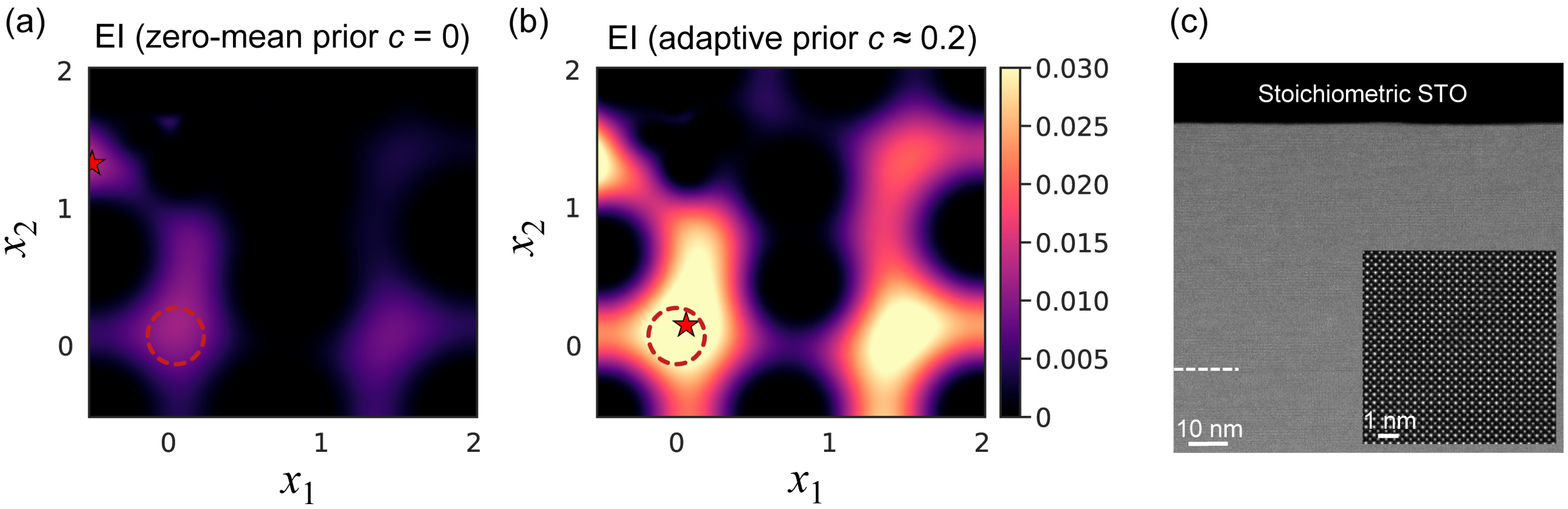

Figure 6. Adaptive prior-mean BO for stoichiometric STO film growth. (a) and (b) EI landscapes for a benchmark function with multiple local optima obtained using different prior means. The red star denotes the next candidate selected by EI, and the dashed circle indicates the global-optimum region. (c) scanning transmission electron microscopy (STEM) image of the optimized STO film. Reproduced from Ref. 59 under a CC BY 4.0 license.

Beyond zero or constant prior means, many materials problems contain known baselines or approximate trends. A prior mean can encode such a baseline so that the GP learns residual deviations from it. Formally, one can write

$$y(\boldsymbol{x}) = m(\boldsymbol{x}) + g(\boldsymbol{x}) + \epsilon, \tag{3-3}$$

where $m(\boldsymbol{x})$ is a known or estimated baseline, $g(\boldsymbol{x})$ is the residual modelled by the GP, and $\epsilon$ is observational noise.[57] This residual-learning view is related to model-discrepancy approaches, where imperfect computational or theoretical models are corrected using data-driven residual terms.[78] For BO, this idea motivates us to employ a physics-based, empirical, or expert-informed model as $m(\boldsymbol{x})$ to capture the broad trend, while the GP $g(\boldsymbol{x})$ corrects for system-specific deviations using actual observations. Related examples include accelerator tuning, where neural-network models trained on beam-dynamics simulations or historical data have been used as prior mean functions in GP-based BO.[79] Detailed materials examples in which physics knowledge is encoded into prior means or residual models are discussed in Section 6.[57,80,81]

Prior design is also one way to transfer information from previous experiments. Historical data from related substrates, chambers, batches or material families can be used to initialize or learn elements of the surrogate, including the prior mean or kernel hyperparameters. Pre-trained GP approaches provide a general formulation of this idea, in which GP priors are learned from related optimization tasks before being adapted to a new campaign.[82] Although this concept is attractive for materials SDLs, convincing experimental demonstrations of transferred GP priors in materials optimization are still lacking. Such transfer may improve sample efficiency when the new campaign is genuinely related to the old one, but the central risk is negative transfer: historical data will reflect a different calibration state, precursor source, chamber history, substrate quality or objective definition. A transferred prior should therefore be treated as uncertain, and its benefit should be checked against a baseline model without transferred prior information.

### 3.3 Kernels

While the prior mean specifies the baseline response, the kernel determines how this response is allowed to vary across the experimental search space. In a GP surrogate, the kernel $k(\boldsymbol{x}, \boldsymbol{x}')$ specifies how strongly the material responses at two experimental conditions $\boldsymbol{x}$ and $\boldsymbol{x}'$ are correlated. In distance-based kernels, this correlation is usually expressed through the assumption that nearby experimental conditions tend to give similar responses. For example, two growth temperatures separated by a few degrees, two compositions differing by a small amount, or two gas-flow ratios close to each other may be expected to give similar responses. In practice, this covariance structure is determined by both the chosen kernel form and its hyperparameters, which are usually estimated from the observed data when constructing the GP surrogate. The resulting covariance structure is then used to compute the posterior predictive mean and uncertainty. As a concrete example of how a kernel encodes such similarity assumptions, consider the radial-basis-function (RBF) kernel for a one-dimensional input variable:[13,14]

$$k_{\mathrm{RBF}}(x, x') = \sigma_f^2 \exp\left(-\frac{(x-x')^2}{2l^2}\right), \tag{3-4}$$

where $l$ is the length scale for the input variable and $\sigma_f^2$ is the signal variance. This expression shows how the distance between two points in the experimental search space is mapped to the correlation between their responses. The length scale determines how far one can move in temperature, composition or growth parameters before the response is expected to change substantially, whereas the signal variance sets the expected magnitude of response variation. A short length scale allows the predicted response to vary locally, whereas a long length scale imposes smoother variation over a wider region. In most BO implementations, these hyperparameters are fitted from data. Therefore, the kernel should be viewed as defining a model class for possible response surfaces rather than a fully fixed prior chosen manually before the experiment.

Common choices in BO include the RBF and Matérn kernels. The RBF kernel assumes that the response varies very smoothly with distance in the experimental search space. Nearby conditions are strongly correlated, and this correlation decays smoothly as the distance between them increases. Matérn kernels provide greater flexibility by controlling smoothness through a parameter denoted by $\nu$:[13,14] smaller values of $\nu$ allow rougher response functions, whereas larger values impose smoother variation. The Matérn kernel is defined as follows:

$$k_{\mathrm{Mat},\nu}(x, x') = \frac{\sigma_f^2}{\Gamma(\nu)2^{\nu-1}}\left(\sqrt{2\nu}\frac{|x-x'|}{l}\right)^{\nu} K_\nu\left(\sqrt{2\nu}\frac{|x-x'|}{l}\right), \tag{3-5}$$

where $\Gamma(\cdot)$ and $K_\nu(\cdot)$ are the gamma function and the modified Bessel function of the second kind, respectively. The exponential kernel corresponds to the Matérn kernel with $\nu$ = 1/2 and represents the roughest commonly used member of this family; its sample functions can vary more abruptly and are not mean-square differentiable. Matérn kernels with $\nu$ = 3/2 and $\nu$ = 5/2 impose progressively stronger smoothness assumptions, allowing once- and twice-mean-square-differentiable functions, respectively. In the limit $\nu \to \infty$, the Matérn kernel approaches the RBF kernel. Because this family spans a range of smoothness assumptions, Matérn kernels are often used as practical default choices in GP-based BO when the very strong smoothness assumption of the RBF kernel is not desired. They can better accommodate moderately rough response surfaces, although genuinely abrupt changes near phase boundaries, growth-mode transitions or degradation thresholds may require nonstationary or regime-aware surrogate models.

The difference between kernel types should not be interpreted independently of their hyperparameters. In particular, the length scale determines the distance over which two experimental conditions are treated as similar, as described above. Thus, the apparent smoothness of the GP prediction can depend strongly on the fitted length scale even within the same kernel type. Figure 7 illustrates this point using a one-dimensional synthetic dataset. Figure 7a compares GP posterior means obtained with different kernel types using the same observed dataset, the same fixed normalized length scale and the same fixed observation-noise variance. The differences are visible but relatively moderate. Figure 7b then fixes the kernel type to RBF and changes only the length scale, showing that the prediction becomes sharper for a shorter length scale and smoother for a longer length scale. Figure 7c shows a more practical setting in which the RBF length scale is optimized from the observed data, while the observation-noise variance is kept fixed to isolate the effect of length-scale fitting. The fitted posterior mean, depicted as the automatically tuned result in Fig. 7c, shows a moderate curve between the two sharp and smooth ones with extreme fixed-length scales, demonstrating that practical BO implementations usually infer the relevant correlation scale from data rather than requiring it to be specified manually in advance. Kernel length scales are commonly learned by maximizing the marginal likelihood of observed data.[14] As a simple alternative or initialization for hyperparameter learning, the median heuristic sets the length scale as the median pairwise distance of input points.[83]

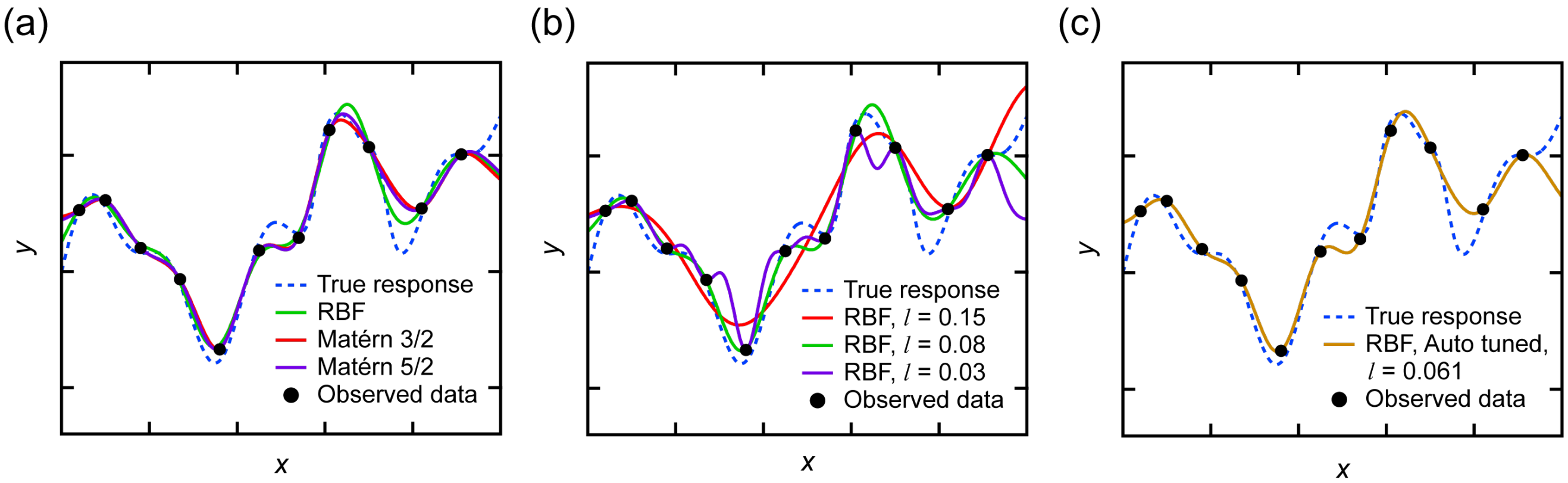

Figure 7. Kernel-type and length-scale effects in one-dimensional GP regression. (a) Posterior means obtained with different kernel types. (b) RBF posterior means obtained with different fixed normalized length scales. (c) RBF posterior mean obtained with a data-fitted normalized length scale. In (a)-(c), the same synthetic response, observed points and fixed observation-noise variance are used.

In a multi-dimensional experimental space, kernels are extended to handle the distance between input vectors that contain a combination of experimental variables. The interpretation of this multi-dimensional distance depends strongly on how the experimental variables are represented and scaled. For example, a gas-phase catalytic experiment may combine a reactor pressure on the order of $10^5$–$10^6$ Pa with a reactant mole fraction ranging from 0.1 to 0.2. If a distance-based kernel is applied directly to these raw values, the pressure coordinate dominates the numerical distance simply because of its larger scale. The input variables should therefore be standardized or normalized before fitting the GP. Even after such scaling, however, different input variables can have very different effects on the material response. In chemical and catalytic reactions, for example, product yield or selectivity can depend much more sensitively on some coordinates, such as electrolyte composition, pH or catalyst composition/loading, than on others within the same reaction regime.[4,84] Similarly, in complex-oxide thin-film growth, the phase, defect concentration, valence state or transport properties can respond sharply to oxygen chemical potential, as illustrated by vanadium oxide and $LaNiO_3$ thin films.[85,86] Thus, even after normalization, equal distances along different input coordinates do not necessarily correspond to equal changes in the material response. For the RBF kernel, a common multi-dimensional extension is the anisotropic, or automatic relevance determination (ARD), form,

$$k_{\mathrm{RBF}}(\boldsymbol{x}, \boldsymbol{x}') = \sigma_f^2 \exp\left(-\frac{1}{2}\sum_{d=1}^{D}\frac{(x_d - x_d{}')^2}{l_d^2}\right), \tag{3-6}$$

where $l_d$ is the length scale for the $d$-th input variable.[16,87] In this formulation, each length scale controls how rapidly the response is allowed to vary as the corresponding experimental variable changes. A short length scale indicates that the surrogate can be sensitive to that variable, whereas a long length scale indicates that the response is treated as relatively insensitive over the explored range. The same idea can also be applied to Matérn kernels:[16] instead of using a single length scale for all input variables, the distance between two experimental conditions is calculated using separate length scales $l_d$ for each variable. Thus, both RBF and Matérn kernels can represent different sensitivities along different experimental variables. More generally, fitted ARD length scales should be interpreted as model-dependent diagnostics of variable sensitivity, because they depend on the chosen kernel, input representation, data coverage and correlations among input variables. In materials optimization, anisotropic or ARD-type GP kernels are widely used, although they are not always explicitly described using the term ARD.[34,88] A benchmark study across various materials systems has further shown that GP surrogates with anisotropic kernels can improve BO performance compared with that of GP surrogates using isotropic kernels.[72] Related studies on high-dimensional materials-synthesis design have also used ARD length scales as a route to identifying relevant variables, while highlighting the difficulty of thresholding such length scales in practical optimization.[88,89]

In their standard forms, RBF and Matérn kernels are stationary, meaning that the correlation between two responses depends on the difference between their inputs rather than on their absolute locations in the search space. This assumption can be too restrictive for materials experiments. Response surfaces may change abruptly across phase boundaries, growth-mode transitions, solubility limits, percolation thresholds, degradation thresholds or device-failure boundaries. In such cases, the response surface may vary sharply near a boundary, and nearby experimental conditions on opposite sides of the boundary may exhibit qualitatively different material responses. Despite its simplicity, a stationary kernel may cause a model misspecification: it may over-smooth the response, underestimate uncertainty near discontinuities or guide BO toward misleading regions. Nonstationary or regime-aware surrogate models can address this problem. Input transformations or input warping can make the effective length scale vary across the search space, thereby making the response more nearly stationary in the transformed input space.[90] Local, partitioned or change-point GP models provide another methodological route by allowing different response regimes to have different covariance structures.[91–93] Although these approaches are well established in the broader GP and BO literature, their application to experimental materials optimization remains limited. In materials workflows, however, abrupt boundaries often reflect not only changes in smoothness but also changes in phase identity, feasibility or measurement validity. Autonomous phase-mapping studies illustrate that phase regions and phase boundaries may themselves need to be represented as part of the decision problem.[24] Similarly, failure-aware and constrained BO studies show that failed syntheses, missing outputs and feasibility boundaries should often be modelled explicitly rather than treated only as irregularities in a single smooth response surface.[34–38] These issues motivate surrogate designs that represent regimes, feasibility or failure in addition to the objective itself; related structured surrogates are introduced in Section 3.4, and failure- and constraint-aware treatments are discussed in more detail in Section 5.

Other structured kernel designs address different types of experimental structure. Periodic kernels are useful when the response contains a physically or experimentally meaningful repeating component.[94] For example, Durrande *et al*. decomposed a GP covariance function into periodic and aperiodic components, allowing the model to separate a periodic signal from its nonperiodic complement, quantify periodicity in noisy data and identify periodically expressed genes in the Arabidopsis genome.[95] Periodic or quasi-periodic GP kernels have also been used in other experimental sciences, including astronomy and geodetic time-series analysis.[96–98] In materials optimization, however, their adoption remains limited; they should therefore be introduced only when the input representation or measured response has a justified cyclic or repeating structure, such as oscillatory signals, angular variables, periodically repeated process cycles or spatially repeating image features.

Deep kernel learning (DKL) provides a further extension by combining neural-network-based representation learning with GP posterior prediction in the resulting feature space.[99] In a standard GP surrogate, the kernel evaluates similarity directly from the original input variables. In DKL, the inputs are first converted by a neural network into a learned feature representation, and the GP kernel then evaluates similarity using this representation. In the materials-BO implementation of Kiyohara and Kumagai,[100] for example, crystal structures were encoded by a crystal graph convolutional neural network, and the resulting learned representation was used as the input to a GP surrogate for BO (Fig. 8). They employed DKL to reduce the need for manual feature engineering while retaining the GP-based BO framework. In their benchmarks, DKL showed comparable or improved BO efficiency relative to standard GP surrogates for oxide property datasets and hybrid organic–inorganic perovskite alloy band gaps, whereas the standard GP performed better for alloy Curie-temperature search when a strongly correlated descriptor was directly available. This result shows that deep kernels are not universally superior to standard GPs, but can be useful when meaningful input representations are difficult to specify manually. Related DKL concepts have also begun to appear in experimental materials workflows, particularly in autonomous microscopy.[101] For example, ensemble DKL has been used in automated piezoresponse force microscopy experiments to identify which imaging channel is most predictive of spectroscopic hysteresis-loop properties.[102] These examples suggest a route for future SDLs that generate rich experimental data: neural networks could preprocess in situ spectra, atomic force microscopy (AFM) images, microscopy images or process logs into learned feature representations, while the GPs provide predictive uncertainty for acquisition-function-based decision making. However, because DKL introduces additional representation choices and trainable parameters, it requires sufficiently informative data, reliable pretrained representations or careful validation against simpler GP baselines.[103]

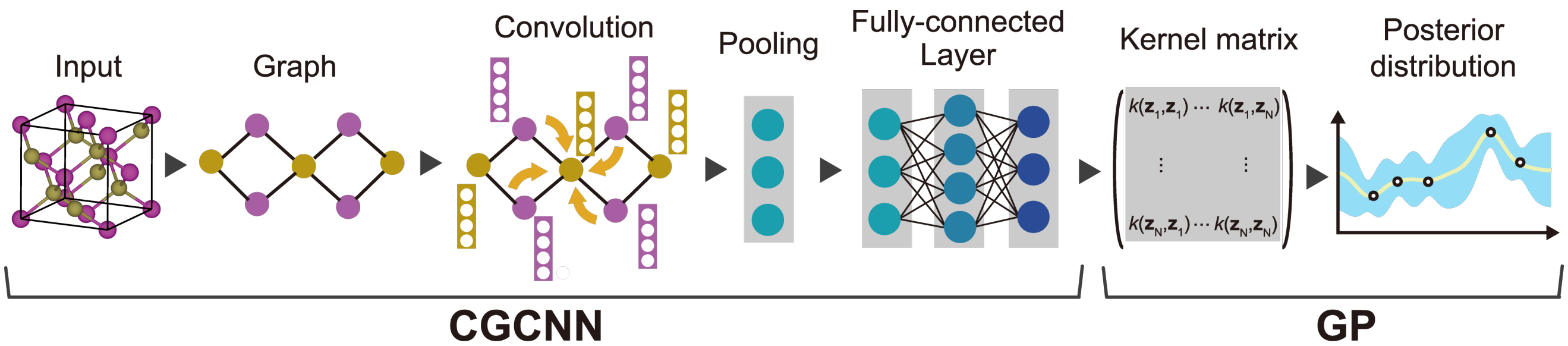

Figure 8. DKL for materials BO. Crystal structures are transformed into learned feature representations by a crystal graph convolutional neural network (CGCNN), from which a GP kernel matrix is constructed to obtain the posterior distribution used for BO. Reproduced from Ref. 100 under a CC BY 4.0 license.

Kernels also provide an important mechanism for incorporating materials knowledge into BO. Similarity can be defined not only in terms of raw experimental parameters, but also through physically meaningful feature representations or descriptors,[56] such as composition-derived features, thermodynamic quantities, lattice mismatch, precursor properties or calculated stability metrics. Physics-informed kernels extend this idea by encoding physical or chemical knowledge about which experiments should produce similar responses. Detailed examples of physics-informed representations and kernels are discussed in Section 6, whereas mixed-variable search spaces involving categorical choices such as substrate, solvent, precursor or processing route are discussed in Section 5.

### 3.4 Structured surrogates for multi-output, multi-fidelity and multi-stage materials workflows

The basic GP setting assumes one scalar objective $y$ observed at one fidelity for each candidate condition. Many materials workflows require a more structured surrogate. The structure should reflect what is actually observed and what decision the BO loop must make next.

In multi-objective BO, each experiment provides a $J$-dimensional response vector, with the responses indexed by $j$ =1, ..., $J$. These responses are collected as $\boldsymbol{y}(\boldsymbol{x}) = [y_1(\boldsymbol{x}), y_2(\boldsymbol{x}), \dots, y_J(\boldsymbol{x})]^{\mathrm{T}}$. For example, $\boldsymbol{y}(\boldsymbol{x})$ may contain activity, stability and selectivity in catalyst optimization; conductivity, transparency and processing temperature in transparent-electrode optimization; or phase purity, growth rate and surface roughness in thin-film growth. The surrogate then represents the corresponding underlying objective functions, $\boldsymbol{f}(\boldsymbol{x}) = [f_1(\boldsymbol{x}), f_2(\boldsymbol{x}), \dots, f_J(\boldsymbol{x})]^{\mathrm{T}}$, from noisy observations, by assuming $y_j(\boldsymbol{x}) = f_j(\boldsymbol{x}) + \epsilon_j$, where $\epsilon_j$ denotes the observational noise for the $j$-th response. A practical baseline for constructing this multi-output surrogate is to model each underlying objective function with an independent GP,

$$f_j(\boldsymbol{x}) \sim \mathcal{GP}(m_j(\boldsymbol{x}), k_j(\boldsymbol{x}, \boldsymbol{x}')),\ j = 1, \dots, J. \tag{3-7}$$

This independent-GP implementation treats the objectives as separate surrogate targets; correlations or trade-offs among objectives are then handled at the acquisition and decision stage rather than through the covariance model. When objectives are correlated through shared physical factors, such as crystallinity affecting both mobility and optical linewidth, a correlated multi-output GP may instead be useful. One common correlated multi-output GP formulation is the intrinsic coregionalization model (ICM), which assumes a separable covariance structure between experimental inputs and objective indices:

$$\mathrm{cov}[f_a(\boldsymbol{x}), f_b(\boldsymbol{x}')] = B_{ab} k(\boldsymbol{x}, \boldsymbol{x}'). \tag{3-8}$$

Here, $k(\boldsymbol{x}, \boldsymbol{x}')$ means the kernel function shared across all objectives that describes the similarity between experimental conditions, whereas $B_{ab}$ is the ($a$,$b$) element of a coregionalization matrix and represents the covariance between the $a$-th and $b$-th objectives.[104,105] Such multi-output GP formulations have also been used for Pareto-front learning with expensive correlated objectives.[106] However, they require enough data to identify inter-objective covariance reliably and can become misleading when correlations are inherently weak or scarcely observable due to limited data coverage.

The Pareto front is the set of best trade-off solutions, where improving one objective requires sacrificing at least one other objective. Pareto-front acquisition functions, such as expected hypervolume improvement (EHVI), use the objective posteriors to select candidates that are expected to expand this front; the detailed workflow of multi-objective acquisition functions is discussed in Section 4.[39–41] The combination of multi-objective surrogates and Pareto-front acquisition functions has been used in materials-optimization and discovery studies, including precipitation-strengthened NiTi shape-memory alloys,[107] NASICON-type solid electrolytes,[108] ion-conductive materials,[109] and combustion-synthesized Pd films in an SDL,[42] as well as constrained multi-objective optimization in SDLs.[50] In terms of surrogate implementation, MacLeod et al. explicitly used independent GPs for the conductivity–processing-temperature Pareto-front search in an SDL,[42] whereas Khatamsaz et al. used a multi-task GP for multi-objective alloy design, with each objective treated as a task in the correlated multi-output model.[110] Independent GPs are usually the safer baseline for small experimental datasets, whereas correlated multi-output GPs may become useful when inter-objective correlations are physically expected and the dataset is large enough to learn them reliably. Systematic materials-SDL benchmarks are still needed to clarify when this additional covariance structure improves, rather than biases, the search.

A simpler alternative for handling multiple objectives is scalarization-based BO, in which multiple objectives are transformed into a single utility $s(\boldsymbol{x})$ before applying a standard single-output surrogate. For example, one simple form assigns a weight $w_j$ to each response $y_j(\boldsymbol{x})$ and takes the following weighted sum:

$$s(\boldsymbol{x}) = \sum_{j=1}^{J} w_j \, y_j(\boldsymbol{x}). \tag{3-9}$$

Other scalarization choices, such as Tchebycheff-type, random-scalarization and desirability-based formulations, can also be used to emphasize different regions of the trade-off surface.[111–114] ParEGO is a representative example of such scalarization-based multi-objective BO.[111] In materials workflows, scalarization is useful when the experimental preference can be encoded as a single evaluation score. For example, scalarized objectives have been used in alloy-design studies for multicomponent alloys and high-entropy-alloy grain-boundary segregation.[115,116] A related thin-film example is BO-guided Metal-Organic Vapor Phase Epitaxy (MOVPE) growth of InGaAs/InP quantum-well structures, where XRD- and PL-derived quantities were combined into one evaluation function before applying GP and EI.[117]

Materials experiments in which some candidate conditions are infeasible or fail to produce usable observations require another layer of structure. If a candidate condition may fail to produce the target phase, cause film peeling, short a device, or violate an unknown process constraint, the workflow should include not only an objective surrogate but also an additional surrogate that predicts whether the proposed experiment will produce a feasible or usable result. Constrained BO methods model the probability that a candidate satisfies unknown constraints or yields a valid observation, and then combine this probability with an objective-related acquisition function.[35–38] Failure-aware strategies may also encode failed observations directly into the objective model when failure can be interpreted as a poor outcome rather than as an unobserved value.[34] These choices are discussed in detail in Section 5. The point here is that feasibility and failure should be represented as modelling targets rather than treated only as missing or discarded data.

Multi-fidelity BO becomes useful when several ways of evaluating a candidate condition are available with different cost, speed and reliability. In materials workflows, a low-fidelity value may come from a simulation, rapid screening measurement, lower-resolution characterization or short-duration test, whereas the high-fidelity value corresponds to the final, more accurate or expensive evaluation of the target property. The purpose is to use cheap but imperfect information to guide the search, while reserving expensive high-fidelity evaluations for candidates that are likely to be informative or promising. In Eq. (3-8), the indices $a$ and $b$ denote different objectives. For multi-fidelity modelling, the same idea can be applied by using indices $r$ and $s$ to label different fidelities or information sources, so that the covariance matrix represents correlations between information sources rather than correlations between objectives. An ICM-type source-indexed surrogate can then be written as

$$\mathrm{cov}[f_r(\boldsymbol{x}), f_s(\boldsymbol{x}')] = B_{rs} k(\boldsymbol{x}, \boldsymbol{x}'), \tag{3-10}$$

where $k(\boldsymbol{x}, \boldsymbol{x}')$ describes similarity between candidate conditions and $B_{rs}$ describes covariance between the $r$-th and $s$-th information sources. This type of source-indexed or multi-information-source surrogate has been used in materials-discovery settings that combine multiple computational or experimental information sources.[44,118]

When the fidelity levels exhibit a clear low-to-high hierarchy, this structure can be naturally incorporated into the model. In autoregressive co-kriging, the high-fidelity response $f_{\text{high}}(\boldsymbol{x})$ is modelled as a corrected version of the low-fidelity response $f_{\text{low}}(\boldsymbol{x})$:

$$f_{\text{high}}(\boldsymbol{x}) = \rho f_{\text{low}}(\boldsymbol{x}) + \delta(\boldsymbol{x}). \tag{3-11}$$

Here, $\rho$ captures the scaling or correlation between low- and high-fidelity responses, whereas $\delta(\boldsymbol{x})$ represents the discrepancy, bias or residual high-fidelity contribution not explained by the low-fidelity source.[43–45] If the low-fidelity model $f_{\text{low}}(\boldsymbol{x})$ and the discrepancy $\delta(\boldsymbol{x})$ are modelled as independent GPs, their linear autoregressive combination, high-fidelity $f_{\text{high}}(\boldsymbol{x})$, is also a GP. The surrogate must therefore represent both correlation and systematic bias between fidelities. If the low-fidelity source is informative, it can guide the search toward promising regions at lower cost; if it is biased or weakly correlated with the final objective, the optimizer must learn when it is worth switching to high-fidelity evaluation. The acquisition function then decides not only which candidate condition to evaluate, but also which fidelity to use. This decision-level aspect is treated in Section 4.

Multi-stage workflows are related to multi-fidelity workflows, but they represent a different type of experimental structure. In multi-fidelity BO, different sources usually provide cheaper or more expensive observations related to the same target quantity. In a multi-stage materials workflow, by contrast, each stage may correspond to a different experimental action and may produce a different type of output. For example, a campaign may involve synthesis, rapid screening, precision characterization, device fabrication and long-term stability testing. These stages are not simply different fidelities of one measurement; they have different costs, risks, timescales and outputs, and the decision at one stage may determine whether the next stage is worth performing. The challenge in modelling the surrogate is therefore to model not only correlations between early and late observations, but also how information propagates through the workflow: an upstream output may become an input or latent state for a downstream stage, while an intermediate measurement may provide only partial or noisy information about the final target. Cascade-type multi-stage BO and proxy-measurement-based multi-stage BO formulations provide representative algorithmic frameworks for this class of problems, with the acquisition- and policy-level aspects discussed in Section 4.[119,120]

More generally, many of these structured surrogates can be formulated as multi-task GPs.[52] The task index may denote an objective, fidelity, substrate, growth chamber, material system or experimental stage, and the covariance function encodes both input-space similarity and task similarity. This enables transfer across related experimental contexts, but can also cause negative transfer when the assumed task relationships are incorrect.[52–55] Structured surrogates therefore provide a bridge between simple BO and realistic SDL operation. They allow the model to represent the structured data produced by real materials experiments, including multiple measured properties, missing or failed observations, low- and high-fidelity evaluations, stage-dependent outputs and related but non-identical experimental tasks. The prediction from such structured surrogates given multiple sources of information forms the basis for acquisition-level decisions, such as which condition, fidelity, information source or experimental stage should be queried next, as discussed in Section 4. Practical issues associated with failure, feasibility and transfer are discussed in more detail in Section 5.

### 3.5 Noise models

Most GP-based BO studies in materials experiments start from a constant-noise assumption, in which the observation-noise variance is treated as independent of the experimental condition, as in Eqs. (2-3) and (2-4). This homoscedastic model is often a robust practical baseline when data are scarce and replicate measurements are limited, and is implemented by adding a diagonal noise variance to the kernel matrix. In scikit-learn,[121] this noise variance is specified by alpha ($\alpha$), whereas BoTorch[122] has a similar adjustable parameter of noise variance. For objective values $y$ standardized to unit variance, very small values such as $\alpha = 10^{-6}$–$10^{-4}$ are employed mainly to improve numerical stability in computations involving the kernel matrix. Noise variances of $\alpha = 10^{-3}$–$10^{-2}$ correspond to assuming that a single measured objective value contains roughly 3–10% observation noise on this normalized $y$ scale, whereas a larger value, $\alpha = 0.1$, corresponds to much noisier observations, with a scatter of about 32% on the same scale.

As illustrated in Fig. 9, $\alpha = 0$ forces the GP to interpolate noisy observations and can overfit small differences among near-repeated measurements. Nonzero $\alpha$ treats each observation as noisy and leaves finite posterior uncertainty at measured points, while repeated or near-repeated observations reduce the uncertainty of the underlying response locally. For nearly deterministic evaluations, such as well-converged simulations or first-principles calculations, $\alpha$ can be set to zero or to a very small value for numerical stability. In experimental BO, however, the noise variance should be chosen to be broadly consistent with the expected experimental scatter. Too small a value can make the GP overfit noisy observations, whereas too large a value can oversmooth the response surface. Replicate measurements, instrument uncertainty and fitting errors provide useful guidance for setting this scale, but the noise variance does not need to be known exactly in advance. When repeated conditions are allowed as candidate actions, BO can use replicates to check reproducibility and reduce uncertainty in the underlying-response estimate by combining multiple noisy observations of the same underlying value.

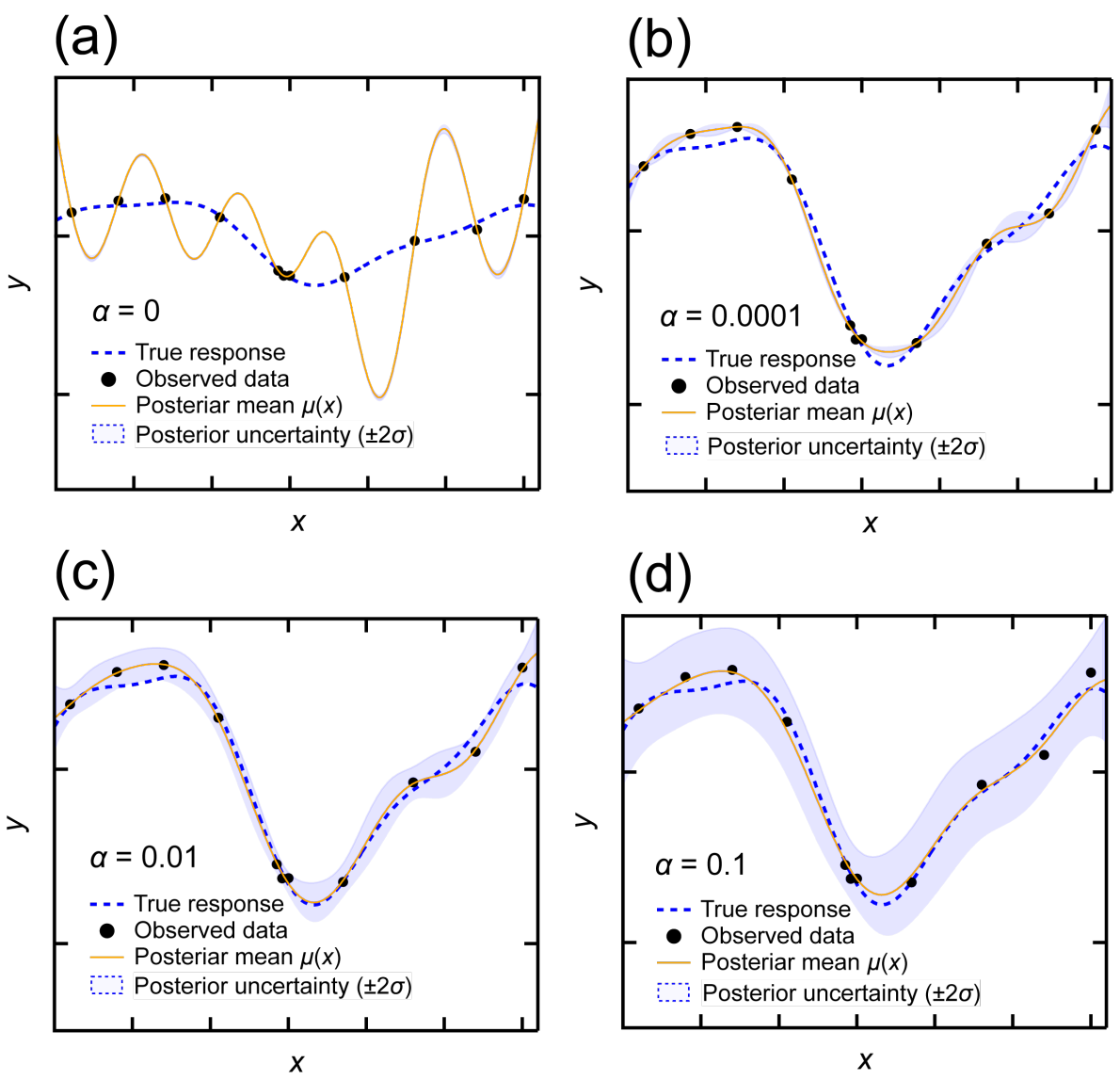

Figure 9. Effect of the diagonal noise variance $\alpha$ in GP regression. GP posterior means and uncertainties are shown for the same synthetic response and observed data using different diagonal noise variances: (a) $\alpha = 0$, (b) $\alpha = 0.0001$, (c) $\alpha = 0.01$ and (d) $\alpha = 0.1$. The dashed curve represents the underlying response, the symbols represent noisy observations, and the shaded regions indicate the posterior uncertainty of the latent response. Nonzero $\alpha$ treats observations as noisy, preventing exact interpolation and leaving finite uncertainty at measured points.

In real materials workflows, however, the reliability and noise level of the extracted objective can vary across the process space. Such an input-dependent noise level can originate from both characterization and synthesis. For example, in optical absorption spectroscopy, the noise level may change across photon-energy or wavelength ranges because different light sources, gratings, filters or detectors are used in different spectral windows. The uncertainty of extracted quantities such as band-gap energy can therefore depend on which spectral region is used for band-gap estimation. In synthesis, the reproducibility of the material response itself can also depend on the process condition; for instance, larger temperature fluctuations at high growth or annealing temperatures may lead to larger run-to-run variations. These effects make the observation noise effectively dependent on the experimental condition, even when the same nominal measurement and analysis protocol is used.

Heteroscedastic noise models, which allow the observation-noise level to vary with the experimental condition, provide one route to representing such input-dependent reliability, but their use in materials BO remains relatively limited. A notable example is autonomous X-ray scattering, where Noack et al. incorporated inhomogeneous measurement noise and anisotropic kernels into GP-driven decision making for polymer-grafted gold-nanorod thin films.[123] Related examples include heteroscedastic GP-based Bayesian experimental design for thermoplastic starch films,[124] noise-aware BO for automated piezoresponse force microscopy in which measurement duration, noise and property discovery are co-optimized,[125] and uncertainty-aware surrogate modelling for high-entropy-alloy datasets containing heteroscedastic, heterotopic and incomplete observations.[126] These studies show that noise-aware modelling can be valuable when measurement reliability is itself structured. At the same time, heteroscedastic modelling introduces a practical identifiability problem: from sparse single-shot experiments, it is difficult to distinguish whether variation in the observed objective reflects a real change in the underlying materials response or a change in the observation noise. Reliable use of such models therefore usually requires replicate measurements at the same or nearby conditions, uncertainty estimates from spectral fitting, detector statistics, or other metadata that directly constrain the noise level. Without such information, a flexible heteroscedastic model can become unstable or poorly calibrated, and a simpler constant-noise model may remain the safer baseline for experimental BO.

### 3.6 Alternative, tree-based and neural surrogates

Although GP is the standard surrogate for materials BO when experimental observations are scarce and expensive, several alternative model classes have also proved useful. Alternative surrogates become attractive when the experimental structure is poorly matched to a single smooth GP defined on a low-dimensional continuous input space. Examples include irregular response surfaces, threshold-like behavior, mixed continuous and categorical variables, and high-dimensional or structured inputs such as spectra, microscopy images, and in situ monitoring signals. This section briefly introduces tree-based models, neural-network-based surrogates and deep Gaussian processes as representative alternatives and extensions. The purpose of using these models is to represent experimental structure that a standard GP does not capture well. Because BO decisions depend on both prediction and uncertainty, the central requirement remains the same: the surrogate must provide uncertainty estimates that are reliable enough to guide acquisition-function-based decisions.

Tree-based models provide one practical class of alternative surrogates. Random forests (RFs),[127] gradient-boosted trees,[128] and XGBoost-type models[129] can represent nonlinear interactions, thresholds, plateaus and conditional structure without imposing a globally smooth covariance function. They are therefore useful when the response surface contains regime-like changes or when the search space contains mixed continuous, discrete and categorical variables.[130] In materials research, Liang et al.[72] benchmarked RF-based BO across multiple experimental materials datasets, including CNT polymer blends, silver nanoparticles, lead-halide perovskites and additively manufactured polymer structures. The authors found that RF-based BO showed performance comparable to GP with ARD, while both approaches outperformed isotropic GP surrogates.[72] Lei et al. explored Bayesian additive regression trees and Bayesian multivariate adaptive regression splines as adaptive non-GP surrogates for automated materials experimental design, showing their usefulness for relatively high-dimensional or non-smooth objective functions.[131] Zhang et al. further compared a random-forest-based uncertainty model with a latent-variable GP for mixed-variable BO in materials design, showing that their relative performance depends on problem dimensionality, complexity, prediction accuracy and uncertainty estimation quality.[132] These studies indicate that tree-based and related non-GP surrogates can be competitive alternatives to GP-based methods in materials BO, particularly for mixed-variable, irregular or non-smooth search spaces. However, their uncertainty estimates are usually obtained from ensembles, quantiles or empirical variability rather than from a calibrated posterior distribution in the GP sense.[133] This distinction matters in BO, because the acquisition function uses uncertainty to decide whether to explore untested regions or exploit promising ones.[15,16] Tree-based models can also be useful for post-hoc interpretation of experimental data through variable importance, partial dependence and interaction analysis.[29,134]

Neural-network-based approaches provide possible alternatives to standard GP, particularly when complex nonlinear mappings or high-dimensional inputs must be represented. Bayesian neural networks, deep ensembles and related neural uncertainty models can process inputs such as composition descriptors, crystal-structure features, spectra, images and process histories. General BO benchmarks have shown that Bayesian neural-network surrogates can be competitive with GP-based surrogates, but their performance depends strongly on the problem setting.[135] In materials BO, however, direct use of neural-network surrogates remains relatively limited. Lim et al. compared GP and neural-network ensemble surrogates on materials datasets and showed that neural-network ensembles can accelerate early convergence in some cases, while optimized GP models can still achieve the best final values.[136] Partially Bayesian neural networks have also been explored for active and transfer learning of molecular and materials properties, providing a related example of uncertainty-aware neural models for sequential data selection rather than a direct BO surrogate.[137] Thus, neural-network-based approaches are most compelling when high-dimensional inputs, learned representations or transferable prior information are central to the problem. For small experimental BO campaigns, however, they should be validated against simpler GP baselines, because flexible neural models can overfit early observations, extrapolate poorly in unexplored regions and produce poorly calibrated uncertainty estimates. Neural-network ensembles can also serve as transfer-learning surrogates by pretraining model parameters on related properties and fine-tuning them using sparse data for the target property;[54] a materials-search example is discussed in Section 5.3.

Deep Gaussian processes provide another extension of the standard GP framework. A deep Gaussian process composes multiple GP mappings, allowing the model to represent hierarchical or nonstationary input–output relationships more flexibly than a single GP.[138] For example, Alvi et al. used prior-guided deep Gaussian processes for uncertainty-aware multi-task prediction of high-entropy-alloy properties, where different but related properties were predicted from incomplete and heterotopic datasets.[126] This example suggests that deep Gaussian processes may be useful when a materials dataset contains multiple correlated targets, missing property values or heterogeneous sources of uncertainty. However, they also introduce additional inference complexity, so their use in experimental BO should be motivated by such structured-data requirements rather than by model complexity alone.

Overall, alternative surrogates should be introduced when motivated by the structure of the experimental problem rather than by model complexity alone. For BO, the decisive issue is not regression accuracy alone, but whether the surrogate provides calibrated uncertainty and decision-relevant structure. For many small-data materials BO campaigns with low- to moderate-dimensional continuous variables, standard GPs therefore remain the most practical baseline, while Table 2 summarizes when more specialized surrogate models may be advantageous.

| Surrogate | Typical use cases | Strength | Consideration |
|---|---|---|---|
| Standard GP | Standard BO baseline; Small data, low- to moderate-dimensional continuous variables | Principled posterior uncertainty; different variable sensitivities can be handled with ARD kernels | Kernel, noise and hyperparameter assumptions matter |
| Structured GP surrogates | Multiple outputs, fidelities, information sources, stages or related tasks are available | Represents experimental structure within a GP framework | Requires enough data to learn structure |
| Tree-based models | Mixed variables, thresholds, conditional recipes or regime-like responses | Handles non-smooth and conditional spaces | Possibly misspecified uncertainty due to ensemble- or quantile-based rather than GP posterior |
| Neural / ensemble models | High-dimensional inputs, spectra/images/rich historical data exist | Learns complex representations and nonlinear mappings | Requires uncertainty calibration; risk of overfitting with small datasets |
| Deep GP | Correlated targets, missing property values or heterogeneous uncertainty sources | Flexible hierarchical or nonstationary modelling | Complex model fitting; unreliable learning with small datasets |

Table 2. Practical selection of surrogate models for materials BO.

## 4. Acquisition Functions and Experimental Design

### 4.1 Acquisition functions as experimental decision rules

Acquisition design is the interface between surrogate prediction and experimental strategy. Based on the surrogate posterior, an acquisition function assigns priorities to possible next actions in the laboratory, such as trying a new synthesis condition, measuring an existing sample, repeating an uncertain experiment, performing a low-cost screening measurement or carrying out a high-fidelity confirmation. The common role of an acquisition function is to balance exploitation and exploration. Exploitation refines regions predicted to perform well, whereas exploration probes regions where the response is uncertain. However, this balance is not determined by the acquisition function alone. It depends directly on how the surrogate posterior represents both the response surface and its uncertainty. Thus, the prior mean, kernel, noise model and task structure discussed in Section 3 all influence the acquisition landscape. If the surrogate prediction or uncertainty is poorly matched to actual material responses, measurement noise or sample-to-sample variation, the optimizer may explore regions that only appear uncertain because of model mismatch, or repeatedly select conditions that only appear promising because of noise. For this reason, acquisition functions should be chosen together with the objective definition, surrogate model and experimental constraints, rather than as isolated algorithmic components. In realistic materials workflows, a well-designed acquisition function can do more than search for a higher scalar objective: it can avoid infeasible regions, allocate measurements across fidelities, or systematically map trade-offs among competing properties in a multi-objective setup. This section therefore treats acquisition functions not only as mathematical criteria, but also as experimental decision rules for materials SDLs.

### 4.2 Basic acquisition functions for scalar objectives

Table 3 summarizes representative acquisition functions for scalar-objective BO. These criteria use the same posterior information in different ways. Among these choices, EI remains a standard starting point because it directly evaluates the expected improvement over the current best value and naturally balances exploitation and exploration, as described in Section 2.[15] Probability of improvement (PI) is simpler, but it considers only the probability of exceeding the current best value, without accounting for the magnitude of the possible improvement, and can therefore become too exploitative.[15] Upper confidence bound (UCB) constructs an optimistic score by adding an uncertainty bonus to the predicted mean, thereby providing direct control over exploration.[16] Larger exploration weights can be useful in early campaigns, rugged response surfaces or situations where missing a distant high-performance region would be costly. Thompson sampling (TS)[139] and information-based acquisitions provide different perspectives on scalar-objective BO. TS is conceptually simple: it samples one plausible response surface from the posterior and selects the best candidate on that sampled surface. This stochasticity can help maintain exploration, especially when multiple candidates are selected in parallel. Entropy search (ES) and max-value entropy search (MES) instead select experiments that are expected to reduce uncertainty about the optimum.[140,141] ES focuses on the location of the optimum, whereas MES focuses on the optimum value. In practice, these criteria may often identify similar informative regions because the optimum location and optimum value are closely related. Knowledge gradient (KG) evaluates the expected increase in the best posterior mean after incorporating the new measurement, and is therefore useful when the value of information is more important than immediate improvement.[142]

In materials applications, EI and closely related improvement-based acquisitions have been widely adopted in BO for materials design and optimization, ranging from solution-processed systems and bulk or powder materials to thin films, devices and process-optimization problems.[61–72] UCB-type acquisitions have also been used when explicit control of exploration is desired, as illustrated by BO-guided CVD growth of CNTs.[60] Entropy-based methods have appeared in materials optimization mainly through constrained or information-oriented formulations, such as active learning of design constraints.[38] TS-type strategies are also becoming relevant for materials SDLs where diverse near-optimal candidates or batch proposals are desired.[143] These examples show that acquisition functions should be chosen according to the goals of the experimental campaign, such as rapid improvement, information gain or generation of diverse batch candidates. Their performance can also depend strongly on the objective definition, input representation, failure handling and surrogate calibration. The adaptive-prior-mean example in Section 3.2 illustrates this point: even with the same EI criterion, changing the prior mean reshapes the posterior and acquisition landscape, thereby altering the selected experiments.[59]

| Acquisition | Representative form | Main idea |
|---|---|---|
| Expected improvement (EI) | $\mathrm{EI}(\boldsymbol{x}) = (\mu(\boldsymbol{x}) - y_{\mathrm{best}} - \xi)\Phi(z) + \sigma(\boldsymbol{x})\phi(z)$, $z = \dfrac{\mu(\boldsymbol{x}) - y_{\mathrm{best}} - \xi}{\sigma(\boldsymbol{x})}$ | Evaluates the expected improvement over the current best value $y_{\mathrm{best}}$. |
| Probability of improvement (PI) | $\mathrm{PI}(\boldsymbol{x}) = \Phi(z), z = \dfrac{\mu(\boldsymbol{x}) - y_{\mathrm{best}} - \xi}{\sigma(\boldsymbol{x})}$ | Selects points with a high probability of improving over the current best value: $P(f(\boldsymbol{x}) \geq y_{\mathrm{best}} + \xi)$. |
| Upper confidence bound (UCB) | $\mathrm{UCB}(\boldsymbol{x}) = \mu(\boldsymbol{x}) + \kappa\sigma(\boldsymbol{x})$ | Selects optimistic candidates by combining predicted mean and uncertainty. |
| Thompson sampling (TS) | Sample $\tilde{f}(\cdot) \sim p(f(\cdot)\|\mathcal{D}_M)$, then choose $\boldsymbol{x}_{M+1} = \operatorname{argmax}_{\boldsymbol{x} \in X} \tilde{f}(\boldsymbol{x})$ | Selects the optimum of one sampled response surface $\tilde{f}(\cdot)$ drawn from the posterior. |
| Entropy search (ES) | Expected reduction in the entropy of $p(\boldsymbol{x}^*\|\mathcal{D}_M)$, where $\boldsymbol{x}^* = \operatorname{argmax}_{\boldsymbol{x} \in X} f(\boldsymbol{x})$ | Selects experiments that are expected to reduce uncertainty about the location of the optimum. |
| Max-value entropy search (MES) | Expected reduction in the entropy of $p(f^*\|\mathcal{D}_M)$, where $f^* = \max_{\boldsymbol{x} \in X} f(\boldsymbol{x})$ | Selects experiments that are expected to reduce uncertainty about the optimum value. |
| Knowledge gradient (KG) | $\mathrm{KG}(\boldsymbol{x})$ is the expected increase in the best posterior mean after measuring at $\boldsymbol{x}$ | Selects experiments with the largest expected value for improving future decisions. |

Table 3. Representative acquisition functions for scalar-objective BO. Here, $\Phi$ and $\phi$ represent the standard normal cumulative distribution function (CDF) and probability density function (PDF), respectively. $\xi \geq 0$ is an improvement margin that defines how much a candidate is required to exceed the current best value, and $\kappa \geq 0$ controls the weight assigned to predictive uncertainty in UCB.

### 4.3 Acquisition functions for multi-objective problems

Many materials problems are inherently multi-objective. A high-performance material may require high activity and stability, high conductivity and transparency, high phase purity and smooth morphology, or high device efficiency and operational stability. In multi-objective BO, weighted sums and other scalarizations are useful when the experimental preference can be encoded as a single utility, as described in Section 3.4. However, reducing several objectives to one scalar value fixes the relative importance of the objectives before the search. As a result, scalarization may hide useful trade-off relationships between objectives. These trade-offs are commonly represented by the Pareto front, which can be viewed as the best achievable trade-offs among competing objectives. More formally, the Pareto front consists of non-dominated combinations of objective values: no other combination is at least as good in all objectives and strictly better in at least one objective. Moving along this front to improve one objective therefore requires sacrificing at least one other objective. Scalarization is therefore appropriate when the relative preferences among objectives can be specified in advance, whereas Pareto-based approaches are more suitable when the trade-offs themselves are of interest.

Figure 10 illustrates these ideas for a two-objective maximization problem. The filled blue points represent the current non-dominated observations, whereas the open circles are evaluated but dominated observations. To quantify how far the Pareto front has advanced, multi-objective BO often uses the hypervolume indicator.[39–41] For example, in the two-objective case shown here, the hypervolume corresponds to the area of objective space dominated by the current Pareto set and bounded by a chosen reference point; in higher-dimensional objective spaces, the corresponding dominated volume is defined analogously. When a new candidate improves the Pareto front, it expands this dominated region; the added area is called the hypervolume improvement, as shown in red in Fig. 10. This idea provides the basis for hypervolume-based multi-objective acquisition functions such as expected hypervolume improvement (EHVI).[39–41] In practice, multi-objective acquisition functions can be evaluated using the posterior distributions of each objective, obtained either from separate single-output surrogates or from a correlated multi-output surrogate, as introduced in Section 3.4.[104–106] Noise-aware variants of EHVI, often referred to as noisy EHVI, extend hypervolume-based acquisition to settings where the observed objectives and the Pareto front are uncertain because of measurement or analysis noise.[41]

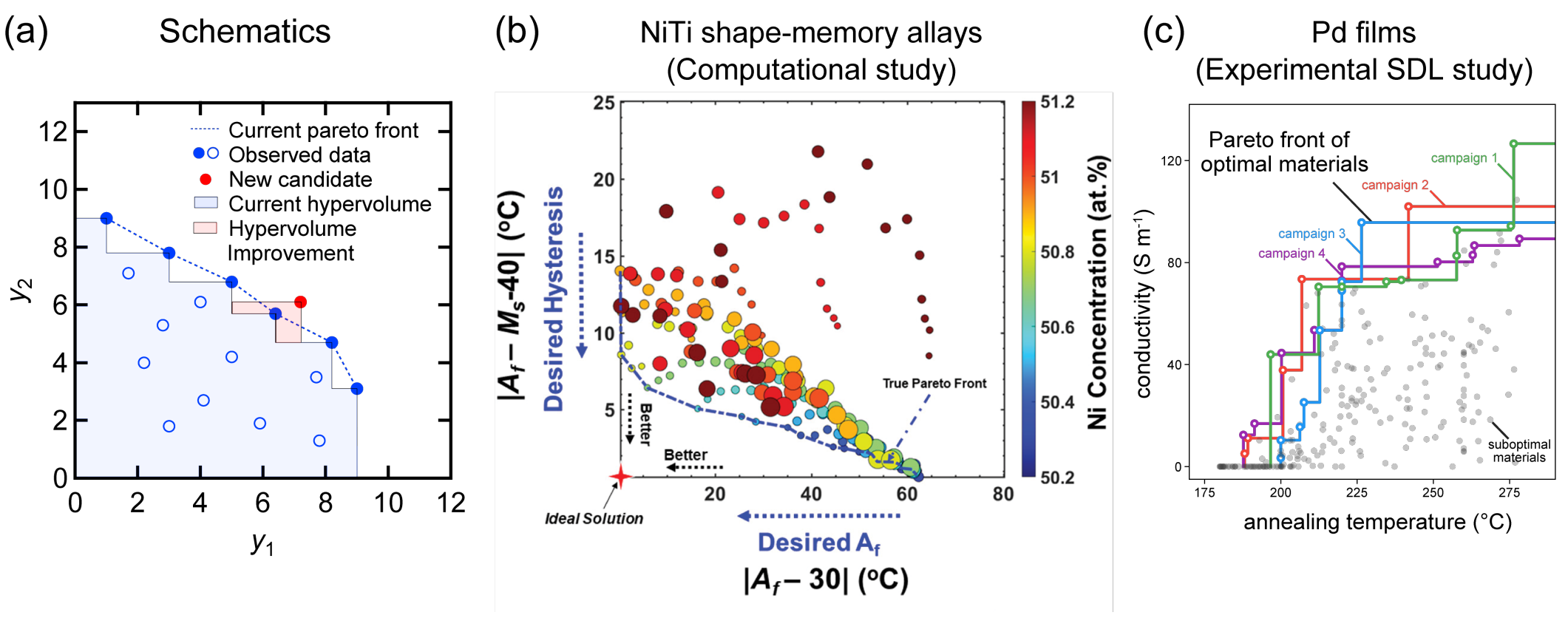

Figure 10. Pareto-front-based multi-objective BO in materials research. (a) Conceptual illustration of the Pareto front, hypervolume and hypervolume improvement in a two-objective optimization problem. Filled blue circles indicate the current non-dominated observations, and the blue shaded region represents the dominated hypervolume relative to a reference point. A new candidate, shown in red, expands this region, giving the hypervolume improvement. (b) Example Pareto front from a computational materials-BO study of NiTi shape-memory alloys. (c) Example Pareto front from an experimental SDL study of Pd films. Panel (b) adapted from Ref. 107 with permission from Elsevier. Panel (c) adapted from Ref. 42 under a CC BY 4.0 license.

Early materials-BO implementations already combined GP-based surrogate modelling with EHVI-type acquisition functions.[107,109,110,115,144,145] For example, Solomou et al. used a GP surrogate and an EHVI acquisition function for multi-objective materials discovery of precipitation-strengthened NiTi shape-memory alloys.[107] MacLeod et al. later provided a representative experimental SDL demonstration, in which separate single-output GP surrogates for each objective were combined with the batch hypervolume-based acquisition function qEHVI to map the Pareto front between conductivity and low-temperature processing in combustion-synthesized Pd films.[42] More recently, multi-objective BO has been demonstrated in constrained SDLs,[50] CNT synthesis,[146] material extrusion,[147] steel heat-treatment design,[148] and high-entropy-alloy discovery.[149] Taken together, these studies indicate that the combination of GP-based surrogates and EHVI-based acquisition functions has become an increasingly standard framework for materials BO when the goal is to optimize experimental or computational candidates while explicitly preserving trade-offs among competing objectives.

### 4.4 Constraint-weighted and feasibility-aware acquisition functions

Constraint handling represents a distinct axis of BO design. Known hard constraints, such as instrument limits, unsafe temperatures, impossible compositions, unavailable substrates or charge-balance requirements, should usually be enforced before acquisition optimization by restricting the parameter search space. Unknown or uncertain constraints, such as target-phase formation, film adhesion, device validity or analysis validity, can instead be modelled probabilistically and incorporated into the acquisition function.[35–38] The constrained optimization problem is to maximize the objective function $f(\boldsymbol{x})$ over $\boldsymbol{x} \in \mathcal{X}$, while satisfying certain constraints indexed by $c = 1, \ldots, C$.

A useful general quantity is the feasibility probability for the $c$-th constraint,

$$p_{\text{feas},c}(\boldsymbol{x}) = P(\text{constraint } c \text{ is satisfied at } \boldsymbol{x}|\mathcal{D}_M), \tag{4-1}$$

where $c = 1, \ldots, C$ indexes the constraints and $\mathcal{D}_M$ is the dataset after $M$ experiments. This probability can be obtained in different ways depending on how the constraint is observed. For some constraints in materials experiments, the available information is naturally binary. For example, after a proposed synthesis or processing condition is evaluated, one may only record whether the target phase formed, whether the film remained attached, whether the device remained electrically usable rather than shorted, or whether the automated analysis returned a valid value. In such cases, the $c$-th constraint can be represented by a binary label,

$$z_c = \begin{cases} 1, & \text{if constraint } c \text{ is satisfied,} \\ 0, & \text{otherwise.} \end{cases} \tag{4-2}$$

A probabilistic classifier, such as a Gaussian process classifier,[14] logistic regression, random forest[150] or a neural network,[151] can then be used to estimate the feasibility probability,

$$p_{\text{feas},c}(\boldsymbol{x}) = P(z_c = 1|\boldsymbol{x}, \mathcal{D}_M). \tag{4-3}$$

This probability directly represents how likely a proposed condition is to satisfy the corresponding feasibility requirement.[35,152,153]

Other constraints are measured as continuous quantities. Examples include requiring a leakage current to remain below an acceptable threshold, a root mean square surface roughness to remain below a specified limit, or an XRD rocking-curve width to be smaller than a target value. In such cases, the $c$-th constraint can be written as a continuous constraint function $g_c(\boldsymbol{x})$, where

$$g_c(\boldsymbol{x}) \leq 0 \tag{4-4}$$

means that the constraint is satisfied. For example, if the leakage current $I_{\text{leak}}(\boldsymbol{x})$ must be smaller than an upper limit $I_{\max}$, one may define

$$g_c(\boldsymbol{x}) = I_{\text{leak}}(\boldsymbol{x}) - I_{\max}. \tag{4-5}$$

If $g_c(\boldsymbol{x})$ is modelled by a GP,

$$g_c(\boldsymbol{x})|\mathcal{D}_M \sim \mathcal{N}(\mu_c(\boldsymbol{x}), {\sigma_c}^2(\boldsymbol{x})), \tag{4-6}$$

then the feasibility probability is obtained from the posterior probability that the constraint function is below zero:

$$p_{\text{feas},c}(\boldsymbol{x}) = P(g_c(\boldsymbol{x}) \leq 0|\mathcal{D}_M) = \Phi(\tfrac{0-\mu_c(\boldsymbol{x})}{\sigma_c(\boldsymbol{x})}), \tag{4-7}$$

where $\Phi$ is the standard normal cumulative distribution function. Thus, $p_{\text{feas},c}(\boldsymbol{x})$ can be estimated either from a binary classifier or from a continuous constraint surrogate, depending on how the constraint is observed. When the constraint is given as a range $\Delta_c^- \leq g_c(\boldsymbol{x}) \leq \Delta_c^+$, its feasibility probability is similarly calculated as follows:

$$p_{\text{feas},c}\left(\Delta_c^- \leq g_c(\boldsymbol{x}) \leq \Delta_c^+|\mathcal{D}_M\right) = \Phi\left(\frac{\Delta_c^+ - \mu_c(\boldsymbol{x})}{\sigma_c(\boldsymbol{x})}\right) - \Phi\left(\frac{\Delta_c^- - \mu_c(\boldsymbol{x})}{\sigma_c(\boldsymbol{x})}\right). \tag{4-8}$$

A constrained acquisition is obtained by multiplying an objective-related acquisition function $a_{\text{BO}}(\boldsymbol{x})$ by the probability that the candidate satisfies all $C$ constraints:

$$a_{\text{con}}(\boldsymbol{x}) = a_{\text{BO}}(\boldsymbol{x}) \prod_{c=1}^{C} p_{\text{feas},c}(\boldsymbol{x}), \tag{4-9}$$

This form gives high priority to candidates that are both promising according to the objective surrogate and likely to satisfy the constraints. A candidate with high predicted performance but low feasibility probability is down-weighted, whereas a moderately promising but reliable candidate may be preferred. Note that Eq. (4-9) assumes all constraints are independent of the objective and all other constraints. When correlations among constraints are important, the product in Eq. (4-9) can be replaced by a structured feasibility model.

An alternative implementation is to filter candidates by feasibility thresholds and then optimize the objective-related acquisition only within the likely feasible region:

$$\boldsymbol{x}_{M+1} = \underset{x \in \mathcal{X}_{\text{hard}}}{\text{argmax}}\, a_{\text{BO}}(\boldsymbol{x}). \tag{4-10}$$

Here, $\mathcal{X}_{\text{hard}}$ denotes the search space after applying known hard constraints. Candidate points are further required to satisfy

$$p_{\text{feas},c}(\boldsymbol{x}) \geq \eta_c \ \ (c = 1, \ldots, C) \tag{4-11}$$

where $0 \leq \eta_c \leq 1$ is a user-defined feasibility threshold for the $c$-th constraint. The weighted form in Eq. (4-9) gives a soft penalty, whereas the thresholded form in Eqs. (4-10) and (4-11) imposes a more explicit probability requirement. The former is useful when constraint violation is undesirable but not critical, while the latter is more appropriate when invalid experiments are expensive, unsafe or difficult to recover from. Safe BO methods provide a related but more conservative framework, in which experiments are restricted to conditions that are predicted to be safe with high confidence. The acquisition policy can then either improve the objective within this safe region or cautiously expand the region by evaluating conditions near its boundary.[154–156]

Representative materials and chemistry applications illustrate these constrained and feasibility-aware strategies. In Si epitaxial thin-film growth, constrained BO was used to maximize the epitaxial growth rate while maintaining five film-quality parameters within acceptable ranges, providing a direct example of threshold-based experimental constraints.[157] Known experimental and design constraints have also been incorporated into BO workflows for chemical optimization.[37] For unknown feasibility constraints, feasibility-aware BO has been demonstrated in the inverse design of hybrid organic–inorganic halide perovskites, where material stability was treated as an initially unknown constraint and learned during optimization.[158] Multi-objective BO has also been applied to refractory multi-principal-element alloy design under composition–property constraints learned through active learning.[38,145] Black-box feasibility constraints have further been incorporated into BO-guided electrospray synthesis of polymeric particles.[159]

In this section, we have focused on the acquisition-level treatment of feasibility, where probabilistic classifiers or continuous constraint surrogates provide feasibility probabilities that weight or filter the objective-related acquisition function. This is distinct from data-level failure treatment, in which failed or missing outcomes are encoded into the surrogate model itself, for example by assigning conservative objective values, imputing invalid outputs, or modelling failure labels together with the objective. Practical choices for handling failed or invalid experiments are discussed in Section 5.1.

### 4.5 Multi-fidelity, cost-aware, batch and asynchronous acquisition functions

Materials research workflows often involve multiple information sources with different levels of accuracy, cost and evaluation time. A simulation is cheaper than synthesis and a rapid optical scan is cheaper than an electrical transport measurement. As illustrated in Fig. 11, multi-fidelity Bayesian optimization (MFBO) combines inexpensive but approximate low-fidelity information with expensive but informative high-fidelity evaluations through a joint probabilistic surrogate that models relationships among the fidelity levels, with the aim of optimizing a black-box objective under a limited overall budget.[160] Letting $\ell$ denote the fidelity or source index, the next action can be written as a pair $(\boldsymbol{x}, \ell)$: which condition to evaluate and at which fidelity. Low-fidelity evaluations are useful when they provide sufficient information about the high-fidelity target relative to their cost. Conversely, repeated evaluation of a weakly correlated or strongly biased low-fidelity source may consume the available budget without substantially improving the final search. The acquisition function should therefore balance the expected improvement or information gain for the high-fidelity objective against the cost of each source and switch to high-fidelity evaluation when the expected value of further low-fidelity sampling becomes small. The surrogate-side problem is to represent the relationships among fidelities, including their cross-fidelity correlation and, depending on the model, systematic bias or discrepancy, as discussed in Section 3.4.

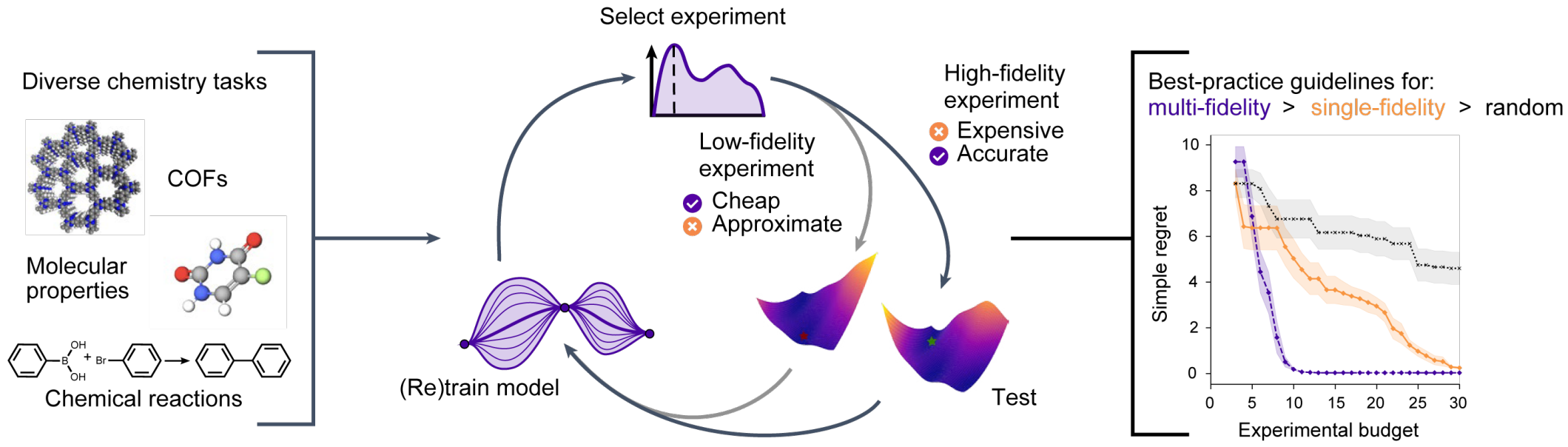

Figure 11. Concept of MFBO for materials and molecular research. MFBO combines inexpensive but approximate low-fidelity information sources with expensive but informative high-fidelity evaluations through a machine-learning model. The optimization loop selects both the candidate condition and the information source to optimize a black-box target under a limited overall budget. Reproduced from Ref. 160 with permission from Springer Nature, copyright 2025.

Representative materials applications illustrate several forms of multi-fidelity and multi-information-source optimization. Multi-information-source BO has been applied to molecular-geometry optimization and binding-energy maximization by combining computational sources with different costs and accuracies.[118] MFBO has also been used to screen covalent organic frameworks (COFs) for xenon–krypton gas separation by combining molecular simulations of different fidelities,[45] and to optimize battery-electrode materials using coin-cell measurements as lower-fidelity approximations of pouch-cell performance.[161] Multi-fidelity multi-objective BO has further been applied to nanophotonic structure design, where coarse- and fine-mesh electromagnetic simulations provide evaluations with different computational costs and accuracies.[162] Related multi-fidelity sequential-learning studies have combined lower-cost semilocal density-functional-theory calculations with experimental measurements, or with more accurate but computationally expensive hybrid-functional calculations, to accelerate the identification of materials with targeted electronic band gaps.[163,164] More recent benchmark work has evaluated MFBO across COF screening, molecular-polarizability prediction and solvation-free-energy optimization tasks.[160] Collectively, these studies demonstrate that MFBO can accelerate materials discovery across molecular, crystalline, electrochemical and photonic systems when lower-fidelity evaluations provide useful information about the high-fidelity target relative to their cost.

Cost-aware BO incorporates such differences by explicitly including the cost of each candidate action in the acquisition-level decision. Let $\boldsymbol{u}$ denote a candidate action, with $\boldsymbol{u} = \boldsymbol{x}$ in a single-fidelity setting and $\boldsymbol{u} = (\boldsymbol{x}, \ell)$ when the fidelity or information source $\ell$ is also selected. A simple cost-normalized acquisition can be written as

$$a_{\mathrm{cost}}(\boldsymbol{u}) = \frac{a_{\mathrm{obj}}(\boldsymbol{u})}{c(\boldsymbol{u})}, \tag{4-12}$$

where $a_{\mathrm{obj}}(\boldsymbol{u}) \geq 0$ is the acquisition value of the action and $c(\boldsymbol{u}) > 0$ is its estimated cost, quantified in terms of the experimental resources consumed and the time required to perform the evaluation. When $a_{\mathrm{obj}}$ is expected improvement, this formulation reduces to expected improvement per unit cost.[165] More generally, $a_{\mathrm{obj}}$ can represent expected improvement, information gain or predictive uncertainty, so that the acquisition prioritizes actions expected to provide the greatest improvement or information per unit cost. Cost-aware BO therefore enables efficient optimization while reducing the experimental time and resources required.

Materials implementations of cost-aware BO include the structure search of face-centered-cubic aluminum grain boundaries, where computational costs vary among orientation-dependent simulation cells because of differences in supercell size,[166] and an in-silico cost-aware batch BO campaign for refractory high-entropy alloys, where different material-property evaluations were assigned different query costs.[167] A direct experimental demonstration has recently been reported for automated nanoindentation of a combinatorial Ta–Ti–Hf–Zr thin-film library, where candidate measurement locations and hold times were selected by favoring measurements with high acquisition value relative to the time required for indentation, stage motion, instrument reconfiguration and drift stabilization.[46]

Batch BO is useful when $q > 1$ samples can be synthesized, processed or characterized in parallel before the surrogate model is updated. Instead of selecting a single condition, batch BO selects $q$ candidate conditions jointly:

$$(\boldsymbol{x}_1^*, \ldots, \boldsymbol{x}_q^*) = \underset{(\boldsymbol{x}_1, \ldots, \boldsymbol{x}_q) \in \mathcal{X}^q}{\operatorname{argmax}} a_q(\boldsymbol{x}_1, \ldots, \boldsymbol{x}_q), \tag{4-13}$$

where $a_q$ evaluates the expected improvement or information obtained from all $q$ candidates while accounting for redundancy among them. Simply choosing the $q$ conditions with the largest individual acquisition values may produce a cluster of similar candidates near the same predicted optimum, resulting in redundant experiments and inefficient use of parallel experimental capacity.

Representative batch-selection strategies address this redundancy in different ways. Sequential local penalization selects candidates one at a time while suppressing the acquisition function around candidates already included in the batch.[49] Posterior sampling, also referred to as Thompson sampling, draws multiple response functions from the surrogate posterior and selects the maximizer of each sampled function as a batch candidate.[168] Parallel predictive entropy search instead selects a batch expected to maximize the information gained about the location of the global optimum.[48] Monte Carlo q-point acquisition functions evaluate a batch jointly under the surrogate posterior rather than treating the $q$ candidates independently. For example, $q$-expected improvement ($q$EI) evaluates the expected improvement in the best outcome among the $q$ evaluations,[169] whereas $q$-expected hypervolume improvement ($q$EHVI) evaluates the expected expansion of the Pareto-dominated hypervolume in multi-objective BO.[40] Parallel knowledge gradient ($q$KG) extends KG to batch selection by evaluating the expected increase in the best posterior mean that would result from observing a batch of $q$ candidates.[170]

Experimental implementations demonstrate several of these batch-selection strategies in materials and chemical SDLs. A posterior-sampling-based extension, nested Thompson sampling, was combined with automated electrochemical experiments to propose batches of 48 electrolyte compositions and collect diverse near-optimal candidates.[143] Batch EHVI strategies have been used to select batches of 24 compositions for refractory-alloy synthesis and characterization[171] and to optimize automated 96-well reaction campaigns using $q$-noisy expected hypervolume improvement ($q$NEHVI).[172] Related parallel experimental operation has also been demonstrated in the Ada spray-coating SDL, where up to four samples were processed simultaneously using a four-mode acquisition policy that cycled among three UCB settings and a space-filling selection.[173] These studies show that batch BO can convert the parallel capacity of high-throughput experimental platforms into faster and more diverse experimental exploration while avoiding unnecessary duplication among simultaneously selected experiments.

Asynchronous BO becomes important when experiments or measurements do not finish at the same time: some complete quickly, whereas others take much longer or have uncertain durations. In synchronous batch BO, the optimizer waits until every experiment in the batch has finished before updating the surrogate and selecting the next candidates. This synchronization can leave instruments, robots or processing stations idle if one evaluation takes substantially longer than the others. Asynchronous BO instead updates the surrogate and acquisition function whenever an individual result becomes available, allowing a new candidate to be assigned immediately to the available experimental resource. The acquisition function must, however, account for pending points, namely conditions that are already being evaluated but whose outcomes remain unknown. Representative strategies include averaging the acquisition function over possible outcomes sampled for evaluations that are still in progress,[16] reducing the acquisition value around ongoing experiments to avoid redundant proposals,[174] and selecting subsequent candidates using asynchronous Thompson sampling or related posterior-sampling methods.[168] Without such treatment, the optimizer may propose near-duplicate conditions while earlier evaluations are still underway.

Experimental implementations of asynchronous BO are beginning to appear in materials research. In a delocalized, asynchronous, closed-loop discovery platform for organic laser emitters, the optimizer continuously updated its ranked recommendations as results became available and, before ranking new candidates, temporarily represented ongoing experiments using plausible outcomes sampled from the current GP.[175] Related algorithmic and simulation studies have examined asynchronous BO using data from a functional-coating SDL[176] and asynchronous multi-fidelity BO for battery-electrode design.[161] These examples illustrate the practical value of asynchronous operation when synthesis, characterization and device testing proceed on different timescales, although explicit experimental demonstrations of general pending-aware BO remain less common than synchronous batch implementations in materials SDLs.

Multi-fidelity, cost-aware, batch and asynchronous acquisitions therefore expand BO from choosing a single experimental condition to choosing a broader experimental action. Depending on the workflow, the algorithm may need to decide not only which condition to test, but also which information source or fidelity to use, which candidates to evaluate in parallel, and which experiment to start when a resource becomes available. In this way, BO becomes part of laboratory scheduling and workflow control.

### 4.6 Multi-stage experimental design

As discussed in Section 3.4, materials development often proceeds through multiple stages, including synthesis, rapid screening, precision characterization, device fabrication and long-term testing. An intermediate proxy, such as a metric derived from XRD or optical measurements, may help identify promising candidates, but the optimum proxy value does not necessarily correspond to the best final device performance. Multi-stage experimental design therefore extends the BO decision from selecting the next experimental condition to selecting which candidate should proceed to which experimental stage next. The surrogate model represents relationships between intermediate observations and the final target, whereas the acquisition policy evaluates possible next actions, such as synthesizing a new sample, performing a low-cost screening measurement, promoting a promising candidate to an expensive downstream test or discontinuing an unpromising candidate. Cascade-type BO models workflows in which the output of one stage becomes an input to the next and allows an unpromising process to be stopped before all stages are completed.[119] More recent multi-stage BO uses intermediate proxy measurements to decide whether a partially evaluated candidate should be advanced to a later, more expensive stage, retained for possible continuation, or discarded.[120] Their applications to materials SDLs remain relatively limited, but they are expected to become increasingly relevant as autonomous workflows connect synthesis with multiple characterization and device-testing stages.

### 4.7 Practical selection of acquisition strategies

No single acquisition function is universally best; its choice should reflect the objective and structure of the experimental workflow. For a scalar objective with scarce data in a low- to moderate-dimensional continuous search space, a GP with EI or UCB provides a practical baseline. More specialized strategies become useful when the optimization problem involves competing objectives, feasibility constraints, multiple fidelities, parallel experiments or multiple experimental stages. Table 4 summarizes representative choices and their main cautions for materials BO and SDL workflows.

| Acquisition strategy | Typical use cases | Strength | Consideration |
|---|---|---|---|
| Scalar-objective acquisition, e.g., EI or UCB | Single scalar objective; scarce data | Simple and practical baseline | The exploration–exploitation balance depends on the acquisition choice and its parameter settings. |
| Multi-objective acquisition | Competing properties | Reveals trade-offs among competing objectives | Scaling and reference-point choices matter |
| Feasibility-aware acquisition | Uncertain constraints or invalid experiments | Prioritizes promising and feasible candidates | Feasibility predictions may be unreliable with scarce data |
| Multi-fidelity / cost-aware acquisition | Information sources with different cost and accuracy | Uses low-cost evaluations to guide expensive tests | Low-fidelity results must be informative about the final target, and costs must reflect actual resource use. |
| Batch / asynchronous acquisition | Parallel experiments or evaluations with different completion times | Improves experimental throughput | Avoid redundant candidates and account for pending experiments |

| Acquisition strategy | Typical use cases | Strength | Consideration |
|---|---|---|---|
| Multi-stage acquisition | Sequential screening and later-stage evaluation | Selects which sample should enter which stage next | Intermediate proxies should reflect the final target |

Table 4. Practical acquisition-function choices for materials BO and SDL workflows.

## 5. Practical deviations from ideal black-box BO

Sections 3 and 4 discussed how experimental responses are modelled by surrogate models and how subsequent experiments are selected by acquisition functions. Real materials experiments, however, often deviate from idealized optimization settings. This section briefly discusses three such issues: failed or missing outcomes, mixed experimental variables, and the reuse of historical information from related experiments, material systems or properties.

### 5.1 Failures, missing data and constraints

In materials experiments, a proposed condition may fail to produce the target material phase, resulting in an unusable objective value, often recorded as NaN (not a number). Simply discarding such synthesis failures ignores information about unfavorable regions and may cause BO to propose similar conditions repeatedly. A simple treatment is floor padding, in which a failed experiment is assigned the worst valid objective value observed so far, allowing the failed experiment to inform the surrogate without requiring a predefined penalty. The floor padding trick converts the original $n$ measurements $y_1, y_2, \ldots, y_n$, which may contain NaNs, to $\tilde{y}_1, \tilde{y}_2, \ldots, \tilde{y}_n$ to train the surrogate model:[34]

$$\tilde{y}_n = \begin{cases} y_n & \text{if } y_n \neq \text{NaN}, \\ \min\limits_{1 \le i < n} \tilde{y}_i & \text{if } y_n = \text{NaN}. \end{cases} \tag{5-1}$$

Wakabayashi et al. compared floor padding alone (F) with floor padding combined with a binary failure classifier (FB) using simulated test functions containing predefined experimental-failure regions (Fig. 12a).[34] The classifier helps avoid conditions likely to fail but can make the search more conservative, particularly when high-performing conditions lie close to a failure boundary. In the corresponding experimental demonstration, floor padding was incorporated into BO-guided MBE growth of the ferromagnetic oxide $SrRuO_3$, producing tensile-strained $SrRuO_3$ films with the highest reported transport quality in only 35 growth runs.[34] Other feasibility-aware approaches based on probabilistic constraint models are discussed in Section 4.4.

Missing outputs caused by measurement, analysis or equipment problems require greater care, because they do not necessarily indicate poor material performance. Such outcomes may instead require remeasurement, an alternative characterization method or separate treatment of measurement validity. Known safety limits, instrument ranges and impossible recipes should be imposed as hard constraints, whereas uncertain feasibility boundaries can be learned from experimental outcomes using the approaches discussed in Section 4.4. Physics-based constraints, such as thermodynamic stability regions, reaction free-energy landscapes, kinetically accessible regions inferred from molecular dynamics or other atomistic simulations, and empirically established process windows, can be incorporated as either hard or soft constraints depending on their reliability and the cost of violating them. Well-established physical limits may reasonably define hard boundaries, whereas approximate or system-dependent knowledge may be introduced as soft guidance to preserve exploratory flexibility. Their physics-informed incorporation is discussed further in Section 6.

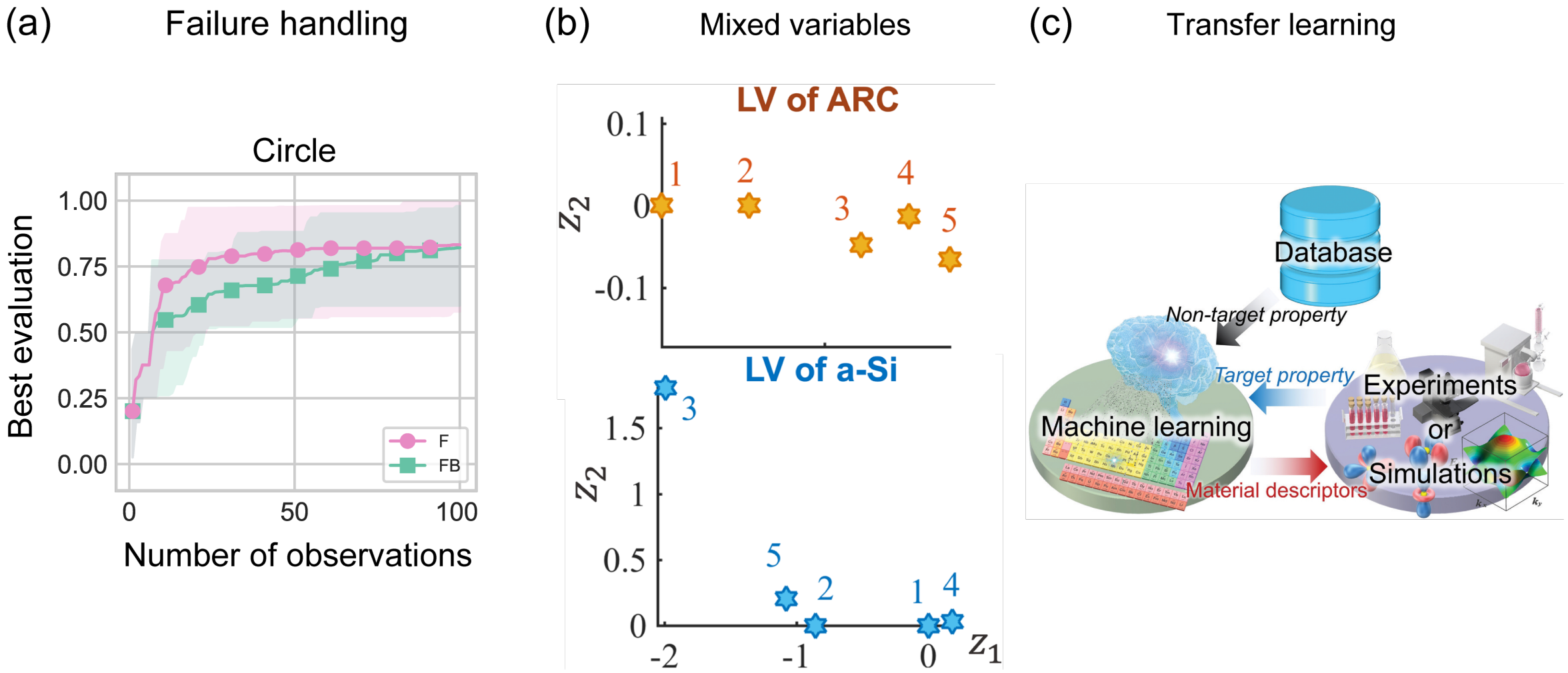

Figure 12. (a) Comparison of failure-handling strategies on simulated test functions containing predefined failure regions. Floor padding (F) improves the objective more rapidly, while floor padding combined with a binary failure classifier (FB) yields a more conservative search. (b) Learned two-dimensional latent-variable representations of categorical anti-reflection coating (ARC) and amorphous-silicon (a-Si) choices. Each numbered symbol denotes a category, and distances between symbols represent their learned similarity in their effects on the objective; the latent axes themselves have no direct physical meaning. (c) Transfer-learning-assisted materials search, in which information learned from non-target properties and material descriptors is transferred to optimization of a target property using data from experiments or simulations. Panels (a)-(c) adapted from Refs. 34, 177 and 54, respectively, under a CC BY 4.0 license.

### 5.2 Discrete, categorical and conditional variables

Materials experiments often combine continuous variables, such as temperature, pressure and concentration, with integer-valued variables, such as the number of deposition cycles or repeated layers, and categorical choices, such as substrate, solvent, precursor or catalyst type. Some variables are also conditional: for example, the concentration and reaction time of a particular precursor are relevant only when that precursor is selected. These variable types require explicit treatment because non-integer values are not experimentally meaningful, categorical labels do not have a natural numerical distance, and some variables are active only under specific experimental choices.

A common way to represent a categorical variable is one-hot encoding, in which each category is represented by a binary vector with one element set to 1 and all others to 0.[177] This makes categorical choices usable by a numerical model, but assigns the same distance to every pair of categories. For example, in a solvent-selection problem, methanol and ethanol may produce similar experimental responses, whereas water may behave very differently. One-hot encoding cannot represent such differences in similarity because all three solvents are treated as equally distinct categories. A latent-variable model instead learns a numerical coordinate for each category from the observed data. Categories that produce similar objective responses are assigned nearby coordinates, whereas categories with different effects are placed farther apart. For example, if methanol and ethanol show similar effects on reaction yield, information obtained for one solvent can help predict the response for the other.

Figure 12b illustrates this idea for a design problem containing categorical choices of anti-reflection coating (ARC) and amorphous-silicon (a-Si) type.[177] In the learned latent space, relative distances between categories reflect similarities in their effects on the objective, whereas the axes $Z_1$ and $Z_2$ do not necessarily correspond to specific physical quantities; the relevant information is primarily the relative distance between categories. In the ARC example in Fig. 12b, the learned arrangement was approximately consistent with the ordering of refractive index, illustrating that a latent representation can sometimes recover a physically interpretable relationship among categories.[177] Tree-based alternatives, particularly for threshold-like or conditional responses, were discussed in Section 3.6.[130,132]

Integer-valued variables should be restricted to allowable whole-number values, such as the number of deposition cycles or repeated layers. When the number of feasible recipes is small, all valid combinations can be listed explicitly and the acquisition function evaluated only over this finite set. For conditional variables, tree-structured dependency models or covariance functions designed for conditional parameter spaces can be used so that only variables active in a given experimental branch contribute to the surrogate.[178,179] For example, annealing temperature and annealing time are active only when post-annealing is selected; when post-annealing is not performed, these variables are ignored rather than treated as ordinary numerical inputs.

### 5.3 Historical data, warm starts and transfer learning

Materials optimization rarely begins without relevant prior experimental data. Previous data may be available from another substrate, composition, instrument or related material system. The simplest reuse strategy is a warm start, in which selected historical observations are added to the initial dataset.[53] This is appropriate when the previous and current experiments have compatible inputs, objectives and measurement scales. When the tasks are related but not identical, information can be transferred through a prior mean, kernel parameters, a learned representation or a multi-task surrogate.[52] Historical data for related, non-target properties can also support transfer learning when the source and target properties depend on shared materials descriptors or underlying physical and chemical factors. For example, historical carrier-mobility data may provide useful source information for predicting semiconductor conductivity because both properties are influenced by composition, defect density and microstructure.

Figure 12c shows a representative example of property-to-property transfer learning.[54] Hwang and Iwasaki[54] used an ensemble neural-network surrogate to search for $B_2$-structured ternary magnetic alloys with high Curie temperature. The model was pretrained using magnetic-moment and/or spin-polarization data and then fine-tuned using a smaller amount of Curie-temperature data. The ensemble mean and variance were used as the prediction and uncertainty, respectively, in a UCB-based autonomous search. Transfer from magnetic-moment data accelerated the early search for alloys with high Curie temperature, whereas transfer from spin polarization alone degraded performance, providing a direct example of negative transfer. The usefulness of transfer should therefore be checked against a baseline model trained only on current-task data. As target-task data accumulate, the benefit of transfer can be reassessed, and source information can be down-weighted when it no longer improves prediction or optimization.[54,55]

Experimental drift creates a related problem. Data obtained even from the same nominal process may become less transferable after changes in chamber history, source condition, precursor lot, calibration or data-analysis procedure. Historical datasets should therefore retain contextual metadata, and reference conditions may need to be repeated to determine whether old and new observations remain comparable. From the BO perspective, drift may be addressed by treating batches or instruments as related tasks,[52] down-weighting or periodically discarding outdated observations,[180] or using nonstationary and change-point surrogate models.[90,93] Because systematic demonstrations of drift-aware strategies in materials SDLs remain scarce, determining whether historical data should be retained, adjusted, down-weighted or excluded remains an important practical challenge.

## 6. Physics-Informed Bayesian Optimization (PIBO)

### 6.1 Why introduce physics knowledge?

Standard BO treats the experimental response surface as a black-box function and guides sequential search using observed data and predictive uncertainty. Materials experiments, however, are rarely complete black-box systems. Thermodynamic models, chemical-reaction mechanisms and kinetic relations, chemical-equilibrium relations, composition–property relations, conservation laws, phase-stability calculations and empirically established process rules may all provide partial information about the response surface or the physically meaningful search space. PIBO refers here to a family of strategies that incorporate physically, chemically or empirically grounded knowledge into BO while retaining data-driven updating and uncertainty-guided experimental selection. This section discusses GP-based BO as the primary framework, though the idea of PIBO is applicable to other classes of surrogate models.

These forms of physics knowledge can enter the BO workflow at several distinct points, as summarized in Fig. 13. On the surrogate-model side, raw experimental set-point parameters can be transformed into physically meaningful representations, quantitative physical relations can be introduced through a prior mean, and physical similarity can be encoded through the kernel. On the acquisition side, physics knowledge can modify the acquisition function or define feasibility constraints. The former approaches shape how the response surface is modelled and how predictions extend across the search space, whereas the latter guide or restrict where BO searches. These approaches are complementary, and several forms of knowledge can be combined within a single workflow. Physics knowledge is particularly valuable when experimental data are scarce and decisions must be made near or beyond the observed domain. In such regimes, it can guide extrapolative prediction and experimental selection and improve interpretability and transferability. However, incorrect or regime-limited assumptions can misguide prediction or exclude unexpected discovery routes. The strength and insertion point of the physics knowledge should therefore reflect its reliability, with robust knowledge imposed more strongly and approximate relations introduced as soft guidance.

In general, physics knowledge can be incorporated into a GP-based BO framework schematically as follows:

$$\boldsymbol{z} = \phi_{\mathrm{phys}}(\boldsymbol{x}) \,, \tag{6-1}$$

$$f(\boldsymbol{z}) \sim \mathcal{GP}(m_{\mathrm{phys}}(\boldsymbol{z}), k_{\mathrm{phys}}(\boldsymbol{z}, \boldsymbol{z}')), \tag{6-2}$$

$$\boldsymbol{x}_{\mathrm{next}} = \underset{\boldsymbol{x} \in \mathcal{X}_{\mathrm{phys}}}{\operatorname{argmax}} \, a_{\mathrm{BO}}(\phi_{\mathrm{phys}}(\boldsymbol{x})) w_{\mathrm{phys}}(\boldsymbol{x}) p_{\mathrm{feas,phys}}(\boldsymbol{x}), \tag{6-3}$$

where $\phi_{\mathrm{phys}}$ is a physics-informed representation, $m_{\mathrm{phys}}$ is a physical or empirical prior mean, $k_{\mathrm{phys}}$ is a physics-informed kernel, $w_{\mathrm{phys}}$ is a weight applied to the acquisition function, and $\mathcal{X}_{\mathrm{phys}}$ is the physically or operationally feasible search space (Fig. 13). Uncertain feasibility constraints can additionally enter the acquisition function through a probabilistic factor $p_{\mathrm{feas,phys}}$. Individual PIBO implementations may use only a subset of these components, as discussed in the following subsections.

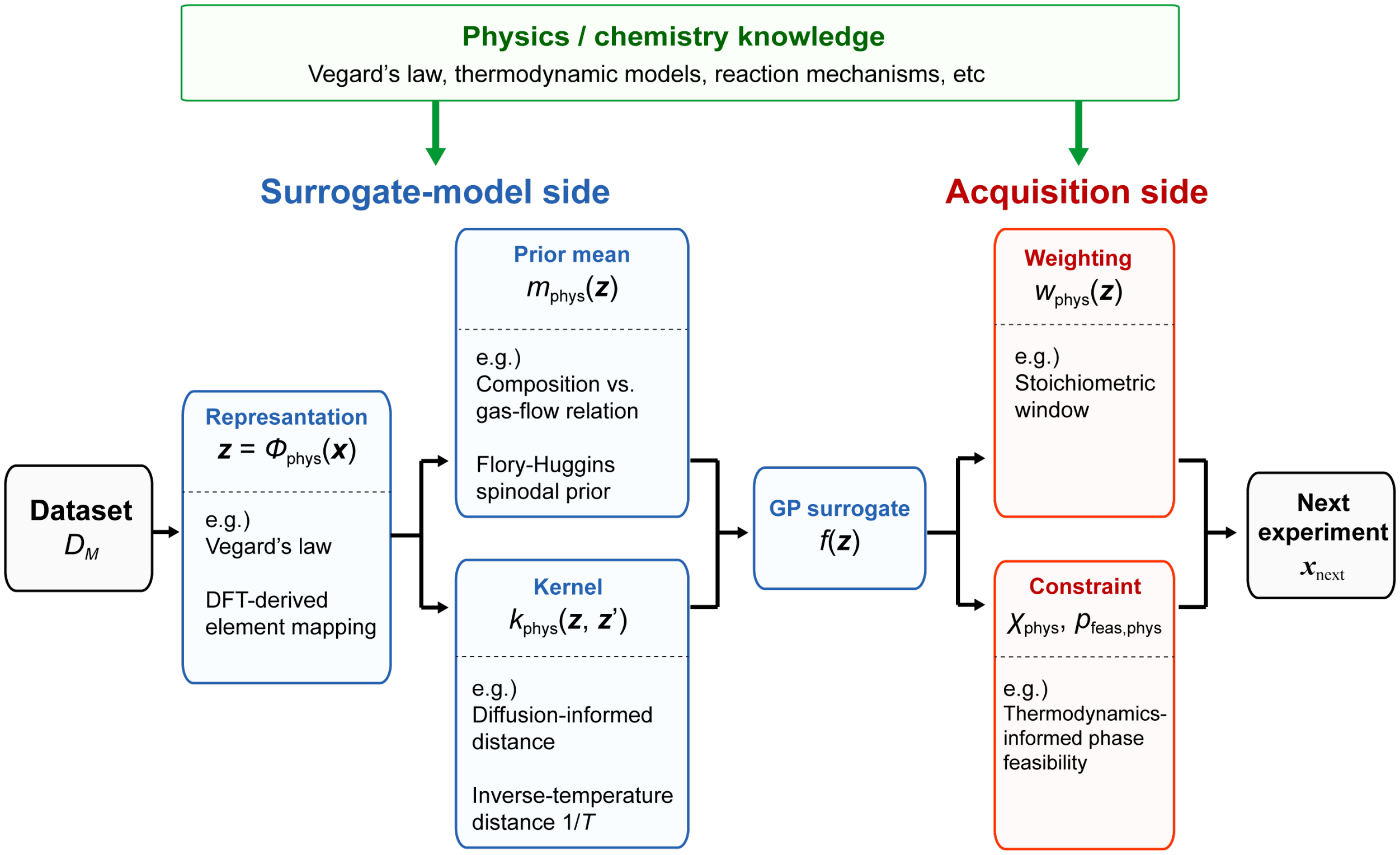

Figure 13. Conceptual framework for PIBO. Physics and chemistry knowledge can be incorporated into the surrogate model through representations, prior means and kernels, or into experimental decision-making through acquisition weighting and feasibility constraints. Representative implementations reported in the literature are shown in each box.

## 6.2 Physics-informed representations

One way to incorporate physics knowledge is through the representation of the experimental problem, including both the experimental conditions and the material quantities to be predicted or targeted. A physics-informed representation maps raw experimental variables or material quantities into coordinates in which the underlying response is expected to vary more smoothly or systematically. For an input-side transformation, this can be written as

$$\boldsymbol{z} = \phi_{\mathrm{phys}}(\boldsymbol{x}) , \tag{6-4}$$

For example, absolute precursor flows may be replaced by flow ratios, nominal elemental choices may be mapped to continuous chemical descriptors, and multiple measured or target quantities may be transformed into common underlying physical variables, such as composition. These transformations change how the experimental problem is represented without prescribing a fixed functional relationship between the transformed variables and the objective. The surrogate model instead learns this relationship from data in coordinates where smoothness and similarity assumptions are expected to be more physically plausible.

Kobayashi et al. used Vegard's law relations[181,182] to transform the target band-gap wavelength $\lambda_{\mathrm{g}}$ and lattice constant $d$ into the corresponding normalized $In_{(1-u)}Ga_uAs_vP_{(1-v)}$ $(0 \leq u \leq 1, 0 \leq v \leq 1)$ composition variables $u$ and $v$, as shown in Fig. 14a. This transformation represented the two target quantities in a common composition space while retaining the correspondence between each target pair and its derived composition. A subsequent physics-informed model related the Ga and As gas flows to these composition variables. Thus, the representation defined the physically meaningful quantities to be predicted, whereas the subsequent model described how they varied with the process conditions.[57]

Representation is also important when the original design variables are discrete. Park et al. developed an element-mapping BO framework for $Na_3V_2(PO_4)_2F_3$-based sodium-ion battery cathodes, in which discrete elemental choices were mapped onto continuous coordinates using density functional theory (DFT)-derived scores of their calculated Na insertion and extraction voltages. This transformation produced a smoother chemical space in which elements with similar effects on the voltage profile were represented more closely, allowing the space to be modelled and searched using a GP.[183]

Representation and kernel design are closely related but conceptually distinct. The representation determines which variables are supplied to the surrogate, whereas the kernel determines how similarity is evaluated within those variables. For example, transforming elemental identity into a continuous physical descriptor is a representation choice, whereas defining a distance or covariance function over those descriptors is a kernel choice. In practice, the two are closely connected because the effectiveness of a distance-based kernel depends strongly on whether the input coordinates reflect physically meaningful variation.

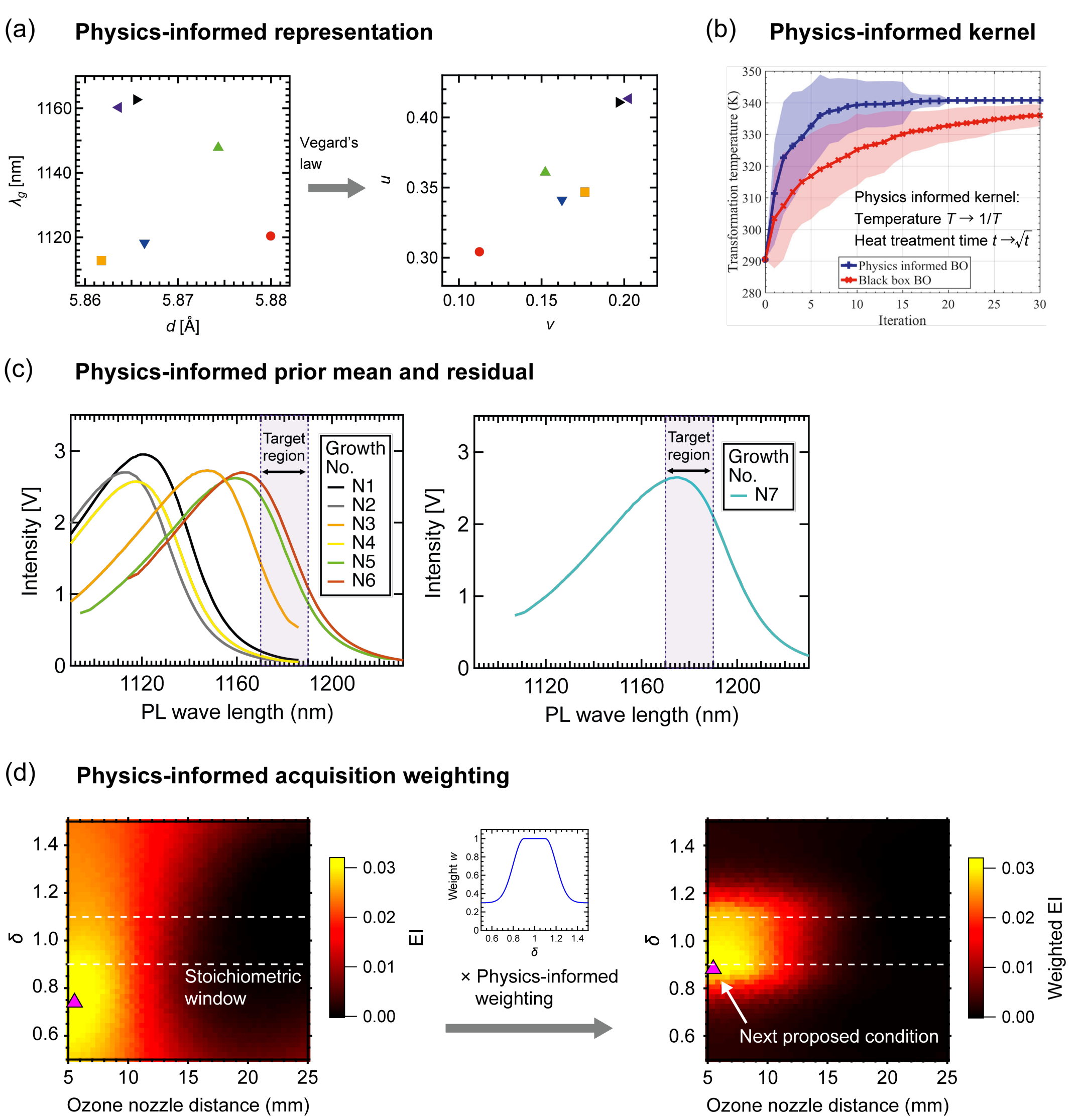

Figure 14. Representative implementations of physics-informed Bayesian optimization in materials synthesis and design. (a) Physics-informed representation for InGaAsP metal-organic chemical vapor deposition, in which the target lattice constant $d$ and band-gap wavelength $\lambda_g$ are transformed using Vegard's law relations into the corresponding normalized composition variables $u$ and $v$. Identical symbols indicate corresponding target and composition pairs. (b) Physics-informed kernel for NiTi shape-memory-alloy design. Similarity between heat-treatment conditions is evaluated using inverse temperature $\frac{1}{T}$ and square-root heat-treatment time $\sqrt{t}$, enabling faster identification of high transformation temperatures than conventional black-box BO. (c) Physics-informed prior-mean and residual modelling for extrapolative InGaAsP growth. The PL wavelengths of the six initial films (N1–N6) lie below the target region, whereas the first additional film selected by PIBO (N7) reaches the target wavelength range. (d) Physics-informed acquisition weighting for $LaAlO_3$ MBE growth. A stoichiometry-based weighting function reshapes the bare-EI landscape toward the expected stoichiometric window while retaining finite acquisition values outside it. The triangles indicate the conditions selected by bare and weighted EI. Panels (a) and (c) adapted from Ref. 57 under a CC BY 4.0 license. Panels (b) and (d) adapted from Refs. 56 and 58, respectively, under a CC BY 4.0 license.

### 6.3 Physics-informed kernels

The GP kernel defines the covariance between response values at two experimental conditions and therefore determines how information is generalized across the search space. A physics-informed kernel evaluates this covariance using physically meaningful similarity rather than relying only on distance in the raw experimental variables. One general form is

$$k_{\mathrm{phys}}(\boldsymbol{x}, \boldsymbol{x}') = k_0(\psi_{\mathrm{phys}}(\boldsymbol{x}), \psi_{\mathrm{phys}}(\boldsymbol{x}')), \tag{6-5}$$

where $\psi_{\mathrm{phys}}$ maps the original variables into physically meaningful variables and $k_0$ is a base kernel that evaluates covariance in the resulting space. Here, the transformation is used specifically to define similarity within the kernel. This differs conceptually from a representation-level transformation, in which the transformed variables are supplied more generally to the surrogate model.

Khatamsaz et al. demonstrated this approach in BO-guided design of NiTi shape-memory alloys, where the objective was to identify alloy compositions and heat-treatment conditions that maximize the martensitic transformation temperature.[56] Because precipitation is diffusion controlled, the heat-treatment time $t$ was transformed to $\sqrt{t}$. The heat-treatment temperature $T$ was transformed to $\frac{1}{T}$ to approximately capture the temperature dependence of the precipitation kinetics, whereas composition was retained in its original form. The kernel distance between two candidate conditions was then calculated using differences in composition, $\sqrt{t}$, and $\frac{1}{T}$, rather than differences in the raw temperature and time variables. These physics-motivated distance measures were used in a modified squared-exponential kernel.

Figure 14b compares the optimization performance of the resulting physics-informed BO with that of conventional black-box BO. The curves show the average maximum transformation temperature found at each iteration over 50 independent optimization runs, with shaded regions indicating 95% confidence intervals. The physics-informed BO approached the theoretical maximum transformation temperature of approximately 340 K within about 15 iterations, whereas the black-box BO did not reach this value within 30 iterations.[56] This example demonstrates that even approximate physical scaling relations can provide a useful covariance structure and substantially improve search efficiency when they capture the dominant variation of the underlying process.

Physics-informed kernels can improve information sharing between physically related conditions and thereby increase data efficiency, as illustrated in Fig. 14b. At the same time, they introduce a relatively strong inductive bias. If the assumed physical scaling is valid only within a limited kinetic or thermodynamic regime, it may produce confidently misleading predictions elsewhere. Their performance should therefore be compared with that of conventional ARD kernels, particularly when the physical scaling is approximate.

### 6.4 Physics-informed prior means and residual models

Unlike representations and kernels, which determine the variables and similarity structure used by the surrogate, a physics-informed prior mean specifies the baseline trend expected for the response. When a physical or empirical model captures the broad behavior of the system but is not sufficiently accurate for direct quantitative prediction, it can be combined with a flexible GP residual:

$$y(\boldsymbol{x}) = m_{\mathrm{phys}}(\boldsymbol{x};\, \theta) + g(\boldsymbol{x}) + \epsilon, \tag{6-6}$$

where $m_{\mathrm{phys}}$ is the available physical or empirical model, $g$ is a GP describing the discrepancy between that model and the actual system, and $\epsilon$ is observational noise.[78] The parameter $\theta$, which denotes the parameters of the physical model, may be fixed from prior knowledge, fitted to the accumulated observations before each GP update, jointly estimated with the GP hyperparameters, or treated probabilistically when their uncertainty is important. This division of roles allows the physical model to provide a physically reasonable global trend, while the GP learns apparatus-specific, sample-specific or otherwise unmodelled deviations.

In materials synthesis, Kobayashi et al. combined the Vegard-law-based representation discussed in Section 6.2 with approximate linear relations between precursor gas flows and InGaAsP compositions.[57] The Ga and As compositions were modelled using linear baseline functions of the corresponding Ga and As gas flows, while GPs learned the residual deviations from these relations. This construction provided a monotonic trend between precursor supply and composition even beyond the observed region, while allowing the GP residual to learn deviations from the assumed linear relationship. Starting from six InGaAsP films with band-gap wavelengths ranging from 1120.4 to 1162.7 nm, the first additional growth selected by PIBO produced a film with a wavelength of 1175 nm, within the target range of 1180±10 nm, as shown in Fig. 14c. X-ray diffraction further confirmed that the film satisfied the target lattice-matching condition. This example shows how a simple physical baseline and a data-driven residual model can together enable experimentally useful extrapolation beyond the observed region.

Physics-informed prior means can also be constructed from more flexible forms of knowledge. Ziatdinov et al. replaced the usual constant GP mean with a structured probabilistic model representing the expected physical behavior.[184] The physical-model parameters were inferred jointly with the GP kernel parameters, allowing uncertainty in the physical model itself to be incorporated into the GP predictions, while the GP captured deviations from the assumed trend. Priyadarshini et al. used physics knowledge to select the model inputs and constructed a data-driven model that served as the GP prior.[185] Sanchez et al. constructed a mean function by combining several candidate models for the composition dependence of the band gap in hybrid perovskites.[186] These examples show that a physics-informed prior need not be a single fixed equation: its parameters may be learned from data, it may itself be constructed using machine learning, or it may combine several candidate physical models.

Hughes et al. experimentally demonstrated a trainable physics-informed prior mean for polymer-blend phase mapping.[81] A Flory–Huggins model of the spinodal curve was incorporated as a trainable structured prior mean, while optical measurements of film cloudiness updated the GP surrogate. The physical prior guided sampling toward the composition–temperature phase boundary, and convergence of the kernel and Flory–Huggins parameters provided an interpretable stopping criterion for the human-in-the-loop campaign.

The main advantage of a physics-informed prior mean is that the expected physical trend is represented explicitly, while the GP captures the observed deviation from that trend. This approach is particularly useful when the direction or broad functional form of a physical relation is more reliable than its absolute numerical predictions, especially outside the region covered by the initial data. The physical baseline should nevertheless be validated against a simpler GP model, while systematic residuals can reveal where the assumed relation breaks down or is only locally valid.

### 6.5 Physics-informed acquisition functions and constraints

Physics knowledge can also be introduced at the candidate-selection stage, complementing its incorporation into representations, kernels and prior means discussed in Sections 6.2–6.4. Because acquisition-level knowledge is applied after the surrogate posterior has been computed, it can be used independently or together with these surrogate-side approaches. A simple implementation multiplies a conventional acquisition function by a non-negative physical weighting function:

$$a_{\mathrm{PI}}(\boldsymbol{x}) = a_{\mathrm{BO}}(\boldsymbol{x}) w_{\mathrm{phys}}(\boldsymbol{x}). \tag{6-7}$$

Here, $a_{\mathrm{BO}}$ is a standard acquisition function such as EI, while $w_{\mathrm{phys}}$ assigns larger values to conditions regarded as physically promising. The GP posterior itself is unchanged; only the acquisition values used to prioritize candidate experiments are modified. Because the weight is decoupled from the surrogate model, it can be added to a standard BO workflow without changing the surrogate model or how its hyperparameters are fitted. Acquisition weighting therefore offers an intuitive and practical way to translate knowledge about preferred regions of the search space directly into candidate selection.

Wakabayashi et al. applied this approach to the MBE growth of perovskite insulator $LaAlO_3$.[58] EI was multiplied by a weighting function derived from the expected La:Al = 1:1 stoichiometric relation. As illustrated in Fig. 14d, the weight was high and nearly constant within the expected stoichiometric window and decreased outside it, thereby reshaping the bare-EI landscape and moving the next proposed condition toward the physically preferred region. A positive minimum weight was retained outside the window so that off-window conditions were penalized rather than excluded. The weighted EI reached a bulk-like lattice constant within 15 growth runs, whereas BO using bare EI required 43 runs to reach a comparably good condition. The final optimum was located slightly outside the initially assumed window, demonstrating that soft acquisition weighting can guide the search without enforcing the prior knowledge as an absolute boundary.

Physics knowledge can also be built into the mathematical definition of the acquisition function itself. Wen et al. developed acquisition functions that measure improvement relative to the currently observed formation-energy convex hull, rather than relative to a single best energy. By prioritizing structures or batches expected to refine this hull, the method directed calculations toward candidate ground-state configurations across the composition space.[187]

Constraints serve a different purpose from acquisition preferences. A known physical, operational or safety constraint can restrict the candidate domain directly:

$$\mathcal{X}_{\text{phys}} = \{\boldsymbol{x} \in \mathcal{X} : g_{\text{phys}}(\boldsymbol{x}) \leq 0\}, \tag{6-8}$$

where $g_{\text{phys}}(\boldsymbol{x})$ represents the physics knowledge as an inequality constraint. Such a hard constraint excludes conditions known to be impossible, unsafe or unacceptable. When feasibility is uncertain a priori but observable as a result of evaluation, it can instead be learned from observations using a probabilistic model $p_{\text{feas,phys}}(\boldsymbol{x})$. The resulting acquisition function can be written as

$$a_{\text{PI}}(\boldsymbol{x}) = a_{\text{BO}}(\boldsymbol{x}) w_{\text{phys}}(\boldsymbol{x}) p_{\text{feas,phys}}(\boldsymbol{x}), \tag{6-9}$$

where the weighting function expresses physical preference, while $p_{\text{feas,phys}}(\boldsymbol{x})$ reduces the priority of candidates that are unlikely to satisfy the required constraints. Here, equation (6-9) has a similar form to the constrained acquisition function in Eq. (4-9). A physics-informed implementation can be obtained by incorporating physics knowledge into the probabilistic feasibility model $p_{\text{feas,phys}}(\boldsymbol{x})$. For example, Hardcastle et al. used computational thermodynamic predictions of solid-solution stability as prior information for GP classifiers and updated these predictions using observed phase labels.[188]

The choice between an acquisition weight and a constraint should reflect the meaning and reliability of the available knowledge. An expected stoichiometric region or preferred process window is generally a search preference and should retain a route to conditions outside it. In contrast, an instrument-damage threshold or a condition known to be physically impossible should be treated as a hard constraint. Boundaries predicted by calculations or inferred from limited data are often better represented by a soft weight or a probabilistic feasibility model rather than imposed as hard constraints. Although acquisition weights and feasibility probabilities can enter the acquisition function in mathematically similar ways, the former typically encode externally specified preferences, whereas the latter are learned or updated from observed feasibility outcomes.

### 6.6 Combining and validating physics-informed strategies

These physics-informed components can be used individually or combined according to the form and role of the available knowledge. A useful design principle is to use representations for physically meaningful variables, prior means for approximate response relations, kernels for similarity assumptions, acquisition weighting for search preferences, and constraints for conditions that must be satisfied. The benefit of physics-informed components should be evaluated against an otherwise comparable physics-agnostic BO baseline. Beyond improvements in optimization efficiency, their broader value lies in providing a natural interface between human expertise and algorithmic decision making. Knowledge accumulated by researchers—such as physically meaningful variables, expected response trends, similarity relations, preferred process regions and practical constraints—can be incorporated through the corresponding components of the BO workflow. Physics-informed BO therefore allows experimental exploration to build on established scientific understanding while new observations continuously refine that knowledge.

## 7. Materials-science and chemical achievements enabled by BO and related active learning

The scientific value of BO-driven materials research should not be assessed only by the number of experiments required to reach an optimum. Its broader scientific impact emerges when closed-loop experimentation enables access to a previously difficult-to-reach synthesis window, accelerates the discovery of new materials or compositions, realizes a targeted material state, or produces samples of sufficient quality to reveal new physical phenomena. Platform-level milestones were introduced in Section 1 and Fig. 1. The present section instead focuses on representative advances in materials science and chemistry enabled by BO and closely related closed-loop active-learning methods. The selected examples span semiconductors, catalysts, chemical reactions, batteries, alloys and quantum materials (Fig. 15).

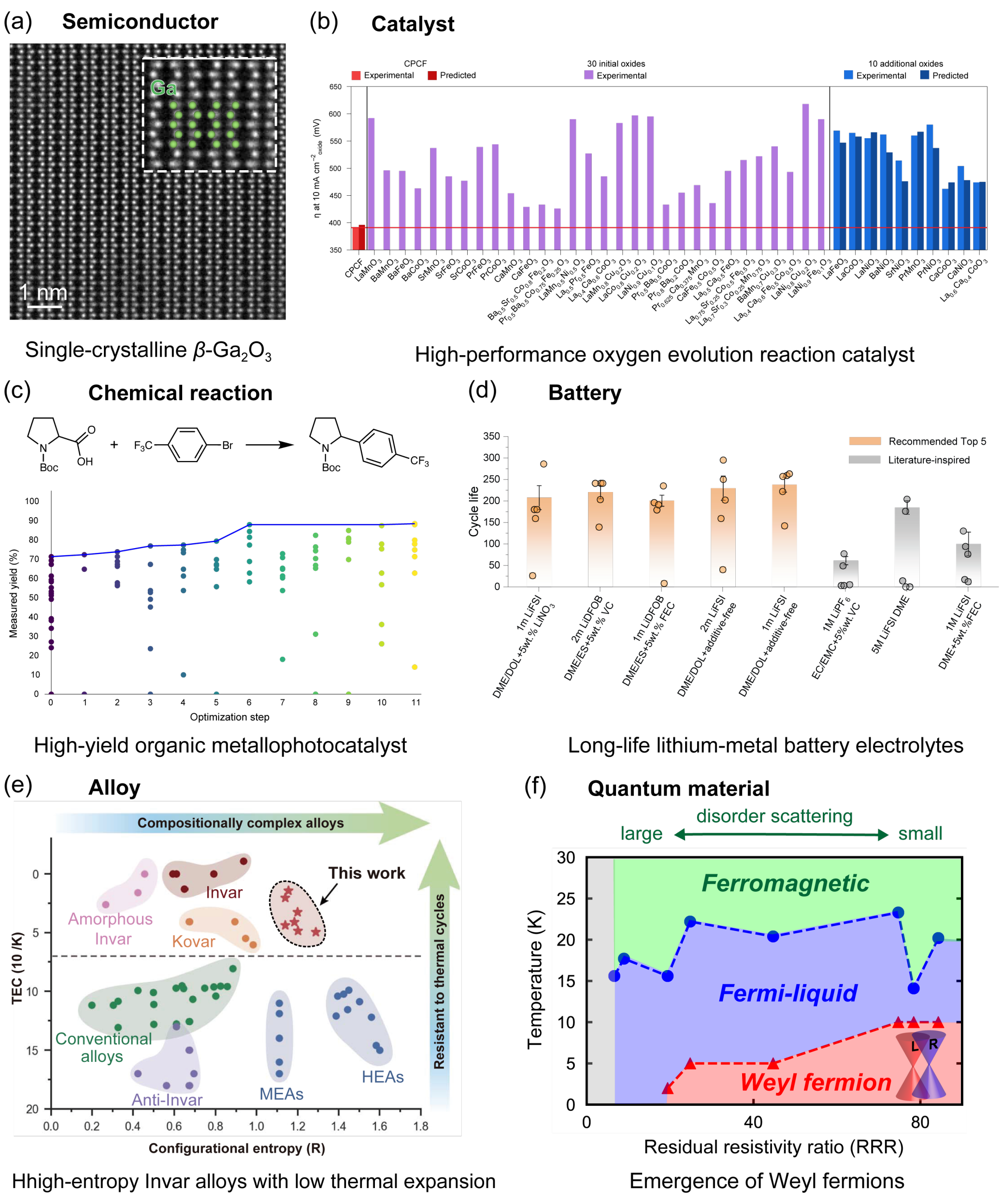

Figure 15. Representative advances in materials science and chemistry enabled by BO and related closed-loop active learning: (a) sputter-grown single-crystalline $\beta$-$Ga_2O_3$ films; (b) a low-overpotential perovskite catalyst; (c) optimized photocatalytic cross-coupling; (d) long-life lithium-metal battery electrolytes; (e) low-thermal-expansion high-entropy Invar alloys; and (f) Weyl fermions in high-quality $SrRuO_3$ films. Panels (a), (c), (d) and (f) reproduced from Refs. 29, 189, 190 and 20, respectively, under a CC BY 4.0 license. Panel (b) reproduced from Ref. 191 with permission from Springer Nature, copyright 2024. Panel (e) reproduced from Ref. 192 with permission from AAAS, copyright 2022.

### 7.1 Semiconductors

Semiconductor growth requires simultaneous control of composition, crystallinity, defects and interfaces, often within narrow and poorly understood process windows. $\beta$-$Ga_2O_3$ is an ultra-wide-band-gap semiconductor with a high critical electric field, making it a promising material for next-generation high-voltage and power-electronic devices.[193,194] Its practical deployment, however, requires scalable deposition methods capable of producing single-crystalline films suitable for device fabrication. Sputtering is attractive for large-area and industrial thin-film processing, but single-crystalline $\beta$-$Ga_2O_3$ films had not previously been realized by this method. An interpretable SDL addressed this practical bottleneck by combining a self-driving sputter system, automated optical characterization and BO.[29] The autonomous closed-loop campaign first searched the growth conditions using the Urbach energy as an optical metric and identified a low-defect heteroepitaxial growth window. The discovered conditions were subsequently transferred to $\beta$-$Ga_2O_3$ substrates without additional optimization, achieving the world's first sputter-grown single-crystalline $\beta$-$Ga_2O_3$ thin films. Atomic-resolution scanning transmission electron microscopy directly confirmed the ordered crystal structure of the homoepitaxial film (Fig. 15a). This result demonstrates that an SDL can overcome a longstanding materials-processing challenge relevant to scalable wide-band-gap semiconductor devices, rather than merely improve an already established deposition process.

The value of the campaign extended beyond the world-first synthesis result. To convert the accumulated closed-loop data into interpretable growth rules, the authors trained a random-forest surrogate and found that the growth landscape was largely captured by additive contributions from the four growth parameters together with a temperature–oxygen interaction. This largely additive structure suggested a human-executable optimization strategy consisting of sequential one-dimensional tuning followed by focused two-dimensional refinement of the temperature and oxygen variables. Human-executed re-optimization validated this strategy by further reducing the Urbach energy to 163 meV. This combination of autonomous synthesis, materials realization and interpretable knowledge extraction illustrates a progression from black-box optimization toward knowledge-generating SDLs.[29]

BO and related machine-learning-assisted growth approaches have also been applied to a range of semiconductor and optoelectronic systems, including InGaAsP alloys,[57] InAs/GaAs quantum dots,[28] InGaAs/InP quantum-well structures,[117] Si epitaxy,[157] III–V thermoelectric thin films,[70] and layered GaSe and $In_2Se_3$ semiconductors.[195,196] These studies address composition and emission-wavelength targeting, simultaneous control of structural and optical properties, constrained epitaxial growth and optimization of transport-related performance.

### 7.2 Catalytic materials

Catalyst discovery involves large composition spaces and strong coupling among composition, structure, surface state and operating conditions. Moon et al. combined catalytic-activity measurements with structural-characterization data and used uncertainty-driven active learning to guide the synthesis and evaluation of four-metal perovskite oxides.[191] Starting from 30 experimentally characterized oxides, the active-learning procedure selected 10 additional compositions to reduce predictive uncertainty across the search space. Rather than directly optimizing catalytic activity as in conventional BO, the active-learning loop selected additional experiments to reduce predictive uncertainty and thereby improve the surrogate model across the candidate space. The resulting model captured the measured activity trends and identified $Ca_{0.8}Pr_{0.2}Co_{0.8}Fe_{0.2}O_{3-\delta}$ (CPCF) as a high-performance catalyst. CPCF exhibited an intrinsic overpotential of 391 mV at 10 mA $cm^{-2}_{oxide}$, the lowest among the 40 perovskite oxides evaluated in the study and among the lowest reported for four-metal perovskite oxides (Fig. 15b). The close agreement between the predicted and experimentally measured overpotentials further demonstrates that structural information can make sparse catalytic datasets sufficiently informative for sequential materials discovery.

Related BO and active-learning studies have addressed other catalyst classes and discovery settings. BO-guided exploration of mesoporous PtPdAu alloys identified non-intuitive compositions with enhanced methanol-oxidation activity using 47 experiments, corresponding to less than 1% of the candidate space.[66] Active machine learning combined with density-functional-theory calculations and experimental validation was also used to identify Cu–Al catalysts for electrochemical $CO_2$ reduction.[197] Collectively, these studies illustrate how BO and closely related active-learning strategies can support catalyst discovery using sparse experiments, computational screening or characterization-enriched datasets.

### 7.3 Chemical reactions

An early systematic demonstration of BO for chemical-reaction optimization was provided by Shields et al., who benchmarked BO against decisions made by expert chemists and engineers for a palladium-catalysed direct arylation reaction.[198] In this benchmark, BO was more consistent and, on average, more experimentally efficient than human decision making. The framework was subsequently applied to practical optimization of Mitsunobu and deoxyfluorination reactions. Li et al. extended closed-loop optimization beyond reaction conditions by treating molecular discovery and reaction formulation as consecutive search problems.[189] In the first workflow, batched BO guided the synthesis and testing of 55 organic photoredox catalysts from a virtual library of 560 molecules. Eighteen carbazole-containing catalysts selected from this stage were then carried into the second campaign in which photocatalyst identity, nickel loading and ligand choice were optimized jointly. Only 107 of 4500 possible formulations were evaluated, yet the maximum cross-coupling yield increased from 71% to 88%, whereas random sampling reached 75% (Fig. 15c). The campaign therefore moved beyond tuning a fixed reaction recipe: the optimizer guided molecular-catalyst selection, catalytic-formulation design and reaction-performance refinement.

Other BO-enabled chemical-reaction workflows include a mobile robotic platform for photocatalyst-formulation optimization,[21] a self-driving catalysis laboratory for mapping Pareto fronts among competing reaction objectives,[199] and a highly parallel automated platform for reaction optimization.[172]

### 7.4 Battery materials

Battery-material optimization often combines discrete choices of electrolyte salts, solvents, additives and electrode constituents with continuous formulation and processing variables.[200,201] Candidate evaluation can also require extended cycling tests, making exhaustive exploration impractical. Electrolyte formulation is particularly challenging because interactions among multiple components can produce discontinuous changes in cell performance. These characteristics illustrate the importance of adapting BO to the structure and practical constraints of the experimental workflow. Hong et al. used a two-stage deep active-learning framework to identify durable electrolytes for lithium-metal batteries.[190] In the first stage, a deep-kernel GP surrogate combined with a Thompson-sampling acquisition function formed a BO loop for searching 720 formulations generated from four lithium salts, six solvents, three additives and three salt concentrations. After three learning iterations, the mean lifetime of the tested Li||Li symmetric cells, which probe the stability of repeated lithium plating and stripping, increased approximately threefold. The five highest-ranked formulations, each evaluated using five independent cells, achieved mean cycle lives of approximately 200–235 cycles, compared with approximately 60–185 cycles for three literature-reported benchmark formulations (Fig. 15d). The campaign also repeatedly evaluated selected formulations to reduce uncertainty arising from substantial cell-to-cell variability, illustrating the importance of replicate measurements in noisy cycling experiments. Target statistic coding was subsequently used to quantify the learned relationships between electrolyte components and performance, enabling this information to be transferred to an expanded space of 5400 formulations and to Li||$LiNi_{0.8}Co_{0.1}Mn_{0.1}O_2$ full-cell optimization. In the latter task, the average 100-cycle capacity retention increased from 58.2% for the initial eight formulations to 84.0% for the second-round top five after one additional learning iteration. This study illustrates how deep-kernel surrogate modelling, Thompson-sampling-based selection, replicate evaluation under noisy experimental outcomes and transfer across related experimental tasks can be combined within a single materials-development workflow.

BO has also been applied to a range of other battery-material problems. Dave et al. coupled robotic electrolyte preparation with BO to optimize ionic conductivity by varying the concentration of lithium hexafluorophosphate and the ratios of three carbonate solvents, and subsequently validated the selected formulations in graphite||$LiNi_{0.5}Mn_{0.3}Co_{0.2}O_2$ pouch cells.[202] BO was also retrospectively benchmarked using an experimentally measured composition library of $LiZr_2(PO_4)_3$-based lithium-ion-conducting solid electrolytes.[108] DFT-derived element mappings have further been used in BO-guided computational design of substituted $Na_3V_2(PO_4)_2F_3$ sodium-ion cathode materials.[183]

### 7.5 Alloys

Alloy design often involves high-dimensional composition spaces coupled to processing-dependent microstructures and properties. This complexity has therefore motivated both physics-informed BO and other closed-loop active-learning frameworks.[203–205] In the NiTi study discussed in Section 6 and Fig. 14,[56] diffusion-informed transformations of heat-treatment temperature and time were incorporated into the GP kernel. The resulting PIBO approached the theoretical maximum martensitic-transformation temperature within approximately 15 iterations, whereas conventional black-box BO did not reach the same value within 30 iterations. This example shows how approximate kinetic knowledge can improve the search even when it is not sufficiently accurate to predict the final property directly.

In an experiment-integrated study, Rao et al. developed a closed-loop active-learning framework for high-entropy Invar alloys.[192] Starting from a database of previously reported Invar alloys, a generative model first proposed unexplored candidate compositions. A fast composition-based ensemble model screened the large candidate pool, after which a smaller set was evaluated using additional descriptors obtained from DFT and thermodynamic calculations. Candidates were then ranked using their predicted thermal-expansion coefficients and model uncertainties. The three highest-ranked compositions were synthesized and characterized, and the measured results were returned to the database to guide the next iteration. After six iterations involving only 17 new alloys from a space containing millions of possible compositions, the campaign identified two high-entropy Invar alloys with thermal-expansion coefficients of approximately $2 \times 10^{-6}$ $K^{-1}$ at 300 K (Fig. 15e). These studies illustrate two ways of incorporating materials knowledge into alloy search: approximate kinetic relations can be encoded directly into a BO kernel, whereas physical descriptors and learned composition representations can guide experimental discovery across vast compositional spaces.

### 7.6 Other functional materials

BO and related active-learning approaches have also been applied to magnetic, superconducting and ferroic materials.[204,206,207] These functionalities can be highly sensitive to subtle variations in elemental composition, atomic-site ordering and interfacial structure. In ultrathin spinel ferrites, for example, Co/Ni substitution and cation inversion strongly modify magnetization, whereas interface-induced changes in cation distribution and Fe valence can produce magnetically dead layers.[208,209] In superconducting TiN, a closed-loop workflow combining BO with metal-organic molecular beam epitaxy identified suitable epitaxial-growth conditions after 11 thin-film experiments, and the resulting films exhibited sharp superconducting transitions above 5 K.[69] For permanent magnets, an active-learning pipeline incorporating machine learning and BO optimized the direct-hot-extrusion conditions of anisotropic Nd–Fe–B magnets, producing high coercivity and remanence and a maximum energy product of approximately 380 kJ $m^{-3}$.[71] In ferroic materials, uncertainty-guided adaptive design identified a lead-free $BaTiO_3$-based piezoelectric composition exhibiting an electrostrain of 0.23%.[67] More broadly, related uncertainty-guided active-learning workflows have combined latent representations of chemical composition and polarization-domain structure with multi-objective evolutionary optimization and closed-loop synthesis of ferroelectric ceramics.[210]

### 7.7 Quantum materials

Quantum materials host electronic states governed by effects such as band topology, quantum geometry, spin–orbit coupling and electronic correlations, often giving rise to unconventional quasiparticles and transport phenomena.[211] Magnetic Weyl semimetals provide a representative example, in which topological band crossings can produce high-mobility carriers and characteristic magnetotransport responses. However, these intrinsic properties can be obscured by disorder and insufficient sample quality.[212] Realizing and experimentally accessing new quasiparticle states therefore often requires materials of sufficiently high quality. A machine-learning-assisted molecular beam epitaxy workflow was therefore implemented for ferromagnetic perovskite $SrRuO_3$, with BO iteratively selecting subsequent growth conditions based on the measured film properties.[19,20] The optimization substantially reduced disorder scattering, as evidenced by an increase in the residual resistivity ratio (RRR). Figure 15f summarizes the resulting relationship between film quality and low-temperature transport. As RRR increases, Fermi-liquid transport becomes established over a wider temperature range, while a low-temperature regime exhibiting quantum transport of Weyl fermions emerges above an RRR of approximately 20 and extends to about 10 K in the highest-quality films. This strong disorder dependence shows that sufficiently clean samples were essential for making the intrinsic quantum transport experimentally accessible. The optimized films exhibited five transport signatures associated with Weyl fermions: unsaturated linear positive magnetoresistance, chiral-anomaly-induced negative magnetoresistance, a non-trivial $\pi$ Berry phase, light cyclotron masses and quantum mobilities of approximately $10^4$ $cm^2$ $V^{-1}$ $s^{-1}$.[20] Together with first-principles calculations and the systematic dependence of these signatures on disorder, these observations established $SrRuO_3$ as a magnetic Weyl semimetal.

The BO-optimized growth process of $SrRuO_3$ films subsequently enabled the observation of high-mobility two-dimensional carriers with quantum mobilities of approximately $3.5 \times 10^3$ $cm^2$ $V^{-1}$ $s^{-1}$.[213] The thickness-dependent phase of the quantum oscillations supported their connection to surface Fermi arcs,[213,214] while measurements up to 52 T revealed chiral-anomaly-related transport in the quantum limit.[213] The resulting capability to reproducibly synthesize high-quality and structurally controlled $SrRuO_3$ subsequently supported studies of highly conductive ferromagnetism down to a single monolayer,[215] spin–orbit-torque-induced partial magnetization switching in a single $SrRuO_3$ layer,[216] disorder- and strain-dependent tuning of anomalous Hall transport,[217,218] and the identification of ligand-O-2*p*-driven magnetic anisotropy and previously unrecognized correlations in ligand-electron states.[219,220] Thus, BO-guided synthesis established not merely a one-time process optimum, but an experimental platform for a sustained sequence of discoveries spanning quantum transport, low-dimensional magnetism, spintronic functionality and correlated electronic states.

## 8. Practical design, implementation and validation of BO-driven SDLs

### 8.1 Problem formulation and BO strategy selection

A BO campaign should begin by defining the scientific goal, one or more optimization objectives, a controllable search space, known constraints, an evaluation budget, and available resources. The dimensionality of the search space should be kept as small as reasonably possible by selecting variables relevant to the objective, because high-dimensional spaces are more difficult to model reliably with sparse experimental data and require a much larger space to be explored. The goal may be to maximize or minimize a material property, approach a target value, identify trade-offs among competing objectives, or optimize performance subject to feasibility constraints. The objective and the procedure used to extract it should be clearly specified, as discussed in Section 2.3. The surrogate model should then be selected according to the structure of the experimental variables and observations, using the practical choices summarized in Table 2 and discussed in Section 3. The acquisition strategy should similarly reflect the experimental decision to be made, using the guidance summarized in Table 4 and discussed in Section 4.

A BO campaign should generally start with the simplest formulation that adequately captures the experimental problem. Tables 2 and 4 provide practical guidance for selecting such a baseline formulation rather than fixed prescriptions. Practical requirements or additional prior knowledge can motivate extensions to the initial simple formulation. Failures, mixed or conditional variables and historical data may require the treatments discussed in Section 5, while reliable physical or chemical knowledge can be incorporated using the PIBO strategies described in Section 6. Any additional model structure should be evaluated against a simpler baseline because sparse experimental data may not support every assumed correlation, interaction or fidelity relationship. Once the optimization problem and BO formulation have been defined, the resulting experimental decisions must be translated into an executable closed-loop workflow. The architecture required for experiment validation, scheduling, measurement analysis and data management is discussed in Section 8.2.

### 8.2 Closed-loop architecture and experimental orchestration

Embedding BO in an SDL requires algorithmic proposals to be reliably connected to experiment execution, measurement analysis and data management. A practical closed-loop architecture can be divided into four functional roles: next-experiment selection, orchestration and execution, measurement analysis, and data management.[221–223] The optimizer proposes candidate conditions within a predefined search space using the accumulated dataset. Known instrument limits, physically invalid combinations and established safety constraints should normally be incorporated into this search space or imposed as hard constraints before acquisition optimization.[37,157] The proposed condition is then passed to an orchestration and execution layer, which translates it into instrument commands, schedules the required operations and checks whether it can be executed under the current laboratory state, including instrument availability, source or reagent availability, calibration validity and hardware interlocks.[221–223]

If a proposed experiment cannot be executed immediately, it may remain queued until the required laboratory state is reached, while another executable experiment proceeds under batch or asynchronous scheduling. Each candidate should therefore have an explicit operational state, such as proposed, queued, running, completed, failed or cancelled. Queued and running experiments should be treated as pending evaluations and communicated to the optimizer to avoid redundant proposals, particularly when synthesis, characterization and analysis are distributed across multiple instruments with different processing times.[174,175,176,222] After execution, the analysis layer converts raw measurements into the scalar or vector objectives used by BO, while the data-management layer stores the proposed and executed conditions, raw and processed data, experimental state and relevant metadata.[221–223] The resulting loop can be summarized as [proposal → validation and scheduling → execution → measurement and analysis → data accumulation → surrogate model update].

Automated measurement analysis is itself becoming an important research topic in SDLs because the objective returned to BO is often not measured directly but inferred from raw characterization data. Depending on the measurement technique, this conversion may rely on physics-based fitting or increasingly on machine-learning-based interpretation. Examples include machine-learning-assisted phase identification from XRD patterns,[224] extraction of band gaps and Urbach energies from optical spectra[29] and of film thickness and optical constants from ellipsometry,[225] and automated extraction of conductivity, carrier density and mobility from electrical and Hall measurements.[226,227] For epitaxial growth, machine-learning-assisted reflection high-energy electron diffraction analysis has been used to predict cation stoichiometry,[228] detect transitions between growth regimes,[229] provide real-time feedback control of quantum-dot growth,[28] and distinguish growth regimes during rapid screening of oxide MBE conditions.[230] Related developments include automated nuclear magnetic resonance interpretation for chemical-mixture analysis,[231] early prediction of battery lifetime from voltage–capacity curves,[232] and autonomous evaluation of reaction yield and selectivity in catalyst optimization.[199]

Implementation may combine software libraries for BO and surrogate modelling, such as BoTorch, COMBO and scikit-learn,[68,121,122] with separate software for laboratory control, workflow orchestration and data management.[221–223,233,234] The software stack should be selected according to the workflow requirements, including constraint handling, batch or asynchronous scheduling, custom objective analysis, experiment logging and instrument interfaces. Before beginning a long autonomous campaign, a pilot workflow should verify instrument communication, objective extraction, failure handling, data logging and emergency procedures over representative regions of the search space. This preliminary validation can reveal implementation errors that would otherwise compromise the reliability of the subsequent BO campaign.

### 8.3 Data and analysis quality, logging and experimental context

Because BO decisions depend directly on the objective values returned by the analysis layer, data- and analysis-quality control is an integral part of the SDL. The preprocessing operations, fitting ranges, feature-extraction methods and validity criteria used to obtain each objective should be defined and recorded. Where possible, each objective should be accompanied by indicators of its reliability, such as fitting uncertainty or signal-to-noise ratio, together with a flag indicating whether the analysis passed predefined validity criteria.

Invalid outcomes should also be classified according to their cause. A synthesis failure, such as failure to form the target phase, may provide information about an unfavorable region of the search space and may be incorporated into the objective or feasibility model, as discussed in Sections 5.1 and 4.4. By contrast, a measurement failure, communication error or analysis failure does not necessarily indicate poor material performance and may instead require remeasurement. Experiments terminated by a hardware interlock or cancelled by an operator represent distinct operational outcomes and should also be recorded separately. These outcomes should not automatically be reduced to the same missing value.

Experimental inputs $\boldsymbol{x}$ are often represented by instrument set points or proxy measurements rather than by the process conditions actually realized at the sample. For example, substrate temperature may be represented by a heater set point or thermocouple reading, even though the substrate-surface temperature can differ substantially.[235,236] Because the relationship between set points and realized conditions may change with calibration drift, chamber history, source depletion or environmental variation, the set points, any available measurements of realized conditions and the relevant calibration state should be retained. Periodic calibration or reference experiments can be used to determine whether observations collected at different stages of a campaign remain comparable and, where feasible, can be incorporated into the SDL workflow automatically.

Instrument states should be recorded but should not automatically be exposed to the optimizer as decision variables. Experimentally controllable variables, such as temperature, gas flow or applied power, normally define the search space. Other states, such as chamber history, source usage time or calibration status, are generally not directly controllable and should remain metadata when they are needed only for operational checks or drift diagnosis. If such a state reproducibly affects the material response, it may instead be included as a contextual input to the surrogate model without being treated as a variable that the acquisition optimizer can freely select. Continuous quantities such as source usage time may be represented as contextual covariates,[237] whereas discrete differences among instruments or batches may be represented as task indices in a multi-task model.[52] This separation limits the number of decision variables while allowing systematic variation across time, batches or instruments to be modelled. To ensure traceability, each objective value should remain linked to the executed experimental conditions, raw measurements, analysis procedure and relevant instrument state.

### 8.4 Reporting practices for reproducibility

Reproducibility requires more than reporting the best objective value as a function of iteration number. A materials-BO study should describe the optimization problem, algorithmic configuration, execution workflow and treatment of non-ideal outcomes. The appropriate level of detail depends on the study; Table 5 summarizes key information that supports reproducibility and comparison among BO-driven materials SDLs.

The distinction between iteration count and wall-clock time is particularly important. An SDL may achieve a similar optimum in approximately the same number of experiments as a manual campaign while substantially reducing elapsed time through unattended or parallel operation. Conversely, an algorithm may reduce the number of trials but provide little practical acceleration if characterization, queueing or manual sample transfer remains the rate-limiting step. Both quantities should therefore be reported whenever possible.

Clear documentation of optimization, orchestration, analysis, and data management helps connect algorithmic decisions to the experiments actually performed. It also links returned objective values to the corresponding raw measurements, analysis procedures, and experimental context. Such traceability supports reliable BO updates, meaningful comparison among SDL studies and future reuse of experimental data and process knowledge.

| Category | Information to report |
|---|---|
| Problem definition | Experimental search space, variable types and units, objective definition, known hard constraints, uncertain constraints and any normalization or transformation of inputs and outputs |
| Campaign design | Initial experimental design, number of initial and BO-selected trials and stopping criterion |
| BO configuration | Surrogate model (including prior mean and kernel for GP surrogates), noise model, acquisition function, acquisition optimizer, hyperparameter fitting procedure, random seeds and relevant software versions |

| Category | Information to report |
| --- | --- |
| Execution and orchestration | How proposals were validated and translated into instrument actions, scheduling procedure, batch size and treatment of pending experiments |
| Objective extraction | Raw measurement data, preprocessing and fitting procedure, quality thresholds and the software or model used to convert measurements into objectives |
| Failures and missing data | Definition of experimental failures and how failed or missing observations were handled in the BO algorithm |
| Operational performance | Wall-clock time, throughput, manual interventions and the baseline for any reported acceleration |
| Traceability and data availability | Calibration and instrument metadata, timestamps, and availability of code, processed data and raw experimental data |

Table 5. Key information recommended for reporting BO-driven materials SDLs.

## 9. Outlook

The next generation of BO-driven materials SDLs should be judged not only by how efficiently they identify promising experimental conditions or materials with desired properties, but also by whether they can remain reliable under experimental drift, learn from multimodal data across partially characterized samples, revise their objectives as new evidence emerges, and extract interpretable and reusable knowledge. BO should therefore evolve from optimizing fixed objectives toward enabling adaptive, knowledge-generating experimentation. Realizing this vision requires addressing several key challenges.

The first challenge is nonstationarity and experimental drift. Long autonomous campaigns do not operate under perfectly stationary conditions because the experimental system itself changes over time. Source materials are gradually depleted, chamber surfaces evolve, catalysts lose activity, precursor batches are replaced, and samples age. As a result, the relationship between nominal experimental conditions and measured outcomes may vary during a campaign, and data collected earlier may become less representative of the current system. Future SDLs will therefore require time-aware, state-aware or change-point-aware surrogate models, together with reference experiments that detect drift and assess whether historical observations can still be reused.[90,93,180,238] Experimental context and relevant instrument metadata should also be recorded and incorporated into the model when necessary.

The second challenge is learning from multimodal and incompletely characterized experimental datasets. Materials experiments generate heterogeneous observations, including spectra, diffraction patterns, microscopy images, electrical measurements and process histories.[239] Routine measurements such as XRD or optical spectra may be available for most samples, whereas microscopy, transport measurements or device tests may be performed only for selected candidates because they are expensive, destructive or time-consuming. The available modalities therefore often differ across the dataset. Structured multi-output, multi-task, multi-fidelity and multi-stage models provide useful starting points for this problem, but future SDLs will need to extend these ideas to heterogeneous multimodal observations and datasets in which different samples are characterized by different subsets of measurements. Future SDLs will require models that integrate these heterogeneous observations, remain robust when some measurements are missing, and provide uncertainty estimates that reflect the reliability of predictions based on incomplete characterization data.[126,240]

The third challenge is adaptive problem formulation and human-in-the-loop scientific reasoning. Conventional BO assumes that the objective, constraints and search space remain fixed throughout a campaign. Real scientific research evolves as new evidence emerges: unexpected observations may motivate a new objective, a revised constraint, an additional characterization method or a different hypothesis. Human researchers must therefore remain able to inspect, redirect and reinterpret the closed loop. Large language models and research agents may help interpret accumulated evidence and propose next steps,[241–243] but their recommendations should remain transparent and reviewable. Instrument control and data processing should remain validated, reproducible and traceable, while human researchers retain responsibility for major scientific decisions.[244]

The most important challenge is to move from predictive optimization toward interpretable, testable and transferable knowledge. Achieving this requires an SDL to determine which variables and interactions control the response, formulate hypotheses for why the observed outcomes arise, identify the assumptions under which the inferred rules remain valid, and test whether those rules transfer across substrates, instruments, laboratories or related material systems. Post-hoc surrogate analysis, sensitivity analysis, physics-informed models, causal reasoning and intervention-oriented experimental design can help generate and test such hypotheses.[29,245,246] Future SDLs should then select and execute, within the synthesis and characterization capabilities available to the platform, experiments that distinguish among competing explanations. The resulting evidence should update both the predictive model and the underlying scientific hypotheses. Closing this loop would advance interpretable SDLs that extract and experimentally validate human-usable process rules[29] toward knowledge-generating SDLs that autonomously formulate hypotheses, select and execute experiments that discriminate among competing explanations, and refine scientific knowledge while evaluating its transferability.

Taken together, the practical BO design principles discussed throughout this review provide an experiment-centered starting point for this transition. The central idea is to translate challenges encountered in materials experiments—such as failed synthesis, competing properties, variable measurement cost, limited transferability and available physics knowledge—into coordinated choices of objective, surrogate model, acquisition function, constraints and experimental workflow. This perspective treats BO not as a collection of isolated algorithms, but as a framework for designing the entire experimental decision loop. Extending this framework to nonstationary, multimodal and adaptively reformulated campaigns will be important for advancing BO-driven SDLs from efficient optimization toward reliable, interpretable and knowledge-generating experimentation.

**Author contributions**

Y. K. W. defined the scope and structure of the review and wrote the original draft. T. O. contributed to discussions and reviewed the technical content. Both authors revised the manuscript and approved the final version.

**Conflicts of interest**

There are no conflicts of interest to declare.

**Data availability**

No new data were generated or analyzed as part of this review.

**References**

1. H. Park, H. Park, K. Song, S. H. Song, S. Kang, K.-H. Ko, D. Eum, Y. Jeon, J. Kim, W. M. Seong, H. Kim, J. Park and K. Kang, In situ multiscale probing of the synthesis of a Ni-rich layered oxide cathode reveals reaction heterogeneity driven by competing kinetic pathways, Nat. Chem., 2022, 14, 614–622.
2. Y. K. Wakabayashi, Y. Ban, S. Ohya and M. Tanaka, Annealing-induced enhancement of ferromagnetism and nanoparticle formation in the ferromagnetic semiconductor GeFe, Phys. Rev. B, 2014, 90, 205209.
3. E. Kuphal and A. Pöcker, Phase diagram for metalorganic vapor phase epitaxy of strained and unstrained InGaAsP/InP, Jpn. J. Appl. Phys., 1998, 37, 632–637.
4. A. Murata and Y. Hori, Product selectivity affected by cationic species in electrochemical reduction of CO2 and CO at a Cu electrode, Bull. Chem. Soc. Jpn., 1991, 64, 123–127.
5. L. Zhu, M. Zhang, Z. Zhou, W. Zhong, T. Hao, S. Xu, R. Zeng, J. Zhuang, X. Xue, H. Jing, Y. Zhang and F. Liu, Progress of organic photovoltaics towards 20% efficiency, Nat. Rev. Electr. Eng., 2024, 1, 581–596.
6. X. Gao, X. Jia, Y. Mo, C. Wang, Q. Li, T. Bu, Y.-B. Cheng and F. Huang, Additive-assisted liquid medium annealing relieving strains in perovskite solar cells for improved stability, Nat. Energy, 2026, 11, 1194–1202.
7. C. Lee, Y.-S. Yoo, B. Ki, M.-H. Jang, S.-H. Lim, H. G. Song, J.-H. Cho, J. Oh and Y.-H. Cho, Interplay of strain and intermixing effects on direct-bandgap optical transition in strained Ge-on-Si under thermal annealing, Sci. Rep., 2019, 9, 11709.
8. C. Wang, Y. Zhao, T. Ma, Y. An, R. He, J. Zhu, C. Chen, S. Ren, F. Fu, D. Zhao and X. Li, A universal close-space annealing strategy towards high-quality perovskite absorbers enabling efficient all-perovskite tandem solar cells, Nat. Energy, 2022, 7, 744–753.
9. Y.-S. Shiah, K. Sim, Y. Shi, K. Abe, S. Ueda, M. Sasase, J. Kim and H. Hosono, Mobility–stability trade-off in oxide thin-film transistors, Nat. Electron., 2021, 4, 800–807.
10. H. J. Kushner, A new method of locating the maximum point of an arbitrary multipeak curve in the presence of noise, J. Basic Eng., 1964, 86, 97–106.
11. J. Mockus, On Bayesian methods for seeking the extremum, Optimization Techniques IFIP Technical Conference, Springer, Berlin, Heidelberg, 1975, 400–404.
12. J. Mockus, Bayesian Approach to Global Optimization: Theory and Applications, Kluwer Academic Publishers, Dordrecht, 1989.
13. C. K. I. Williams and C. E. Rasmussen, Gaussian processes for regression, Adv. Neural Inf. Process. Syst., 1996, 8, 514–520.
14. C. E. Rasmussen and C. K. I. Williams, Gaussian Processes for Machine Learning, MIT Press, Cambridge, MA, 2006.
15. D. R. Jones, M. Schonlau and W. J. Welch, Efficient global optimization of expensive black-box functions, J. Glob. Optim., 1998, 13, 455–492.
16. J. Snoek, H. Larochelle and R. P. Adams, Practical Bayesian optimization of machine learning algorithms, Adv. Neural Inf. Process. Syst., 2012, 25, 2951–2959.
17. P. Nikolaev, D. Hooper, F. Webber, R. Rao, K. Decker, M. Krein, J. Poleski, R. Barto and B. Maruyama, Autonomy in materials research: a case study in carbon nanotube growth, npj Comput. Mater., 2016, 2, 16031.
18. B. P. MacLeod, F. G. L. Parlane, T. D. Morrissey, F. Häse, L. M. Roch, K. E. Dettelbach, R. Moreira, L. P. E. Yunker, M. B. Rooney, J. R. Deeth, V. Lai, G. J. Ng, H. Situ, R. H. Zhang, M. S. Elliott, T. H. Haley, D. J. Dvorak, A. Aspuru-Guzik, J. E. Hein and C. P. Berlinguette, Self-driving laboratory for accelerated discovery of thin-film materials, Sci. Adv., 2020, 6, eaaz8867.
19. Y. K. Wakabayashi, T. Otsuka, Y. Krockenberger, H. Sawada, Y. Taniyasu and H. Yamamoto, Machine-learning-assisted thin-film growth: Bayesian optimization in molecular beam epitaxy of SrRuO3 thin films, APL Mater., 2019, 7, 101114.
20. K. Takiguchi, Y. K. Wakabayashi, H. Irie, Y. Krockenberger, T. Otsuka, H. Sawada, S. A. Nikolaev, H. Das, M. Tanaka, Y. Taniyasu and H. Yamamoto, Quantum transport evidence of Weyl fermions in an epitaxial ferromagnetic oxide, Nat. Commun., 2020, 11, 4969.
21. B. Burger, P. M. Maffettone, V. V. Gusev, C. M. Aitchison, Y. Bai, X. Wang, X. Li, B. M. Alston, B. Li, R. Clowes, N. Rankin, B. Harris, R. S. Sprick and A. I. Cooper, A mobile robotic chemist, Nature, 2020, 583, 237–241.

22. P. M. Attia, A. Grover, N. Jin, K. A. Severson, T. M. Markov, Y.-H. Liao, M. H. Chen, B. Cheong, N. Perkins, Z. Yang, P. K. Herring, M. Aykol, S. J. Harris, R. D. Braatz, S. Ermon and W. C. Chueh, Closed-loop optimization of fast-charging protocols for batteries with machine learning, Nature, 2020, 578, 397–402.
23. N. J. Szymanski, B. Rendy, Y. Fei, R. E. Kumar, T. He, D. Milsted, M. J. McDermott, M. Gallant, E. D. Cubuk, A. Merchant, H. Kim, A. Jain, C. J. Bartel, K. Persson, Y. Zeng and G. Ceder, An autonomous laboratory for the accelerated synthesis of inorganic materials, Nature, 2023, 624, 86–91.
24. A. G. Kusne, H. Yu, C. Wu, H. Zhang, J. Hattrick-Simpers, B. DeCost, S. Sarker, C. Oses, C. Toher, S. Curtarolo, A. V. Davydov, R. Agarwal, L. A. Bendersky, M. Li, A. Mehta and I. Takeuchi, On-the-fly closed-loop materials discovery via Bayesian active learning, Nat. Commun., 2020, 11, 5966.
25. J. Chen, S. R. Cross, L. J. Miara, J.-J. Cho, Y. Wang and W. Sun, Navigating phase diagram complexity to guide robotic inorganic materials synthesis, Nat. Synth., 2024, 3, 606–614.
26. R. Shimizu, S. Kobayashi, Y. Watanabe, Y. Ando and T. Hitosugi, Autonomous materials synthesis by machine learning and robotics, APL Mater., 2020, 8, 111110.
27. S. B. Harris, A. Biswas, S. J. Yun, K. M. Roccapriore, C. M. Rouleau, A. A. Puretzky, R. K. Vasudevan, D. B. Geohegan and K. Xiao, Autonomous synthesis of thin film materials with pulsed laser deposition enabled by in situ spectroscopy and automation, Small Methods, 2024, 8, 2301763.
28. C. Shen, W. Zhan, K. Xin, M. Li, Z. Sun, H. Cong, C. Xu, J. Tang, Z. Wu, B. Xu, Z. Wei, C. Xue, C. Zhao and Z. Wang, Machine-learning-assisted and real-time-feedback-controlled growth of InAs/GaAs quantum dots, Nat. Commun., 2024, 15, 2724.
29. Y. K. Wakabayashi, Y. Ogawa, F. B. Romero, C. Wagner, T. Otsuka and Y. Taniyasu, Interpretable self-driving sputtering epitaxy reveals human-usable growth rules for $\beta$-$Ga_2O_3$ films, Nat. Commun., 2026, DOI: 10.1038/s41467-026-76533-0.
30. C. W. Coley, N. S. Eyke and K. F. Jensen, Autonomous discovery in the chemical sciences part I: progress, Angew. Chem. Int. Ed., 2020, 59, 22858–22893.
31. M. M. Flores-Leonar, L. M. Mejía-Mendoza, A. Aguilar-Granda, A. Sánchez-Lengeling, H. Tribukait and A. Aspuru-Guzik, Materials acceleration platforms: on the way to autonomous experimentation, Curr. Opin. Green Sustain. Chem., 2020, 25, 100370.
32. M. Abolhasani and E. Kumacheva, The rise of self-driving labs in chemical and materials sciences, Nat. Synth., 2023, 2, 483–492.
33. N. Yoshikawa, Y. Asano, D. N. Futaba, K. Harada, T. Hitosugi, G. N. Kanda, S. Matsuda, Y. Nagata, K. Nagato, M. Naito, T. Natsume, K. Nishio, K. Ono, H. Ozaki, W. Shin, J. Shiomi, K. Shizume, K. Takahashi, S. Takeda, I. Takeuchi, R. Tamura, K. Tsuda and Y. Ushiku, Self-driving laboratories in Japan, Digit. Discov., 2025, 4, 1384–1403.
34. Y. K. Wakabayashi, T. Otsuka, Y. Krockenberger, H. Sawada, Y. Taniyasu and H. Yamamoto, Bayesian optimization with experimental failure for high-throughput materials growth, npj Comput. Mater., 2022, 8, 180.
35. M. A. Gelbart, J. Snoek and R. P. Adams, Bayesian optimization with unknown constraints, Proceedings of the Thirtieth Conference on Uncertainty in Artificial Intelligence (UAI), 2014, 250–259.
36. J. M. Hernández-Lobato, M. A. Gelbart, M. W. Hoffman, R. P. Adams and Z. Ghahramani, Predictive entropy search for Bayesian optimization with unknown constraints, Proc. ICML, 2015, 37, 1699–1707.
37. R. J. Hickman, M. Aldeghi, F. Häse and A. Aspuru-Guzik, Bayesian optimization with known experimental and design constraints for chemistry applications, Digit. Discov., 2022, 1, 732–744.
38. D. Khatamsaz, B. Vela, P. Singh, D. D. Johnson, D. Allaire and R. Arróyave, Bayesian optimization with active learning of design constraints using an entropy-based approach, npj Comput. Mater., 2023, 9, 49.
39. M. Emmerich, K. Giannakoglou and B. Naujoks, Single- and multiobjective evolutionary optimization assisted by Gaussian random field metamodels, IEEE Trans. Evol. Comput., 2006, 10, 421–439.
40. S. Daulton, M. Balandat and E. Bakshy, Differentiable expected hypervolume improvement for parallel multi-objective Bayesian optimization, Adv. Neural Inf. Process. Syst., 2020, 33, 9851–9864.
41. S. Daulton, M. Balandat and E. Bakshy, Parallel Bayesian optimization of multiple noisy objectives with expected hypervolume improvement, Adv. Neural Inf. Process. Syst., 2021, 34, 2187–2200.
42. B. P. MacLeod, F. G. L. Parlane, C. C. Rupnow, K. E. Dettelbach, M. S. Elliott, T. D. Morrissey, T. H. Haley, O. Proskurin, M. B. Rooney, N. Taherimakhsousi, D. J. Dvorak, H. N. Chiu, C. E. B. Waizenegger, K. Ocean, M.

Mokhtari and C. P. Berlinguette, A self-driving laboratory advances the Pareto front for material properties, Nat. Commun., 2022, 13, 995.

43. K. Kandasamy, G. Dasarathy, J. Schneider and B. Póczos, Multi-fidelity Bayesian optimisation with continuous approximations, Proc. ICML, 2017, 70, 1799–1808.
44. M. Poloczek, J. Wang and P. I. Frazier, Multi-information source optimization, Adv. Neural Inf. Process. Syst., 2017, 30, 4288–4298.
45. N. Gantzler, A. Deshwal, J. R. Doppa and C. M. Simon, Multi-fidelity Bayesian optimization of covalent organic frameworks for xenon/krypton separations, Digit. Discov., 2023, 2, 1937–1956.
46. V. Chawla, S. Puplampu, H. Zhu, P. D. Rack, D. Penumadu and S. Kalinin, Cost-aware Bayesian optimization of real-world nanoindentation workflows for accelerated mechanical characterization, Digit. Discov., 2026, DOI: 10.1039/D5DD00518C.
47. A. Desautels, A. Krause and J. Burdick, Parallelizing exploration-exploitation tradeoffs in Gaussian process bandit optimization, J. Mach. Learn. Res., 2014, 15, 4053–4103.
48. A. Shah and Z. Ghahramani, Parallel predictive entropy search for batch global optimization of expensive objective functions, Adv. Neural Inf. Process. Syst., 2015, 28, 3330–3338.
49. J. González, Z. Dai, P. Hennig and N. Lawrence, Batch Bayesian optimization via local penalization, Proc. AISTATS, 2016, 51, 648–657.
50. A. K. Y. Low, F. Mekki-Berrada, A. Gupta, A. Ostudin, J. Xie, E. Vissol-Gaudin, Y.-F. Lim, Q. Li, Y. S. Ong, S. A. Khan and K. Hippalgaonkar, Evolution-guided Bayesian optimization for constrained multi-objective optimization in self-driving labs, npj Comput. Mater., 2024, 10, 104.
51. C. Tamura, H. M. Job, H. Chang, W. Wang, Y. Liang and S. Sun, Autonomous organic synthesis for redox flow batteries via flexible batch Bayesian optimization, Digit. Discov., 2025, 4, 2737–2751.
52. K. Swersky, J. Snoek and R. P. Adams, Multi-task Bayesian optimization, Adv. Neural Inf. Process. Syst., 2013, 26, 2004–2012.
53. M. Poloczek, J. Wang and P. I. Frazier, Warm starting Bayesian optimization, 2016 Winter Simulation Conference (WSC), 2016, 770–781.
54. J. Hwang and Y. Iwasaki, Improving efficiency of autonomous material search via transfer learning from nontarget properties, Sci. Technol. Adv. Mater.: Methods, 2023, 3, 2254202.
55. N. Yoshida, Y. Iwabuchi, Y. Igarashi and Y. Iwasaki, Networking autonomous material exploration systems through transfer learning, npj Comput. Mater., 2025, 11, 362.
56. D. Khatamsaz, R. Neuberger, A. M. Roy, S. H. Zadeh, R. Otis and R. Arróyave, A physics informed Bayesian optimization approach for material design: application to NiTi shape memory alloys, npj Comput. Mater., 2023, 9, 221.
57. W. Kobayashi, T. Otsuka, Y. K. Wakabayashi and G. Tei, Physics-informed Bayesian optimization suitable for extrapolation of materials growth, npj Comput. Mater., 2025, 11, 36.
58. Y. K. Wakabayashi, T. Otsuka, Y. Krockenberger and Y. Taniyasu, Physics-informed acquisition weighting for stoichiometry-constrained Bayesian optimization of oxide thin-film growth, NPG Asia Mater., 2026, DOI: 10.1038/s41427-026-00668-1.
59. Y. K. Wakabayashi, T. Otsuka, Y. Krockenberger, H. Sawada, Y. Taniyasu and H. Yamamoto, Stoichiometric growth of SrTiO3 films via Bayesian optimization with adaptive prior mean, APL Mach. Learn., 2023, 1, 026104.
60. J. Chang, P. Nikolaev, J. Carpena-Núñez, R. Rao, K. Decker, A. E. Islam, J. Kim, M. A. Pitt, J. I. Myung and B. Maruyama, Efficient closed-loop maximization of carbon nanotube growth rate using Bayesian optimization, Sci. Rep., 2020, 10, 9040.
61. P. I. Frazier and J. Wang, Bayesian optimization for materials design, in Information Science for Materials Discovery and Design, ed. T. Lookman, F. J. Alexander and K. Rajan, Springer, Cham, 2016, 45–75.
62. P. V. Balachandran, D. Xue, J. Theiler, J. Hogden and T. Lookman, Adaptive strategies for materials design using uncertainties, Sci. Rep., 2016, 6, 19660.
63. A. Seko, A. Togo, H. Hayashi, K. Tsuda, L. Chaput and I. Tanaka, Prediction of low-thermal-conductivity compounds with first-principles anharmonic lattice-dynamics calculations and Bayesian optimization, Phys. Rev. Lett., 2015, 115, 205901.
64. S. Ju, T. Shiga, L. Feng, Z. Hou, K. Tsuda and J. Shiomi, Designing nanostructures for phonon transport via Bayesian optimization, Phys. Rev. X, 2017, 7, 021024.

65. D. Xue, P. V. Balachandran, J. Hogden, J. Theiler, D. Xue and T. Lookman, Accelerated search for materials with targeted properties by adaptive design, Nat. Commun., 2016, 7, 11241.
66. A. S. Nugraha, G. Lambard, J. Na, M. S. A. Hossain, T. Asahi, W. Chaikittisilp and Y. Yamauchi, Mesoporous trimetallic PtPdAu alloy films toward enhanced electrocatalytic activity in methanol oxidation: unexpected chemical compositions discovered by Bayesian optimization, J. Mater. Chem. A, 2020, 8, 13532–13540.
67. R. Yuan, Z. Liu, P. V. Balachandran, D. Xue, Y. Zhou, X. Ding, J. Sun, D. Xue and T. Lookman, Accelerated discovery of large electrostrains in BaTiO3-based piezoelectrics using active learning, Adv. Mater., 2018, 30, 1702884.
68. T. Ueno, T. D. Rhone, Z. Hou, T. Mizoguchi and K. Tsuda, COMBO: An efficient Bayesian optimization library for materials science, Mater. Discov., 2016, 4, 18–21.
69. I. Ohkubo, Z. Hou, J. N. Lee, T. Aizawa, M. Lippmaa, T. Chikyow, K. Tsuda and T. Mori, Realization of closed-loop optimization of epitaxial titanium nitride thin-film growth via machine learning, Mater. Today Phys., 2021, 16, 100296.
70. T. Ishiyama, K. Nozawa, T. Nishida, T. Suemasu and K. Toko, Bayesian optimization-driven enhancement of the thermoelectric properties of polycrystalline III–V semiconductor thin films, NPG Asia Mater., 2024, 16, 17.
71. G. Lambard, T. T. Sasaki, K. Sodeyama, T. Ohkubo and K. Hono, Optimization of direct extrusion process for Nd-Fe-B magnets using active learning assisted by machine learning and Bayesian optimization, Scr. Mater., 2022, 209, 114341.
72. Q. Liang, A. E. Gongora, Z. Ren, A. Tiihonen, Z. Liu, S. Sun, J. R. Deneault, D. Bash, F. Mekki-Berrada, S. A. Khan, K. Hippalgaonkar, B. Maruyama, K. A. Brown, J. Fisher III and T. Buonassisi, Benchmarking the performance of Bayesian optimization across multiple experimental materials science domains, npj Comput. Mater., 2021, 7, 188.
73. S. Suzuki, T. Dazai, Y. Yamamoto, H. Koinuma and R. Takahashi, Accelerating the combinatorial optimization process for phosphor materials by Bayesian optimization, Jpn. J. Appl. Phys., 2023, 62, 117001.
74. K. Nagai, T. Osa, G. Inoue, T. Tsujiguchi, T. Araki, Y. Kuroda, M. Tomizawa and K. Nagato, Sample-efficient parameter exploration of the powder film drying process using experiment-based Bayesian optimization, Sci. Rep., 2022, 12, 1615.
75. Y. K. Wakabayashi, T. Otsuka, Y. Taniyasu, H. Yamamoto and H. Sawada, Improved adaptive sampling method utilizing Gaussian process regression for prediction of spectral peak structures, Appl. Phys. Express, 2018, 11, 112401.
76. T. Ueno, H. Hino, A. Hashimoto, Y. Takeichi, M. Sawada and K. Ono, Adaptive design of an X-ray magnetic circular dichroism spectroscopy experiment with Gaussian process modelling, npj Comput. Mater., 2018, 4, 4.
77. G. De Ath, J. E. Fieldsend and R. M. Everson, What do you Mean? The Role of the Mean Function in Bayesian Optimisation, GECCO ’20 Companion, 2020, 1623–1631.
78. M. C. Kennedy and A. O’Hagan, Bayesian calibration of computer models, J. R. Stat. Soc. Ser. B Stat. Methodol., 2001, 63, 425–464.
79. T. Boltz, J. L. Martinez, C. Xu, K. R. L. Baker, Z. Zhu, J. Morgan, R. Roussel, D. Ratner, B. Mustapha and A. L. Edelen, Leveraging prior mean models for faster Bayesian optimization of particle accelerators, Sci. Rep., 2025, 15, 12232.
80. Z. Hu, J. Li, H. Shao, R. Chen, L. Filipovic and L. Li, Physics-informed Bayesian optimization framework for etching rate and surface roughness co-optimization, 2025 International Conference on Simulation of Semiconductor Processes and Devices (SISPAD), 2025, 1–4.
81. J. C. Hughes, D. J. York, K. G. Yager, C. O. Osuji and R. J. Composto, Using Flory–Huggins-informed human-in-the-loop Bayesian optimization to map the phase diagram of polymer blends, Digit. Discov., 2026, 5, 1675–1688.
82. Z. Wang, G. E. Dahl, K. Swersky, C. Lee, Z. Nado, J. Gilmer, J. Snoek and Z. Ghahramani, Pre-trained Gaussian processes for Bayesian optimization, J. Mach. Learn. Res., 2024, 25, 1–83.
83. D. Garreau, W. Jitkrittum and M. Kanagawa, Large sample analysis of the median heuristic, arXiv, 2017, arXiv:1707.07269.
84. Z. Zhai, Y. Lu, L. Ouyang, J. Lu, W.-L. Ding, B. Cao, Y. Wang, F. Huo, Q. Zhao, W. Wang, S. Zhang and H. He, Modulating product selectivity in lignin electroreduction with a robust metallic glass catalyst, Nat. Commun., 2025, 16, 3414.

85. S. Lee, T. L. Meyer, S. Park, T. Egami and H. N. Lee, Growth control of the oxidation state in vanadium oxide thin films, Appl. Phys. Lett., 2014, 105, 223515.
86. M. Golalikhani, Q. Lei, R. U. Chandrasena, L. Kasaei, H. Park, J. Bai, P. Orgiani, J. Ciston, G. E. Sterbinsky, D. A. Arena, P. Shafer, E. Arenholz, B. A. Davidson, A. J. Millis, A. X. Gray and X. X. Xi, Nature of the metal-insulator transition in few-unit-cell-thick LaNiO3 films, Nat. Commun., 2018, 9, 2206.
87. D. J. C. MacKay, Bayesian nonlinear modeling for the prediction competition, ASHRAE Trans., 1994, 100, 1053–1062.
88. B. Xu, T. A. Sultana, K. Kitai, J. Guo, T. Seki, R. Tamura, K. Tsuda and J. Shiomi, Experiment-in-loop interactive optimization of polymer composites for “5G-and-beyond” communication technologies, Mater. Horiz., 2025, 12, 3332–3340.
89. R. Masui, U. Lee, R. Nakayama and T. Hitosugi, Sparse modeling based Bayesian optimization for experimental design, Mater. Adv., 2025, 6, 4062–4069.
90. J. Snoek, K. Swersky, R. S. Zemel and R. P. Adams, Input warping for Bayesian optimization of non-stationary functions, Proc. ICML, 2014, 32, 1674–1682.
91. R. B. Gramacy and H. K. H. Lee, Bayesian treed Gaussian process models with an application to computer modeling, J. Am. Stat. Assoc., 2008, 103, 1119–1130.
92. R. B. Gramacy and D. W. Apley, Local Gaussian process approximation for large computer experiments, J. Comput. Graph. Stat., 2015, 24, 561–578.
93. Y. Saatçi, R. D. Turner and C. E. Rasmussen, Gaussian process change point models, Proc. ICML, 2010, 927–934.
94. A. G. Wilson and R. P. Adams, Gaussian process kernels for pattern discovery and extrapolation, Proc. ICML, 2013, 28, 1067–1075.
95. N. Durrande, J. Hensman, M. Rattray and N. D. Lawrence, Detecting periodicities with Gaussian processes, PeerJ Comput. Sci., 2016, 2, e50.
96. V. Rajpaul, S. Aigrain, M. A. Osborne, S. Reece and S. J. Roberts, A Gaussian process framework for modelling stellar activity signals in radial velocity data, Mon. Not. R. Astron. Soc., 2015, 452, 2269–2291.
97. B. A. Nicholson and S. Aigrain, Quasi-periodic Gaussian processes for stellar activity: from physical to kernel parameters, Mon. Not. R. Astron. Soc., 2022, 515, 5251–5265.
98. A. Koulali and P. J. Clarke, Modelling quasi-periodic signals in geodetic time-series using Gaussian processes, Geophys. J. Int., 2021, 226, 1705–1714.
99. A. G. Wilson, Z. Hu, R. Salakhutdinov and E. P. Xing, Deep kernel learning, Proc. AISTATS, 2016, 51, 370–378.
100. S. Kiyohara and Y. Kumagai, Bayesian optimization with Gaussian processes assisted by deep learning for material designs, J. Phys. Chem. Lett., 2025, 16, 5244–5251.
101. A. Biswas, Y. Liu, N. Creange, Y.-C. Liu, S. Jesse, J.-C. Yang, S. V. Kalinin, M. A. Ziatdinov and R. K. Vasudevan, A dynamic Bayesian optimized active recommender system for curiosity-driven partially human-in-the-loop automated experiments, npj Comput. Mater., 2024, 10, 29.
102. Y. Liu, R. K. Vasudevan, K. P. Kelley, H. Funakubo, M. Ziatdinov and S. V. Kalinin, Learning the right channel in multimodal imaging: automated experiment in piezoresponse force microscopy, npj Comput. Mater., 2023, 9, 34.
103. S. W. Ober, C. E. Rasmussen and M. van der Wilk, The promises and pitfalls of deep kernel learning, Proc. UAI, 2021, 161, 1206–1216.
104. E. V. Bonilla, K. M. Chai and C. K. I. Williams, Multi-task Gaussian process prediction, Adv. Neural Inf. Process. Syst., 2007, 20, 153–160.
105. M. A. Álvarez, L. Rosasco and N. D. Lawrence, Kernels for vector-valued functions: a review, Found. Trends Mach. Learn., 2012, 4, 195–266.
106. A. Shah and Z. Ghahramani, Pareto frontier learning with expensive correlated objectives, Proc. ICML, 2016, 48, 1919–1927.
107. A. Solomou, G. Zhao, S. Boluki, J. K. Joy, X. Qian, I. Karaman, R. Arróyave and D. C. Lagoudas, Multi-objective Bayesian materials discovery: application on the discovery of precipitation-strengthened NiTi shape memory alloys through micromechanical modeling, Mater. Des., 2018, 160, 810–827.
108. M. Harada, H. Takeda, S. Suzuki, K. Nakano, N. Tanibata, M. Nakayama, M. Karasuyama and I. Takeuchi, Bayesian-optimization-guided experimental search of NASICON-type solid electrolytes for all-solid-state Li-ion batteries, J. Mater. Chem. A, 2020, 8, 15103–15109.

109. M. Karasuyama, H. Kasugai, T. Tamura and K. Shitara, Computational design of stable and highly ion-conductive materials using multi-objective Bayesian optimization: case studies on diffusion of oxygen and lithium, Comput. Mater. Sci., 2020, 184, 109927.
110. D. Khatamsaz, B. Vela and R. Arróyave, Multi-objective Bayesian alloy design using multi-task Gaussian processes, Mater. Lett., 2023, 351, 135067.
111. J. Knowles, ParEGO: a hybrid algorithm with on-line landscape approximation for expensive multiobjective optimization problems, IEEE Trans. Evol. Comput., 2006, 10, 50–66.
112. B. Paria, K. Kandasamy and B. Póczos, A flexible framework for multi-objective Bayesian optimization using random scalarizations, Proc. UAI, 2020, 115, 766–776.
113. T. Chugh, Scalarizing functions in Bayesian multiobjective optimization, Proc. IEEE Congress Evol. Comput., 2020, 1–8.
114. G. Derringer and R. Suich, Simultaneous optimization of several response variables, J. Qual. Technol., 1980, 12, 214–219.
115. O. Mamun, M. Bause and B. S. M. E. Hai, Accelerated development of multi-component alloys in discrete design space using Bayesian multi-objective optimisation, Mach. Learn.: Sci. Technol., 2025, 6, 015001.
116. S. Das, N. Oyeniran, J. Walter, A. Gesch and C. Hu, Bayesian optimization of grain-boundary segregation in high-entropy alloys, npj Comput. Mater., 2025, 11, 371.
117. R. Nakao, Y. K. Wakabayashi, T. Otsuka, T. Otsuka, T. Sato and H. Sugiyama, Gas flow sequence optimization of InGaAs/InP structure by using machine learning, JSAP Annual Meetings Extended Abstracts, 2020, DOI: 10.11470/jsapmeeting.2020.2.0_2028.
118. H. C. Herbol, M. Poloczek and P. Clancy, Cost-effective materials discovery: Bayesian optimization across multiple information sources, Mater. Horiz., 2020, 7, 2113–2123.
119. S. Kusakawa, S. Takeno, Y. Inatsu, K. Kutsukake, S. Iwazaki, T. Nakano, T. Ujihara, M. Karasuyama and I. Takeuchi, Bayesian optimization for cascade-type multistage processes, Neural Comput., 2022, 34, 2408–2431.
120. L. Torresi and P. Friederich, Multi-stage Bayesian optimisation for dynamic decision-making in self-driving labs, Digit. Discov., 2026, 5, 1900–1912.
121. F. Pedregosa, G. Varoquaux, A. Gramfort, V. Michel, B. Thirion, O. Grisel, M. Blondel, P. Prettenhofer, R. Weiss, V. Dubourg, J. Vanderplas, A. Passos, D. Cournapeau, M. Brucher, M. Perrot and E. Duchesnay, Scikit-learn: Machine Learning in Python, J. Mach. Learn. Res., 2011, 12, 2825–2830.
122. M. Balandat, B. Karrer, D. R. Jiang, S. Daulton, B. Letham, A. G. Wilson and E. Bakshy, BoTorch: A Framework for Efficient Monte-Carlo Bayesian Optimization, Adv. Neural Inf. Process. Syst., 2020, 33, 21524–21538.
123. M. M. Noack, G. S. Doerk, R. Li, J. K. Streit, R. A. Vaia, K. G. Yager and M. Fukuto, Autonomous materials discovery driven by Gaussian process regression with inhomogeneous measurement noise and anisotropic kernels, Sci. Rep., 2020, 10, 17663.
124. G. M. White, A. P. Siegel and A. Tovar, Optimizing thermoplastic starch film with heteroscedastic Gaussian processes in Bayesian experimental design framework, Materials, 2024, 17, 5345.
125. B. N. Slautin, Y. Liu, J. Dec, V. V. Shvartsman, D. C. Lupascu, M. A. Ziatdinov and S. V. Kalinin, Measurements with noise: Bayesian optimization for co-optimizing noise and property discovery in automated experiments, Digit. Discov., 2025, 4, 1066–1074.
126. S. M. A. A. Alvi, M. Mulukutla, N. Flores, D. Khatamsaz, J. Janssen, D. Perez, D. Allaire, V. Attari and R. Arróyave, Accurate and uncertainty-aware multi-task prediction of HEA properties using prior-guided deep Gaussian processes, npj Comput. Mater., 2025, 11, 306.
127. L. Breiman, Random forests, Mach. Learn., 2001, 45, 5–32.
128. J. H. Friedman, Greedy function approximation: a gradient boosting machine, Ann. Stat., 2001, 29, 1189–1232.
129. T. Chen and C. Guestrin, XGBoost: a scalable tree boosting system, Proceedings of the 22nd ACM SIGKDD International Conference on Knowledge Discovery and Data Mining, 2016, 785–794.
130. F. Hutter, H. H. Hoos and K. Leyton-Brown, Sequential model-based optimization for general algorithm configuration, Learning and Intelligent Optimization, Springer, Berlin, Heidelberg, 2011, 507–523.
131. B. Lei, T. Q. Kirk, A. Bhattacharya, D. Pati, X. Qian, R. Arroyave and B. K. Mallick, Bayesian optimization with adaptive surrogate models for automated experimental design, npj Comput. Mater., 2021, 7, 194.
132. H. Zhang, W. W. Chen, A. Iyer, D. W. Apley and W. Chen, Uncertainty-aware mixed-variable machine learning for materials design, Sci. Rep., 2022, 12, 19760.

133. N. Meinshausen, Quantile regression forests, J. Mach. Learn. Res., 2006, 7, 983–999.
134. J. H. Friedman and B. E. Popescu, Predictive learning via rule ensembles, Ann. Appl. Stat., 2008, 2, 916–954.
135. Y. L. Li, T. G. J. Rudner and A. G. Wilson, A study of Bayesian neural network surrogates for Bayesian optimization, Proc. ICLR, 2024.
136. Y.-F. Lim, C. K. Ng, U. S. Vaitesswar and K. Hippalgaonkar, Extrapolative Bayesian optimization with Gaussian process and neural network ensemble surrogate models, Adv. Intell. Syst., 2021, 3, 2100101.
137. S. I. Allec and M. Ziatdinov, Active and transfer learning with partially Bayesian neural networks for materials and chemicals, Digit. Discov., 2025, 4, 1284–1297.
138. A. Damianou and N. D. Lawrence, Deep Gaussian processes, Proc. AISTATS, 2013, 31, 207–215.
139. D. Russo, B. Van Roy, A. Kazerouni, I. Osband and Z. Wen, A tutorial on Thompson sampling, Found. Trends Mach. Learn., 2018, 11, 1–96.
140. P. Hennig and C. J. Schuler, Entropy search for information-efficient global optimization, J. Mach. Learn. Res., 2012, 13, 1809–1837.
141. Z. Wang and S. Jegelka, Max-value entropy search for efficient Bayesian optimization, Proc. ICML, 2017, 70, 3627–3635.
142. P. Frazier, W. Powell and S. Dayanik, The knowledge-gradient policy for correlated normal beliefs, INFORMS J. Comput., 2009, 21, 599–613.
143. R. Shibukawa, S. Matsuda, K. Nakamura, R. Tamura and K. Tsuda, Collecting diverse near-optimal samples via nested Thompson sampling, npj Comput. Mater., 2026, 12, 197.
144. A. Biswas, A. N. Morozovska, M. Ziatdinov, E. A. Eliseev and S. V. Kalinin, Multi-objective Bayesian optimization of ferroelectric materials with interfacial control for memory and energy storage applications, J. Appl. Phys., 2021, 130, 204102.
145. D. Khatamsaz, B. Vela, P. Singh, D. D. Johnson, D. Allaire and R. Arróyave, Multi-objective materials Bayesian optimization with active learning of design constraints: design of ductile refractory multi-principal-element alloys, Acta Mater., 2022, 236, 118133.
146. R. Waelder, W. Kim, M. A. Pitt, J. I. Myung and B. Maruyama, Multi-objective Bayesian optimization of carbon nanotube yield and diameter control at synthesis, APL Mach. Learn., 2025, 3, 026114.
147. J. I. Myung, J. R. Deneault, J. Chang, I. Kang, B. Maruyama and M. A. Pitt, Multi-objective Bayesian optimization: a case study in material extrusion, Digit. Discov., 2025, 4, 464–476.
148. K. Sawai, T. Ogawa, T.-T. Chen, F. Sun and Y. Adachi, Improvement of strength–ductility balance in tempered martensite steel using multi-objective Bayesian optimization, ISIJ Int., 2025, 65, 2260–2268.
149. T. Hastings, M. Mulukutla, D. Khatamsaz, D. Salas, W. Xu, D. Lewis, N. Person, M. Skokan, B. Miller, J. Paramore, B. Butler, D. Allaire, V. Attari, I. Karaman, G. Pharr, A. Srivastava and R. Arróyave, Accelerated multi-objective alloy discovery through efficient Bayesian methods: application to the FCC high entropy alloy space, Acta Mater., 2025, 297, 121173.
150. T. Hastie, R. Tibshirani and J. Friedman, The elements of statistical learning, 2nd ed., Springer-Verlag, New York, NY, 2009.
151. K. P. Murphy, Probabilistic machine learning: An introduction, MIT Press, Cambridge, MA, 2022.
152. H. Lee, R. Gramacy, C. Linkletter and G. Gray, Optimization subject to hidden constraints via statistical emulation, Pac. J. Optim., 2011, 7, 467–478.
153. F. Bachoc, C. Helbert and V. Picheny, Gaussian process optimization with failures: classification and convergence proof, J. Glob. Optim., 2020, 78, 483–506.
154. Y. Sui, A. Gotovos, J. Burdick and A. Krause, Safe exploration for optimization with Gaussian processes, Proc. ICML, 2015, 37, 997–1005.
155. F. Berkenkamp, A. Krause and A. P. Schoellig, Bayesian optimization with safety constraints: safe and automatic parameter tuning in robotics, Mach. Learn., 2023, 112, 3713–3747.
156. Y. Sui, V. Zhuang, J. Burdick and Y. Yue, Stagewise safe Bayesian optimization with Gaussian processes, Proc. ICML, 2018, 80, 4781–4789.
157. K. Osada, K. Kutsukake, J. Yamamoto, S. Yamashita, T. Kodera, Y. Nagai, T. Horikawa, K. Matsui, I. Takeuchi and T. Ujihara, Adaptive Bayesian optimization for epitaxial growth of Si thin films under various constraints, Mater. Today Commun., 2020, 25, 101538.

158. R. J. Hickman, G. Tom, Y. Zou, M. Aldeghi and A. Aspuru-Guzik, Anubis: Bayesian optimization with unknown feasibility constraints for scientific experimentation, Digit. Discov., 2025, 4, 2104–2122.
159. F. Wang, M. Parhizkar, A. Harker and M. Edirisinghe, Constrained composite Bayesian optimization for rational synthesis of polymeric particles, Digit. Discov., 2025, 4, 3066–3077.
160. V. Sabanza-Gil, R. Barbano, D. Pacheco Gutiérrez, J. S. Luterbacher, J. M. Hernández-Lobato, P. Schwaller and L. Roch, Best practices for multi-fidelity Bayesian optimization in materials and molecular research, Nat. Comput. Sci., 2025, 5, 572–581.
161. J. P. Folch, R. M. Lee, B. Shafei, D. Walz, C. Tsay, M. van der Wilk and R. Misener, Combining multi-fidelity modelling and asynchronous batch Bayesian optimization, Comput. Chem. Eng., 2023, 172, 108194.
162. J. Kim, M. Li, Y. Li, A. Gómez, O. Hinder and P. W. Leu, Multi-BOWS: multi-fidelity multi-objective Bayesian optimization with warm starts for nanophotonic structure design, Digit. Discov., 2024, 3, 381–391.
163. A. Palizhati, S. B. Torrisi, M. Aykol, S. K. Suram, J. S. Hummelshøj and J. H. Montoya, Agents for sequential learning using multiple-fidelity data, Sci. Rep., 2022, 12, 4694.
164. R. Jacobs, P. E. Goins and D. Morgan, Role of multifidelity data in sequential active learning materials discovery campaigns: case study of electronic bandgap, Mach. Learn.: Sci. Technol., 2023, 4, 045060.
165. E. H. Lee, V. Perrone, C. Archambeau and M. Seeger, Cost-aware Bayesian optimization, arXiv, 2020, arXiv:2003.10870.
166. T. Yonezu, T. Tamura, I. Takeuchi and M. Karasuyama, Knowledge-transfer-based cost-effective search for interface structures: A case study on fcc-Al [110] tilt grain boundary, Phys. Rev. Mater., 2018, 2, 113802.
167. S. M. A. A. Alvi, B. Vela, V. Attari, J. Janssen, D. Perez, D. Allaire and R. Arróyave, Deep Gaussian process-based cost-aware batch Bayesian optimization for complex materials design campaigns, npj Comput. Mater., 2026, 12, 105.
168. K. Kandasamy, A. Krishnamurthy, J. Schneider and B. Póczos, Parallelised Bayesian optimisation via Thompson sampling, Proc. AISTATS, 2018, 84, 133–142.
169. J. Wang, S. C. Clark, E. Liu and P. I. Frazier, Parallel Bayesian global optimization of expensive functions, Oper. Res., 2020, 68, 1850–1865.
170. J. Wu and P. I. Frazier, The Parallel Knowledge Gradient Method for Batch Bayesian Optimization, Adv. Neural Inf. Process. Syst., 2016, 29, 3126–3134.
171. J. D. Paramore, T. Hastings, B. G. Butler, M. T. Hurst, D. O. Lewis, E. Norris, B. Barkai, J. Cline, B. Miller, J. Cortes, I. Karaman, G. M. Pharr and R. Arróyave, Two-shot optimization of compositionally complex refractory alloys, Acta Mater., 2025, 289, 120820.
172. J. W. Sin, S. L. Chau, R. P. Burwood, K. Püntener, R. Bigler and P. Schwaller, Highly parallel optimisation of chemical reactions through automation and machine intelligence, Nat. Commun., 2025, 16, 6464.
173. C. C. Rupnow, B. P. MacLeod, M. Mokhtari, K. Ocean, K. E. Dettelbach, D. Lin, F. G. L. Parlane, H. N. Chiu, M. B. Rooney, C. E. B. Waizenegger, E. I. de Hoog, A. Soni and C. P. Berlinguette, A self-driving laboratory optimizes a scalable process for making functional coatings, Cell Rep. Phys. Sci., 2023, 4, 101411.
174. A. Alvi, B. Ru, J.-P. Calliess, S. Roberts and M. A. Osborne, Asynchronous batch Bayesian optimisation with improved local penalisation, Proc. ICML, 2019, 97, 253–262.
175. F. Strieth-Kalthoff, H. Hao, V. Rathore, J. Derasp, T. Gaudin, N. H. Angello, M. Seifrid, E. Trushina, M. Guy, J. Liu, X. Tang, M. Mamada, W. Wang, T. Tsagaantsooj, C. Lavigne, R. Pollice, T. C. Wu, K. Hotta, L. Bodo, S. Li, M. Haddadnia, A. Wołos, R. Roszak, C. T. Ser, C. Bozal-Ginesta, R. J. Hickman, J. Vestfrid, A. Aguilar-Granda, E. L. Klimareva, R. C. Sigerson, W. Hou, D. Gahler, S. Lach, A. Warzybok, O. Borodin, S. Rohrbach, B. Sanchez-Lengeling, C. Adachi, B. A. Grzybowski, L. Cronin, J. E. Hein, M. D. Burke and A. Aspuru-Guzik, Delocalized, asynchronous, closed-loop discovery of organic laser emitters, Science, 2024, 384, eadk9227.
176. H. Wen, J. Zeitler and C. Rupnow, Search strategies for self-driving laboratories with pending experiments, arXiv, 2023, arXiv:2312.03466.
177. Y. Zhang, D. W. Apley and W. Chen, Bayesian optimization for materials design with mixed quantitative and qualitative variables, Sci. Rep., 2020, 10, 4924.
178. R. Jenatton, C. Archambeau, J. González and M. Seeger, Bayesian optimization with tree-structured dependencies, Proc. ICML, 2017, 70, 1655–1664.
179. X. Ma and M. Blaschko, Additive tree-structured covariance function for conditional parameter spaces in Bayesian optimization, Proc. AISTATS, 2020, 108, 1015–1025.

180. I. Bogunovic, J. Scarlett and V. Cevher, Time-varying Gaussian process bandit optimization, Proc. AISTATS, 2016, 51, 314–323.
181. R. E. Nahory, M. A. Pollack, W. D. Johnston and R. L. Barns, Band gap versus composition and demonstration of Vegard's law for In1−xGaxAsyP1−y lattice matched to InP, Appl. Phys. Lett., 1978, 33, 659–661.
182. R. Hill, Spin-orbit splitting in semiconductor alloys, J. Phys. C: Solid State Phys., 1974, 7, 516.
183. S. Park, Y. Shim, J. Hur, S. Ji, D. Jeon, J. M. Yuk and C.-W. Lee, Element mapping-based Bayesian optimization framework enabling direct materials design: a case study on NASICON-type cathode materials, npj Comput. Mater., 2026, 12, 92.
184. M. A. Ziatdinov, A. Ghosh and S. V. Kalinin, Physics makes the difference: Bayesian optimization and active learning via augmented Gaussian process, Mach. Learn.: Sci. Technol., 2022, 3, 015003.
185. M. S. Priyadarshini, O. Romiluyi, Y. Wang, K. Miskin, C. Ganley and P. Clancy, PAL 2.0: a physics-driven Bayesian optimization framework for material discovery, Mater. Horiz., 2024, 11, 781–791.
186. S. L. Sanchez, E. Foadian, M. Ziatdinov, J. Yang, S. V. Kalinin, Y. Liu and M. Ahmadi, Physics-driven discovery and bandgap engineering of hybrid perovskites, Digit. Discov., 2024, 3, 1577–1590.
187. D. Wen, V. Tucker and M. S. Titus, Bayesian optimization acquisition functions for accelerated search of cluster expansion convex hull of multi-component alloys, npj Comput. Mater., 2024, 10, 210.
188. C. Hardcastle, R. O'Mullan, R. Arróyave and B. Vela, Physics-informed Gaussian process classification for constraint-aware alloy design, Digit. Discov., 2025, 4, 1884–1900.
189. X. Li, Y. Che, L. Chen, T. Liu, K. Wang, L. Liu, H. Yang, E. O. Pyzer-Knapp and A. I. Cooper, Sequential closed-loop Bayesian optimization as a guide for organic molecular metallophotocatalyst formulation discovery, Nat. Chem., 2024, 16, 1286–1294.
190. X. Hong, X. Wang, S. J. Harris, H. Zhao, J. Meng, Q. Jia, Q. Zhao, K. Xu, Q. Pang and B. Jiang, Deep active learning and knowledge transfer for rapid discovery of lithium metal battery electrolytes, Nat. Commun., 2026, 17, 5146.
191. J. Moon, W. Beker, M. Siek, J. Kim, H. S. Lee, T. Hyeon and B. A. Grzybowski, Active learning guides discovery of a champion four-metal perovskite oxide for oxygen evolution electrocatalysis, Nat. Mater., 2024, 23, 108–115.
192. Z. Rao, P.-Y. Tung, R. Xie, Y. Wei, H. Zhang, A. Ferrari, T. P. C. Klaver, F. Körmann, P. T. Sukumar, A. K. da Silva, Y. Chen, Z. Li, D. Ponge, J. Neugebauer, O. Gutfleisch, S. Bauer and D. Raabe, Machine learning-enabled high-entropy alloy discovery, Science, 2022, 378, 78–85.
193. M. Higashiwaki, K. Sasaki, A. Kuramata, T. Masui and S. Yamakoshi, Gallium oxide ($Ga_2O_3$) metal–semiconductor field-effect transistors on single-crystal $\beta$-$Ga_2O_3$ (010) substrates, Appl. Phys. Lett., 2012, 100, 013504.
194. M. Higashiwaki, K. Sasaki, H. Murakami, Y. Kumagai, A. Koukitu, A. Kuramata, T. Masui and S. Yamakoshi, Recent progress in $Ga_2O_3$ power devices, Semicond. Sci. Technol., 2016, 31, 034001.
195. M. Yu, I. A. Moses, W. F. Reinhart and S. Law, Multimodal machine learning analysis of GaSe molecular beam epitaxy growth conditions, ACS Appl. Mater. Interfaces, 2025, 17, 34707–34716.
196. R. Trice, M. Yu, E. Welp, M. Applegate, W. F. Reinhart and S. Law, Machine-learning-guided polymorph selection in molecular beam epitaxy of $In_2Se_3$, J. Vac. Sci. Technol. A, 2026, 44, 042708.
197. M. Zhong, K. Tran, Y. Min, C. Wang, Z. Wang, C.-T. Dinh, P. De Luna, Z. Yu, A. S. Rasouli, P. Brodersen, S. Sun, O. Voznyy, C.-S. Tan, M. Askerka, F. Che, M. Liu, A. Seifitokaldani, Y. Pang, S.-C. Lo, A. Ip, Z. Ulissi and E. H. Sargent, Accelerated discovery of $CO_2$ electrocatalysts using active machine learning, Nature, 2020, 581, 178–183.
198. B. J. Shields, J. Stevens, J. Li, M. Parasram, F. Damani, J. I. Martinez Alvarado, J. M. Janey, R. P. Adams and A. G. Doyle, Bayesian reaction optimization as a tool for chemical synthesis, Nature, 2021, 590, 89–96.
199. J. A. Bennett, N. Orouji, M. Khan, S. Sadeghi, J. Rodgers and M. Abolhasani, Autonomous reaction Pareto-front mapping with a self-driving catalysis laboratory, Nat. Chem. Eng., 2024, 1, 240–250.
200. S. Liu, D. Dong, Z. Xiao, D. Bedrov, K. Xu, L. Xing and W. Li, Machine learning in electrolyte design: Accelerating discovery and multidimensional optimization for next-generation batteries, Chem, 2026, 12, 103024.
201. J. Li, J. Fleetwood, W. B. Hawley and W. Kays, From Materials to Cell: State-of-the-Art and Prospective Technologies for Lithium-Ion Battery Electrode Processing, Chem. Rev., 2022, 122, 903–956.

202. A. Dave, J. Mitchell, S. Burke, H. Lin, J. Whitacre and V. Viswanathan, Autonomous optimization of non-aqueous Li-ion battery electrolytes via robotic experimentation and machine learning coupling, Nat. Commun., 2022, 13, 5454.
203. Y. Iwasaki, J. Hwang, Y. Sakuraba, M. Kotsugi and Y. Igarashi, Efficient autonomous material search method combining ab initio calculations, autoencoder, and multi-objective Bayesian optimization, Sci. Technol. Adv. Mater.: Methods, 2022, 2, 365–371.
204. Y. Iwasaki, R. Toyama, T. Yamazaki, Y. Igarashi, M. Kotsugi and Y. Sakuraba, Autonomous search for half-metallic materials with B2 structure, Sci. Technol. Adv. Mater.: Methods, 2024, 4, 2403966.
205. K. Nakamura, D. Furuya, Y. Miura, T. Yamazaki, A. L. Foggiatto, Y. Iwasaki and M. Kotsugi, Efficient discovery of large magnetic anisotropy with thermodynamically stable materials via multi-objective Bayesian optimization, Sci. Technol. Adv. Mater.: Methods, 2025, 5, 2485025.
206. D. Furuya, T. Miyashita, Y. Miura, Y. Iwasaki and M. Kotsugi, Autonomous synthesis system integrating theoretical, informatics, and experimental approaches for large-magnetic-anisotropy materials, Sci. Technol. Adv. Mater.: Methods, 2022, 2, 280–293.
207. R. Toyama, R. Tamura, S. Matsuda, Y. Iwasaki and Y. Sakuraba, Autonomous closed-loop exploration of composition-spread films for the anomalous Hall effect, npj Comput. Mater., 2025, 11, 329.
208. Y. K. Wakabayashi, Y. Nonaka, Y. Takeda, S. Sakamoto, K. Ikeda, Z. Chi, G. Shibata, A. Tanaka, Y. Saitoh, H. Yamagami, M. Tanaka, A. Fujimori and R. Nakane, Cation distribution and magnetic properties in ultrathin (Ni1−xCox)Fe2O4 (x = 0–1) layers on Si(111) studied by soft x-ray magnetic circular dichroism, Phys. Rev. Mater., 2018, 2, 104416.
209. Y. K. Wakabayashi, Y. Nonaka, Y. Takeda, S. Sakamoto, K. Ikeda, Z. Chi, G. Shibata, A. Tanaka, Y. Saitoh, H. Yamagami, M. Tanaka, A. Fujimori and R. Nakane, Electronic structure and magnetic properties of magnetically dead layers in epitaxial CoFe2O4/Al2O3/Si(111) films studied by x-ray magnetic circular dichroism, Phys. Rev. B, 2017, 96, 104410.
210. Z. Xi, Z. Wang, C. Guo, K. Xu, W. Zhao, Z. Li, J. Bao, H. Zhou, C. Zou, H. Huang and D. Zhou, Active learning in latent spaces enables rapid inverse design of ferroelectric ceramics for energy storage, Nat. Commun., 2026, 17, 4281.
211. B. Keimer and J. E. Moore, The physics of quantum materials, Nat. Phys., 2017, 13, 1045–1055.
212. Y. K. Wakabayashi, Y. Krockenberger, T. Otsuka, H. Sawada, Y. Taniyasu and H. Yamamoto, Intrinsic physics in magnetic Weyl semimetal SrRuO3 films addressed by machine-learning-assisted molecular beam epitaxy, Jpn. J. Appl. Phys., 2023, 62, SA0801.
213. S. Kaneta-Takada, Y. K. Wakabayashi, Y. Krockenberger, T. Nomura, Y. Kohama, S. A. Nikolaev, H. Das, H. Irie, K. Takiguchi, S. Ohya, M. Tanaka, Y. Taniyasu and H. Yamamoto, High-mobility two-dimensional carriers from surface Fermi arcs in magnetic Weyl semimetal films, npj Quantum Mater., 2022, 7, 102.
214. S. Kaneta-Takada, Y. K. Wakabayashi, Y. Krockenberger, S. Ohya, M. Tanaka, Y. Taniyasu and H. Yamamoto, Thickness-dependent quantum transport of Weyl fermions in ultra-high-quality $SrRuO_3$ films, Appl. Phys. Lett., 2021, 118, 092408.
215. Y. K. Wakabayashi, M. Kobayashi, Y. Krockenberger, T. Takeda, K. Yamagami, H. Yamamoto and Y. Taniyasu, Single monolayer ferromagnetic perovskite $SrRuO_3$ with high conductivity and strong ferromagnetism, Appl. Phys. Lett., 2025, 127, 022402.
216. H. Horiuchi, Y. Araki, Y. K. Wakabayashi, J. Ieda, M. Yamanouchi, Y. Sato, S. Kaneta-Takada, Y. Taniyasu, H. Yamamoto, Y. Krockenberger, M. Tanaka and S. Ohya, Single-layer spin-orbit-torque magnetization switching due to spin Berry curvature generated by minute spontaneous atomic displacement in a Weyl oxide, Adv. Mater., 2025, 37, 2416091.
217. S. Kaneta-Takada, Y. K. Wakabayashi, Y. Krockenberger, H. Irie, S. Ohya, M. Tanaka, Y. Taniyasu and H. Yamamoto, Scattering-dependent transport of $SrRuO_3$ films: From Weyl fermion transport to hump-like Hall effect anomaly, Phys. Rev. Mater., 2023, 7, 054406.
218. H. Shiratani, Y. K. Wakabayashi, Y. Krockenberger, M. Kobayashi, K. Yamagami, T. Takeda, S. Ohya, M. Tanaka and Y. Taniyasu, Intrinsic low-spin state and strain-tunable anomalous Hall scaling in high-quality $SrRuO_3$ (111) films, APL Mater., 2026, 14, 051113.
219. Y. K. Wakabayashi, M. Kobayashi, Y. Seki, Y. Kotani, T. Ohkochi, K. Yamagami, M. Kitamura, Y. Taniyasu, Y. Krockenberger and H. Yamamoto, Magnetic anisotropy driven by ligand in 4d transition-metal oxide $SrRuO_3$, APL Mater., 2024, 12, 041119.

220. Y. Seki, Y. K. Wakabayashi, T. Takeda, K. Inagaki, S.-i. Fujimori, Y. Takeda, A. Fujimori, Y. Taniyasu, H. Yamamoto, Y. Krockenberger, M. Tanaka and M. Kobayashi, Correlated ligand electrons in the transition-metal oxide $SrRuO_3$, Phys. Rev. Lett., 2025, 135, 046402.
221. M. Sim, M. G. Vakili, F. Strieth-Kalthoff, H. Hao, R. J. Hickman, S. Miret, S. Pablo-García and A. Aspuru-Guzik, ChemOS 2.0: An orchestration architecture for chemical self-driving laboratories, Matter, 2024, 7, 2959–2977.
222. Y. Fei, B. Rendy, R. Kumar, O. Dartsi, H. P. Sahasrabuddhe, M. J. McDermott, Z. Wang, N. J. Szymanski, L. N. Walters, D. Milsted, Y. Zeng, A. Jain and G. Ceder, AlabOS: a Python-based reconfigurable workflow management framework for autonomous laboratories, Digit. Discov., 2024, 3, 2275–2288.
223. W. Zhang, L. Hao, V. Lai, R. Corkery, J. Jessiman, J. Zhang, J. Liu, Y. Sato, M. Politi, M. E. Reish, R. Greenwood, N. Depner, J. Min, R. El-khawaldeh, P. Prieto, E. Trushina and J. E. Hein, IvoryOS: an interoperable web interface for orchestrating Python-based self-driving laboratories, Nat. Commun., 2025, 16, 5182.
224. N. J. Szymanski, C. J. Bartel, Y. Zeng, M. Diallo, H. Kim and G. Ceder, Adaptively driven X-ray diffraction guided by machine learning for autonomous phase identification, npj Comput. Mater., 2023, 9, 31.
225. J. Liu, D. Zhang, D. Yu, M. Ren and J. Xu, Machine learning powered ellipsometry, Light Sci. Appl., 2021, 10, 55.
226. H. Castro, J. Galvis and S. Castro, Automated setup for Van der Pauw Hall measurements, IEEE Trans. Instrum. Meas., 2011, 60, 198–205.
227. R. Toyama, Y. Iwasaki, P. D. Kulkarni, H. Suto, T. Nakatani and Y. Sakuraba, High-throughput materials exploration system for the anomalous Hall effect using combinatorial experiments and machine learning, npj Comput. Mater., 2025, 11, 269.
228. S. B. Harris, P. T. Gemperline, C. M. Rouleau, R. K. Vasudevan and R. B. Comes, Deep learning with reflection high-energy electron diffraction images to predict cation ratio in Sr2xTi2(1−x)O3 thin films, Nano Lett., 2025, 25, 5867–5874.
229. T. C. Kaspar, S. Akers, H. W. Sprueill, A. H. Ter-Petrosyan, J. A. Bilbrey, D. Hopkins, A. Harilal, J. Christudasjustus, P. Gemperline and R. B. Comes, Machine-learning-enabled on-the-fly analysis of RHEED patterns during thin film deposition by molecular beam epitaxy, J. Vac. Sci. Technol. A, 2025, 43, 032702.
230. S. Schaefer, D. Febba, K. Egbo, G. Teeter, A. Zakutayev and B. Tellekamp, Rapid screening of molecular beam epitaxy conditions for monoclinic (InxGa1−x)2O3 alloys, J. Mater. Chem. A, 2024, 12, 5508–5519.
231. J. Wagner, K. Münnemann, T. Specht, H. Hasse and F. Jirasek, Deep set model for the automated NMR fingerprinting of unknown mixtures, Digit. Discov., 2026, 5, 1325–1339.
232. K. A. Severson, P. M. Attia, N. Jin, N. Perkins, B. Jiang, Z. Yang, M. H. Chen, M. Aykol, P. K. Herring, D. Fraggedakis, M. Z. Bazant, S. J. Harris, W. C. Chueh and R. D. Braatz, Data-driven prediction of battery cycle life before capacity degradation, Nat. Energy, 2019, 4, 383–391.
233. R. J. Hickman, M. Sim, S. Pablo-García, G. Tom, I. Woolhouse, H. Hao, Z. Bao, P. Bannigan, C. Allen, M. Aldeghi and A. Aspuru-Guzik, Atlas: a brain for self-driving laboratories, Digit. Discov., 2025, 4, 1006–1029.
234. R. Tamura, K. Tsuda and S. Matsuda, NIMS-OS: an automation software to implement a closed loop between artificial intelligence and robotic experiments in materials science, Sci. Technol. Adv. Mater.: Methods, 2023, 3, 2232297.
235. W. S. Lee, G. W. Yoffe, D. G. Schlom and J. S. Harris, Accurate measurement of MBE substrate temperature, J. Cryst. Growth, 1991, 111, 131–135.
236. K. Haberland, A. Kaluza, M. Zorn, M. Pristovsek, H. Hardtdegen, M. Weyers, J.-T. Zettler and W. Richter, Real-time calibration of wafer temperature, growth rate and composition by optical in situ techniques during AlxGa1−xAs growth in MOVPE, J. Cryst. Growth, 2002, 240, 87–97.
237. A. Krause and C. S. Ong, Contextual Gaussian process bandit optimization, Adv. Neural Inf. Process. Syst., 2011, 24, 2447–2455.
238. Y. Deng, X. Zhou, B. Kim, A. Tewari, A. Gupta and N. Shroff, Weighted Gaussian process bandits for non-stationary environments, Proc. AISTATS, 2022, 151, 6909–6932.
239. S. Muroga, H. Nakajima, T. Shimizu, K. Kobashi and K. Hata, Multimodal machine learning for integrating heterogeneous analytical systems, Anal. Sci., 2026, 42, 635–643.
240. Y. Wu, M. Ding, H. He, Q. Wu, S. Jiang, P. Zhang and J. Ji, A versatile multimodal learning framework bridging multiscale knowledge for material design, npj Comput. Mater., 2025, 11, 276.

241. D. A. Boiko, R. MacKnight, B. Kline and G. Gomes, Autonomous chemical research with large language models, Nature, 2023, 624, 570–578.
242. A. Cissé, M. E. Cooper, M. Zhu, X. Evangelopoulos and A. I. Cooper, Can we automate scientific reasoning in closed-loop experiments using large language models?, Digit. Discov., 2026, 5, 1132–1160.
243. A. Vriza, M. H. Prince, T. Zhou, H. Chan and M. J. Cherukara, Operating advanced scientific instruments with AI agents that learn on the job, npj Comput. Mater., 2026, 12, 160.
244. I. Mandal, J. Soni, M. Zaki, M. M. Smedskjaer, K. Wondraczek, L. Wondraczek, N. N. Gosvami and N. M. A. Krishnan, Evaluating large language model agents for automation of atomic force microscopy, Nat. Commun., 2025, 16, 9104.
245. V. Aglietti, X. Lu, A. Paleyes and J. González, Causal Bayesian optimization, Proc. AISTATS, 2020, 108, 3155–3164.
246. J. Zhang, L. Cammarata, C. Squires, T. P. Sapsis and C. Uhler, Active learning for optimal intervention design in causal models, Nat. Mach. Intell., 2023, 5, 1066–1075.